\documentclass[preprint,preprintnumbers,amsmath,amssymb]{revtex4-1}

\usepackage{graphicx}% Include figure files
\usepackage{dcolumn}% Align table columns on decimal point
\usepackage{bm}% bold math
\usepackage{subfigure}
\usepackage{amsmath}
\usepackage[version=3]{mhchem}
\usepackage{multirow}
\usepackage{xr}

\usepackage{float}
\usepackage{bm}
\usepackage{longtable}
\usepackage{array}

\newcolumntype{P}[1]{>{\centering\arraybackslash}p{#1}}
\newcolumntype{M}[1]{>{\centering\arraybackslash}m{#1}}
\begin{document}

\title{A Performance and Cost Assessment of Machine Learning Interatomic Potentials}

\author{Yunxing Zuo}
\affiliation{Department of NanoEngineering, University of California San Diego, 9500 Gilman Dr, Mail Code 0448, La Jolla, CA 92093-0448, United States}
\author{Chi Chen}
\affiliation{Department of NanoEngineering, University of California San Diego, 9500 Gilman Dr, Mail Code 0448, La Jolla, CA 92093-0448, United States}
\author{Xiangguo Li}
\affiliation{Department of NanoEngineering, University of California San Diego, 9500 Gilman Dr, Mail Code 0448, La Jolla, CA 92093-0448, United States}
\author{Zhi Deng}
\affiliation{Department of NanoEngineering, University of California San Diego, 9500 Gilman Dr, Mail Code 0448, La Jolla, CA 92093-0448, United States}
\author{Yiming Chen}
\affiliation{Department of NanoEngineering, University of California San Diego, 9500 Gilman Dr, Mail Code 0448, La Jolla, CA 92093-0448, United States}
\author{J{\"o}rg Behler}
\affiliation{Universit{\"a}t G{\"o}ttingen, Institut f{\"u}r Physikalische Chemie, Theoretische Chemie, Tammannstra{\ss}e 6, 37077 G{\"o}ttingen, Germany}
\author{G{\'a}bor Cs{\'a}nyi}
\affiliation{Department of Engineering, University of Cambridge, Trumpington Street, Cambridge, CB2 1PZ, United Kingdom}
\author{Alexander V. Shapeev}
\affiliation{Skolkovo Institute of Science and Technology, Skolkovo Innovation Center, Building 3, Moscow, 143026, Russia}
\author{Aidan P. Thompson}
\affiliation{Center for Computing Research, Sandia National Laboratories, Albuquerque, New Mexico 87185, United States}
\author{Mitchell A. Wood}
\affiliation{Center for Computing Research, Sandia National Laboratories, Albuquerque, New Mexico 87185, United States}
\author{Shyue Ping Ong}
\email{ongsp@eng.ucsd.edu}
\affiliation{Department of NanoEngineering, University of California San Diego, 9500 Gilman Dr, Mail Code 0448, La Jolla, CA 92093-0448, United States}

\begin{abstract}

Machine learning of the quantitative relationship between local environment descriptors and the potential energy surface of a system of atoms has emerged as a new frontier in the development of interatomic potentials (IAPs). Here, we present a comprehensive evaluation of ML-IAPs based on four local environment descriptors --- Behler-Parrinello symmetry functions, smooth overlap of atomic positions (SOAP), the Spectral Neighbor Analysis Potential (SNAP) bispectrum components, and moment tensors --- using a diverse data set generated using high-throughput density functional theory (DFT) calculations. The data set comprising bcc (Li, Mo) and fcc (Cu, Ni) metals and diamond group IV semiconductors (Si, Ge) is chosen to span a range of crystal structures and bonding. All descriptors studied show excellent performance in predicting energies and forces far surpassing that of classical IAPs, as well as predicting properties such as elastic constants and phonon dispersion curves. We observe a general trade-off between accuracy and the degrees of freedom of each model, and consequently computational cost. We will discuss these trade-offs in the context of model selection for molecular dynamics and other applications.

\end{abstract}

\maketitle

\section{Introduction}

A fundamental input for atomistic simulations of materials is a description of the potential energy surface (PES) as a function of atomic positions. While quantum mechanics-based descriptions, such as those based on Kohn-Sham density functional theory (DFT)\cite{Kohn_Sham_equation, Sham_bandgap}, are accurate and transferable across chemistries, their high cost and poor scaling (typically $O(n_e^3)$ or higher, where $n_e$ is the number of electrons)\cite{DFT_Scaling_N^3, DFT_Scaling_N^4, DFT_Scaling_N^5} limits simulations to $\sim$ 1000 atoms and hundreds of picoseconds. Hence, large-scale and long-time simulations traditionally rely on interatomic potentials (IAPs), which to date are in most cases empirical parameterizations of the PES based on physical functional forms that depend only on the atomic degrees of freedom.\cite{uff, Dreiding, reaxff} IAPs gain linear scaling with respect to the number of atoms at the cost of accuracy and transferability.

In recent years, a modern alternative has emerged in the form of machine-learned IAPs (ML-IAPs), where the PES is described as a function of local environment descriptors that are invariant to translation, rotation and permutation of homonuclear atoms\cite{nnp_perspective, GAP_SOAP}. Examples of such potentials include the high-dimensional neural network potential (NNP)\cite{NNP_Si, NNP_review_2011}, the Gaussian approximation potential (GAP)\cite{GAP_SOAP, GAP_tungsten, GAP_Fe}, the Spectral Neighbor Analysis Potential (SNAP)\cite{SNAP, SNAP_Mo, SNAP_Nickel_Copper, eSNAP_Li3N}, moment tensor potentials (MTP),\cite{MTP, MTP_activelearning, MTP_alloy_search} among others\cite{AGNI_Al, AGNI_Al_force, AGNI_Al_outlook, oganov_ML_potentials, devita_ML_potentials, chmiela_ML_potentials, ML-HK-molecule-IAPs, rupp_ML_potentials, Wang_ML_potentials, Alex_representation}. A typical approach to training such potentials involves the generation of a sufficiently large and diverse data set of atomic configurations with corresponding energies, forces and stresses from DFT calculations, which are then used in the training of the ML-IAP based on one or several target metrics, such as minimizing the mean absolute or squared errors in predicted energies, forces, stresses or derived properties (e.g. elastic constants). ML-IAPs have been shown to be a remarkable improvement over traditional IAPs, in general, achieving near-DFT accuracy in predicting energies and forces across diverse chemistries and atomic configurations. Nevertheless, a critical gap that remains is a rigorous assessment of the relative strengths and weaknesses of ML-IAPs across a standardized data set, similar to what has been done for classical IAPs.\cite{comparison_classical_Si, o1986comparison, godet2003comparison}.

In this work, we present a comprehensive performance comparison of four major ML-IAPs --- GAP, MTP, NNP and SNAP. The four IAPs were evaluated in terms of their accuracy in reproducing DFT energies and forces, as well as material properties such as the equations of state, lattice parameter and elastic constants. An attempt was also made to assess the training data requirements of each ML-IAP and the relative computational cost based on the best-available current implementations. To ensure a fair comparison, standardized DFT data sets of six elements (Li, Mo, Cu, Ni, Si and Ge) with the same training/test sampling and similar fitting approaches was used. The elements were chosen to span diverse chemistries and bonding, e.g., bcc and fcc metals, main group and transition metals, and group IV semiconductors.

\section{Methods}

\subsection{Machine learning interatomic potentials}

The four ML-IAPs investigated in this work have already been extensively discussed in previous works and reviews\cite{NNP_Si, nnp_perspective, NNP_review_2011, NNP_review_2015, NNP_review_2017, GAP_SOAP, GAP_tungsten, GAP_Fe, GAP_multi_component, SNAP, SNAP_Mo, SNAP_Nickel_Copper, eSNAP_Li3N, SNAP_W-Be, MTP, MTP_activelearning, MTP_alloy_search}. All ML-IAPs express the potential energy as a sum of atomic energies that are a function of the local environment around each atom, but differ in the descriptors for these local environments and the ML approach/functional expression used to map the descriptors to the potential energy. The detailed formalism of all four ML-IAPs are provided in the Supplementary Information. Here, only a concise summary of the key concepts and model parameters behind the ML-IAPs in chronological order of development, is provided to aid the reader in following the remainder of this paper.

\begin{enumerate}

 \item \textbf{High-dimensional neural network potential (NNP).} The NNP uses atom-centered symmetry functions (ACSF)\cite{Atom-centered-symmetry-functions} to represent the atomic local environments and fully connected neural networks to describe the PES with respect to symmetry functions\cite{NNP_Si, NNP_review_2011}. A separate neural network is used for each atom. The neural network is defined by the number of hidden layers and the nodes in each layer, while the descriptor space is given by the following symmetry functions:
    \begin{eqnarray}
        G_{i}^{\rm atom, \rm rad} & = & \sum_{j\neq i}^{N_{\rm atom}} e^{-\eta(R_{ij}-R_{s})^{2}} \cdot f_{c}(R_{ij}), \\
        G_{i}^{\rm atom, \rm ang} & = & 2^{1-\zeta}\sum_{j,k \neq i}^{N_{\rm atom}} (1 + \lambda \cos \theta_{ijk})^{\zeta} \cdot e^{-\eta^{\prime}(R_{ij}^{2}+R_{ik}^{2}+R_{jk}^{2})}\cdot f_{c}(R_{ij}) \cdot f_{c}(R_{ik}) \cdot f_{c}(R_{jk}),
    \end{eqnarray}
    where $R_{ij}$ is the distance between atom $i$ and neighbor atom $j$, $\eta$ is the width of the Gaussian and $R_{s}$ is the position shift over all neighboring atoms within the cutoff radius $R_{c}$, $\eta^{\prime}$ is the width of the Gaussian basis and $\zeta$ controls the angular resolution. $f_{c}(R_{ij})$ is a cutoff function, defined as follows:
    \begin{equation}
        f_{c}(R_{ij}) =  
        \begin{cases} 
        \ 0.5 \cdot [\cos{(\frac{\pi R_{ij}}{R_{c}})} + 1], & \text{for } R_{ij} \leq R_{c} \\
        \ 0.0,                                      & \text{for } R_{ij} > R_{c}.
        \end{cases}
    \end{equation}
    These hyperparameters were optimized to minimize the mean absolute errors of energies and forces for each chemistry. The NNP model has shown great performance for Si\cite{NNP_Si}, \ce{TiO2}\cite{NNP_TiO2}, water\cite{NNP_van_water} and solid-liquid interfaces\cite{NNP_water_ZnO_interface}, metal-organic frameworks\cite{NNP_MOF}, and has been extended to incorporate long-range electrostatics for ionic systems such as \ce{ZnO}\cite{NNP_ZnO} and \ce{Li3PO4}\cite{NNP_Li3PO4}.
    
    \item \textbf{Gaussian Approximation Potential (GAP).} The GAP calculates the similarity between atomic configurations based on a smooth-overlap of atomic positions (SOAP)\cite{GAP_review_2010, GAP_SOAP} kernel, which is then used in a Gaussian process model. In SOAP, the Gaussian-smeared atomic neighbor densities $\rho_{i}(\boldsymbol{R})$ are expanded in spherical harmonics as follows:
    \begin{eqnarray}
        \rho_{i}(\boldsymbol{R}) = \sum_{j} f_{c}(R_{ij}) \cdot \exp(-\frac{|\boldsymbol{R} - \boldsymbol{R}_{ij}|^{2}}{2\sigma_{\rm atom}^{2}}) = \sum_{nlm} c_{nlm} \ g_{n}(R) Y_{lm}(\hat{\boldsymbol{R}}),
    \end{eqnarray}
    The spherical power spectrum vector, which is in turn the square of expansion coefficients,
    \begin{eqnarray}
        p_{n_{1}n_{2}l} (\boldsymbol{R}_{i}) = \sum_{m = -l}^{l} c_{n_{1}lm}^{\ast} c_{n_{2}lm},
    \end{eqnarray}
    can be used to construct the SOAP kernel while raised to a positive integer power $\zeta$ (which is 4 in present case) to accentuate the sensitivity of the kernel\cite{GAP_SOAP},
    \begin{eqnarray}
        \boldsymbol{K}(\boldsymbol{R}, \boldsymbol{R^{\prime}}) = \sum_{n_{1}n_{2}l} (p_{n_{1}n_{2}l}(\boldsymbol{R}) p_{n_{1}n_{2}l}(\boldsymbol{R^{\prime}}))^{\zeta},
    \end{eqnarray}
    In the above equations, $\sigma_{\rm atom}$ is a smoothness controlling the Gaussian smearing, and $n_{\text{max}}$ and $l_{\text{max}}$ determine the maximum powers for radial components and angular components in spherical harmonics expansion, respectively\cite{GAP_SOAP}. These hyperparameters, as well as the number of reference atomic configurations used in Gaussian process, are optimized in the fitting procedure to obtain optimal performance. The GAP has been developed for transition metals\cite{GAP_tungsten, GAP_Fe}, main group elements\cite{GAP_Boron, GAP_C, GAP_graphene}, diamond semiconductors\cite{GAP_Si, GAP_Si_network} as well as multi-component systems\cite{GAP_multi_component}.

    \item \textbf{Spectral Neighbor Analysis Potential (SNAP).} The SNAP uses the coefficients of the bispectrum of the atomic neighbor density functions\cite{GAP_SOAP} as descriptors. In the original formulation of SNAP, a linear model between energies and bispectrum components is assumed\cite{SNAP}. Recently, a quadratic model (denoted as qSNAP in this work)\cite{qSNAP} has been developed, which extends the linear SNAP energy model to include all distinct pairwise products of bispectrum components. In this work, both linear and quadratic SNAP models were investigated. The key hyperparameters influencing model performance are the cutoff radius and $J_{\text{max}}$, which limits the indices $j_{1}$, $j_{2}$, $j$ in Clebsch-Gordan coupling coefficients $H\substack{j m m\prime \\ j_{1} m_{1} m_{1}^{\prime} \\ j_{2} m_{2} m_{2}^{\prime}}$ in construction of the bispectrum components:
    \begin{equation}
        \begin{split}
        B_{j_{1}, j_{2}, j} &= \sum_{m_{1}, m_{1}^{\prime}=-j_{1}}^{j_{1}} 
                       \sum_{m_{2}, m_{2}^{\prime}=-j_{2}}^{j_{2}}
                       \sum_{m, m^{\prime}=-j}^{j} (u_{m, m^{\prime}}^{j})^{\ast}
                    \\&
        \times H\substack{j m m\prime \\ j_{1} m_{1} m_{1}^{\prime} \\ j_{2} m_{2} m_{2}^{\prime}} u_{m_{1}, m_{1}^{\prime}}^{j_{1}} u_{m_{2}, m_{2}^{\prime}}^{j_{2}},
        \end{split}
    \end{equation}
    where $u_{m, m^{\prime}}^{j}$ are coefficients in 4-dimensional hyper-spherical harmonics expansion of neighbor density function:
    \begin{equation}
        \rho_{i}(\textbf{R}) = \sum_{j=0}^{\infty} \sum_{m, m^{\prime}=-j}^{j} u_{m, m^{\prime}}^{j} U_{m, m^{\prime}}^{j},
    \end{equation}
    The SNAP model as well as qSNAP model has demonstrated great success in transition metals\cite{SNAP, SNAP_Mo, SNAP_Nickel_Copper, qSNAP} as well as binary systems\cite{SNAP_Nickel_Copper, eSNAP_Li3N, SNAP_W-Be}.
    
    \item \textbf{Moment Tensor Potential (MTP)}. The MTP\cite{MTP} devises rotationally-covariant tensors
    \begin{equation}
        M_{\mu, \nu}(\boldsymbol{R}) = \sum_{j} f_{\mu}(R_{ij})\ \underbrace{\boldsymbol{R}_{ij} \otimes \dots \otimes \boldsymbol{R}_{ij}}_{\nu\ times},
    \end{equation}
    to describe the atomic local environments. Here $f_{\mu}$ are the radial functions, and $\boldsymbol{R}_{ij} \otimes \dots \otimes \boldsymbol{R}_{ij}$ are tensors of rank $\nu$ encoding angular information about the atomic environment. The rank $\nu$ can be large enough to approximate any arbitrary interactions. MTP then contracts these tensors to a scalar yields rotationally-invariant basis functions, and applies linear regression to correlate the energies with the basis functions. The performance of MTP is controlled by the polynomial power-like metric, which defines what tensors and how many times are contracted. The MTP model has been successfully applied to metals\cite{MTP, MTP_activelearning, MTP_diffusion}, boron\cite{MTP_boron}, binary and ternary alloys\cite{MTP_alloy_search} as well as gas-phase chemical reactions\cite{MTP_RPMD}.

\end{enumerate}

\subsection{DFT Data Sets}

A comprehensive DFT data set was generated for six elements - Li, Mo, Ni, Cu, Si and Ge. These elements were chosen to span a variety of chemistries (main group metal, transition metal and semiconductor), crystal structures (bcc, fcc, and diamond) and bonding types (metallic and covalent). For each element, we generated a set of structures with diverse coverage of atomic local environment space, as follows: 
\begin{enumerate}
    \item[(1)] The ground-state crystal for each element.
    \item[(2)] Strained structures constructed by applying strains of $-10\%$ to 10\% at 2\% intervals to the bulk supercell in six different modes, as described in the work by \citet{Elastic-Property-MaterialsProject}. The supercells used are the $3\times 3 \times 3$, $3\times 3 \times 3$ and $2 \times 2\times2$ of the conventional bcc, fcc and diamond unit cells, respectively. 
    \item[(3)] Slab structures up to a maximum Miller index of three, including (100), (110), (111), (210), (211), (310), (311), (320), (321), (322), (331), and (332), as obtained from the Crystalium database\cite{Surface_energies, crystallium}.
    \item[(4)] \textit{NVT} \textit{ab initio} molecular dynamics (AIMD) simulations of the bulk supercells (similar to those in (2)) performed at 300 K and $0.5\times$, $0.9\times$, $1.5\times$, $2.0\times$ of the melting point of each element. A total of 20 snapshots were obtained from each AIMD simulation at an interval of 0.1 ps, unless otherwise stated.
    \item[(5)] \textit{NVT} AIMD simulations of the bulk supercells (similar to those in (2)) with a single vacancy performed at 300 K and $2.0\times$ of the melting point of each element. A total of 40 snapshots were obtained from each AIMD simulation at an interval of 0.1 ps, unless otherwise stated.
\end{enumerate}

All DFT calculations were carried out using the Vienna \textit{ab initio} simulation package (VASP)\cite{VASP} version 5.4.1 within the projector augmented wave approach\cite{PAW}. The Perdew-Burke-Ernzerhof (PBE) generalized gradient approximation (GGA)\cite{GGA} was adopted for the exchange-correlation functional. The kinetic-energy cutoff was set to 520 eV and the $k$-point mesh was $4\times4\times4$ for the Mo, Ni, Cu, Si and Ge supercells, and $3\times3\times3$ for the Li supercells. The electronic energy and atomic force components were converged to within $10^{-5}$ eV and 0.02 eV/{\AA}, respectively, in line with previous works\cite{SNAP_Mo, SNAP_Nickel_Copper}. The AIMD simulations were carried out with a single $\Gamma$ \textit{k} point and were non-spin-polarized, but static calculations using the same parameters as the rest of the data were carried out on the snapshots to obtain consistent energies and forces. All structure manipulations and analyses of DFT computations were carried out using Python Materials Genomics (Pymatgen)\cite{pymatgen} library, and the automation of calculations was performed using the Fireworks software\cite{fireworks}.

\subsection{Optimization scheme}

\begin{figure}[H]
    \centering
    \includegraphics[width=0.9\textwidth]{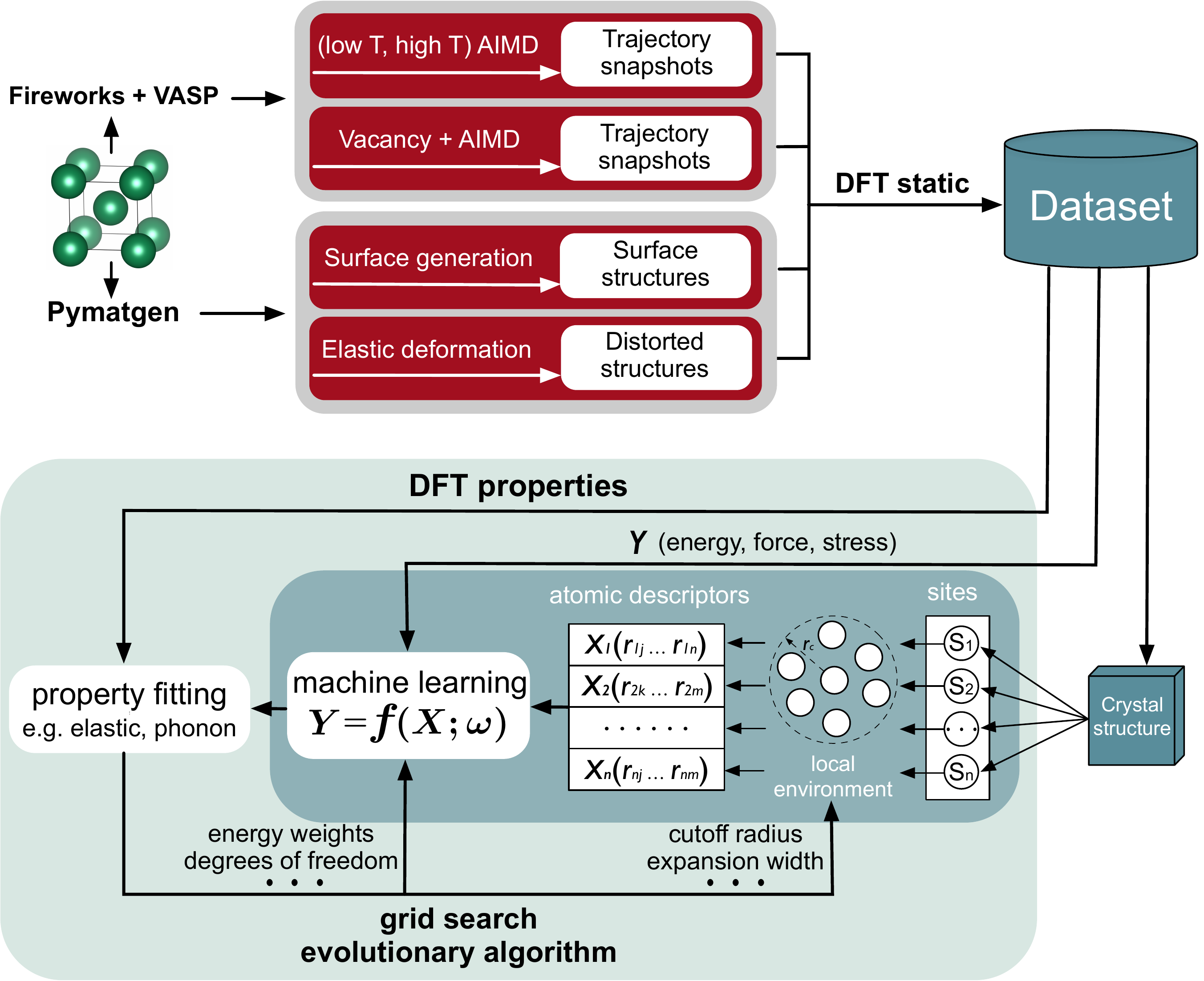}
    \caption{Machine-learning interatomic potential development workflow.}
    \label{fig:workflow}
\end{figure}

Figure \ref{fig:workflow} provides an overview of the general data generation and potential development scheme. The training data set was first generated via DFT static calculations on the four categories of structures. The optimization procedure comprised two loops. In the inner loop, sampled structures in the database were transformed into atomic descriptors (e.g., bispectrum components for SNAP and symmetry functions for NNP), which were then fed into the corresponding ML model together with the DFT energies, forces, and stresses as the targets of training. The data was apportioned into training and test sets with a 90:10 split. The parameters of the ML models were optimized during the training process. In the outer loop, the ML model trained in the inner loop was used to predict basic material properties (e.g., elastic tensors), and the differences between the predicted and reference values were then used to determine the optimal hyperparameters for each ML-IAPs. In this work, we adopted a combination of the grid search algorithm and differential evolution algorithm to perform hyperparameters optimization for different ML-IAPs.

\subsection{Data and code availability}
To facilitate the reuse and reproduction of our results, the code, data and optimized ML models in this work are published open-source on Github (\url{https://github.com/materialsvirtuallab/mlearn}). The code includes high-level Python interfaces for ML-IAPs development as well as LAMMPS material properties calculators. 

\section{Results}

\subsection{Optimized model parameters}

The optimized coefficients and hyperparameters for each ML-IAP are reported in Supplementary Information (see Table S1--Table S10). Here, we will limit our discussions to a parameter that is common to all ML-IAPs - the cutoff radius - and present a convergence study of each ML-IAP with the number of degrees of freedom of the model.

The cutoff radius determines the maximum range of interatomic interactions, and hence, has a critical effect on the prediction performance of ML-IAPs. Table \ref{table:optimized_rc} provides the optimized cutoff radii of different ML-IAPs across different chemistries. Different ML-IAPs yield similar optimized cutoff radii for the same elemental system. The optimized cutoff radii are between the second nearest neighbor (2NN) and 3NN distance for fcc elements (Cu, Ni), between 3NN and 4NN distances for the bcc (Li, Mo) and diamond (Ge and Si) elements. These observations are consistent with those from previous traditional and ML IAP development efforts, where typically 2NN interactions are found to suffice for fcc metals,\cite{2NN_fcc, effective_charge_fcc} while contributions from 3NN cannot be ignored for bcc metals\cite{3NN_bcc_Li, 3NN_V, GAP_tungsten, MTP, MTP_activelearning} and diamond systems\cite{GAP_molecule_bulk, MEAM_semiconductors}.

\begin{center}
\begin{table}
\begin{tabular}{  M{3cm}  M{1.2cm}  M{1.2cm}  M{1.2cm}  M{1.2cm}  M{1.2cm}  M{1.2cm} }
\hline
\hline
\noalign{\smallskip}
    &   \multicolumn{2}{c}{\textbf{fcc}}    &   \multicolumn{2}{c}{\textbf{bcc}}    &   \multicolumn{2}{c}{\textbf{diamond}} \\
\noalign{\smallskip}
cutoff radius ({\AA})    &   Ni  &   Cu  &   Li  &   Mo  &   Si  &   Ge  \\
\hline
\noalign{\smallskip}
GAP     &   3.9 &   3.9 &   4.8     &   5.2 &   5.4 &   5.4     \\
\noalign{\smallskip}
MTP     &   4.0 &   3.9 &   5.1     &   5.2 &   4.7 &   5.1     \\
\noalign{\smallskip}
NNP     &   3.9 &   4.1 &   5.2     &   5.2 &   5.2 &   5.6     \\
\noalign{\smallskip}
SNAP     &   3.9 &   4.1 &   5.1     &   4.6 &   4.9 &   5.5     \\
\noalign{\smallskip}
qSNAP     &   3.8 &   3.9 &   5.1     &   5.2 &   4.8 &   4.9     \\
\noalign{\smallskip}
\hline
\end{tabular}
\caption{Optimized cutoff radius for each element for each ML-IAP.}
\label{table:optimized_rc}
\end{table}
\end{center}

\begin{figure}
    \centering
    \includegraphics[width=0.7\textwidth]{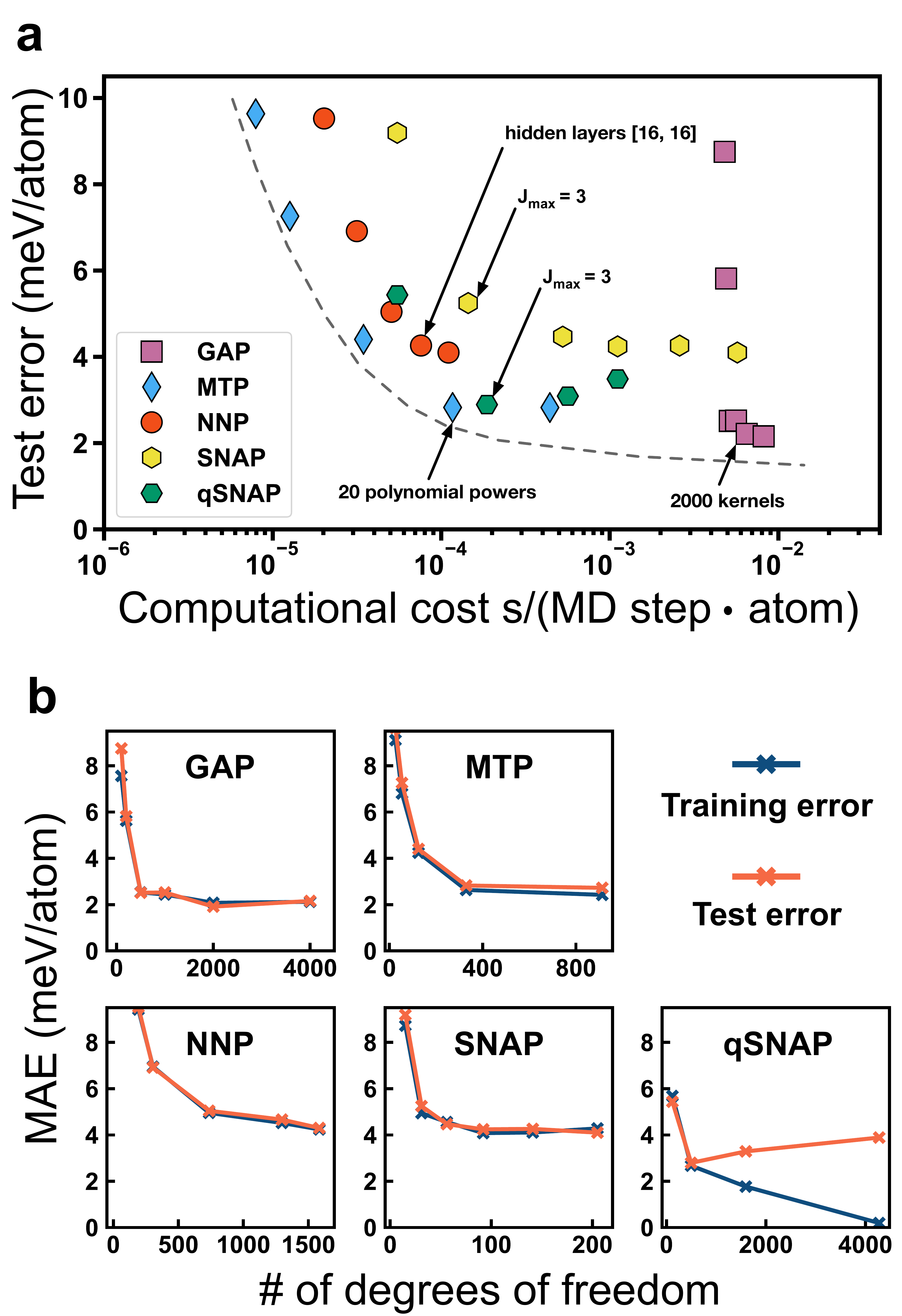}
    \caption{\label{fig:time_benchmark} (a) Test error versus computational cost for the Mo system. The black dash line indicates a Pareto frontier representing an optimal trade-off between accuracy and computational cost. Timings were performed by LAMMPS calculations on a single CPU core of Intel i7-6850k 3.6 GHz. Black arrows denote the ``optimal'' configuration for each ML-IAP that was used in subsequent comparisons. (b) Plots of the training and test errors versus the number of degrees of freedom for each ML-IAP.}
\end{figure}

The number of degrees of freedom (DOF), e.g., the number of weights and biases for the NNP and number of representative points in GAP, has a strong effect on the accuracy and computational cost of each ML-IAP. Figure \ref{fig:time_benchmark} illustrates the trade-off between computational cost and test error under varying DOFs for each fitted Mo ML-IAP. Similar results are obtained for other systems (see Figure S7--Figure S11). It should be noted that the relative computational costs are based on the most efficient available implementations\cite{n2p2, GAP_SOAP, MTP, SNAP, qSNAP} of each ML-IAP at this time in LAMMPS\cite{LAMMPS} and performed on a single CPU core of Intel i7-6850k 3.6 GHz with $18\times18\times18$ bulk supercell containing 23,328 atoms for Mo system. Future implementations may improve on these results. A Pareto frontier is drawn in Figure \ref{fig:time_benchmark}a to represent points at which better accuracy can only be attained at the price of greater computational cost\cite{pareto_curve}, and the black arrows indicate ``optimal'' configurations for each model in terms of the trade-off between test error and computational cost. These ``optimal'' configurations were used for subsequent accuracy comparisons in energies, forces and properties. We find that the ``optimal'' MTP, NNP, SNAP and qSNAP models tend to be two orders of magnitude less computationally expensive than the ``optimal'' GAP model. The MTP models generally lie close to the Pareto frontier, exhibiting an excellent balance between model accuracy and computational efficiency. For the SNAP and qSNAP models, the descriptor space (i.e., bispectrum components) is determined by the parameter $J_{max}$. We find that the rate-limiting step is the calculation of bispectrum and the computation of quadratic terms in qSNAP has only a small effect on the computational cost.\cite{qSNAP}. However, we find that the substantial expansion in the number of fitted coefficients in the qSNAP model results in a greater likelihood of over-fitting, especially for $J_{\text{max}} > 3$ (see Figure \ref{fig:time_benchmark}b). For the GAP model, the computational cost is linearly related to the number of kernels used in Gaussian process regression.\cite{GAP_tungsten}.

\subsection{Accuracy in energies and forces}

\begin{figure}
    \centering
    \subfigure[Mean absolute errors in predicted energies]{\label{fig:energy_error}\includegraphics[width=0.48\textwidth]{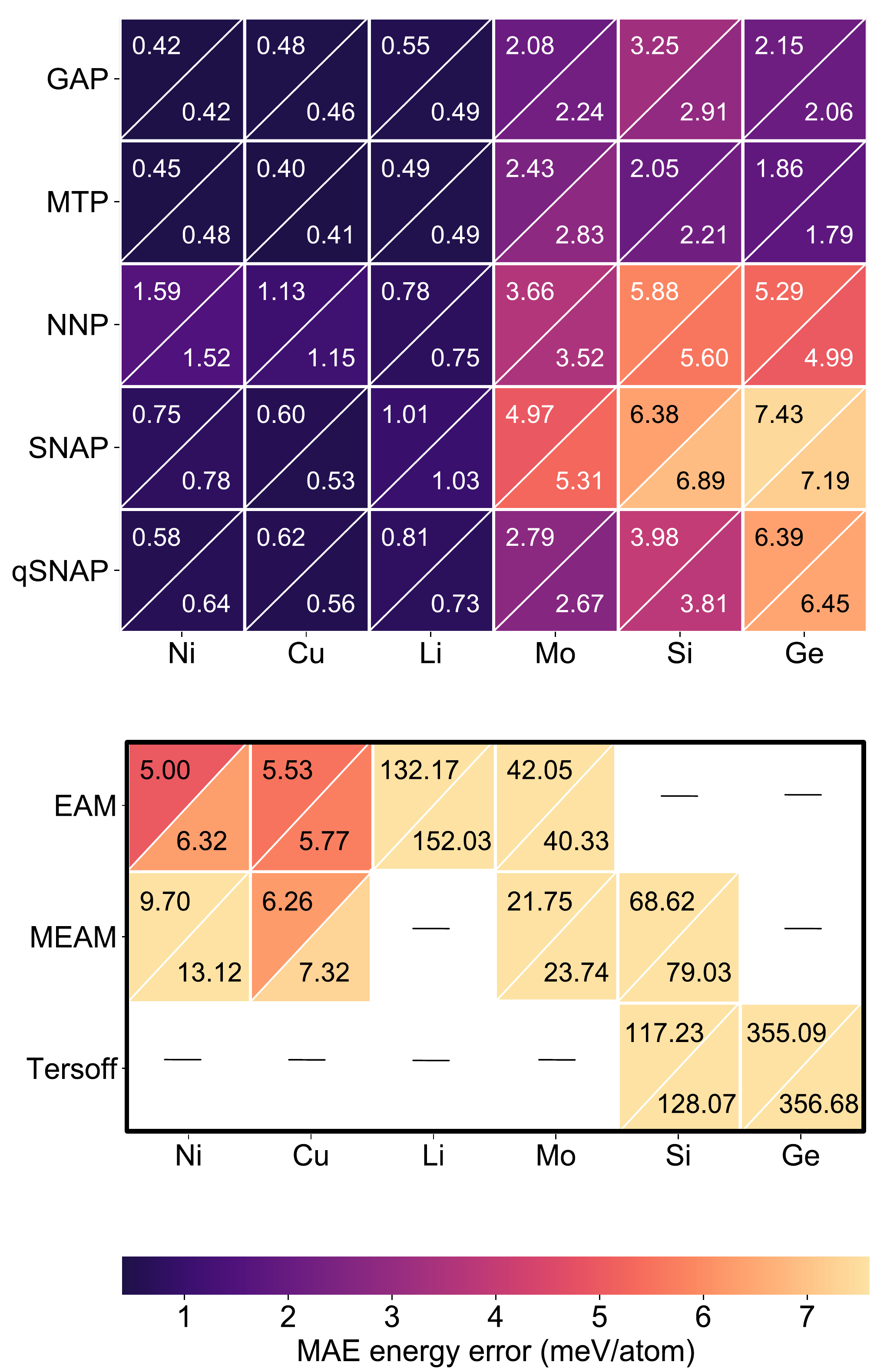}}
    \subfigure[Mean absolute errors in predicted forces]{\label{fig:force_error}\includegraphics[width=0.48\textwidth]{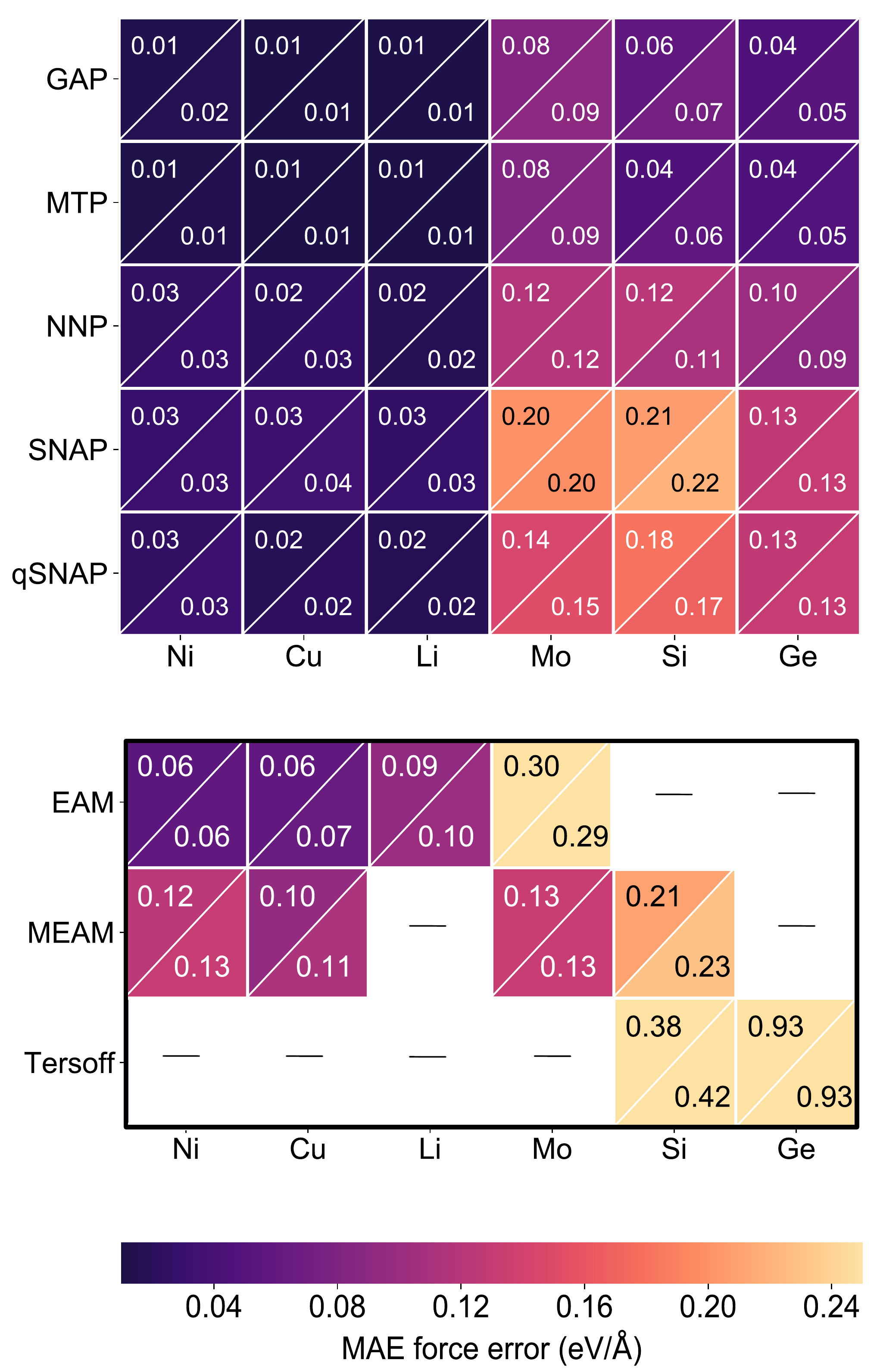}}
    \caption{Mean absolute errors in (a) predicted energies (b) predicted forces for all four ML-IAPs as well as traditional IAPs (EAM\cite{EAM_Mo, EAM_Li}, MEAM\cite{MEAM_Cu, MEAM_Mo, MEAM_Si}, Tersoff\cite{tersoff_Si, tersoff_Ge}). The upper left and lower right triangles within each cell represent training and test errors, respectively.}
    \label{fig:error}
\end{figure}

Figure \ref{fig:error} provides a comparison of the MAEs in energies and forces for the four ML-IAPs and best-available classical IAPs relative to DFT. All ML-IAPs show extremely good performance across all elements studied, achieving MAEs in energies and forces that are far lower than best-available traditional IAPs for each element. It should be noted that differences in MAEs between ML-IAPs are on the scale of meV atom$^{-1}$ in energies and 0.1 eV \AA$^{-1}$ in forces; hence, any subsequent discussion on the relative performances of the ML-IAPs should be viewed in the context that even the largest differences in accuracy between the ML-IAPs are already close to the limits of DFT error. In all cases, the training and test errors are similar, indicating no over-fitting for the optimized ML-IAPs.

The GAP and MTP models generally have the lowest MAEs in energies and forces. The highest MAEs in energies are observed for the SNAP models and NNP models. It is well-known that neural network-based models often require larger data sets for best performance; previous NNP models have been trained on thousands or tens of thousands of structures\cite{benchmark_water_cluster, NNP_water}, while only hundreds of structures are used in training the current ML-IAPs. Nevertheless, the NNP models still show surprisingly good performance for bcc systems. The qSNAP models' performances are between those of the GAP and NNP. In general, the qSNAP models have moderately lower MAEs than the linear SNAP, though at the expense of a large expansion in the number of parameters.

In terms of chemistries, we find that the lowest MAEs in energies are observed for the fcc systems, followed by the bcc systems, and the highest MAEs are observed for the diamond systems. Very low MAEs in forces are observed across all ML IAPs for Cu, Ni and Li, while significantly higher MAEs in forces are observed for Mo, a metal with higher modulus and larger force distributions. Higher MAEs in forces are also observed for the diamond semiconductors. These trends are generally consistent across all ML-IAPs studied.

\begin{figure}
    \centering
    \includegraphics[width=1.0\textwidth]{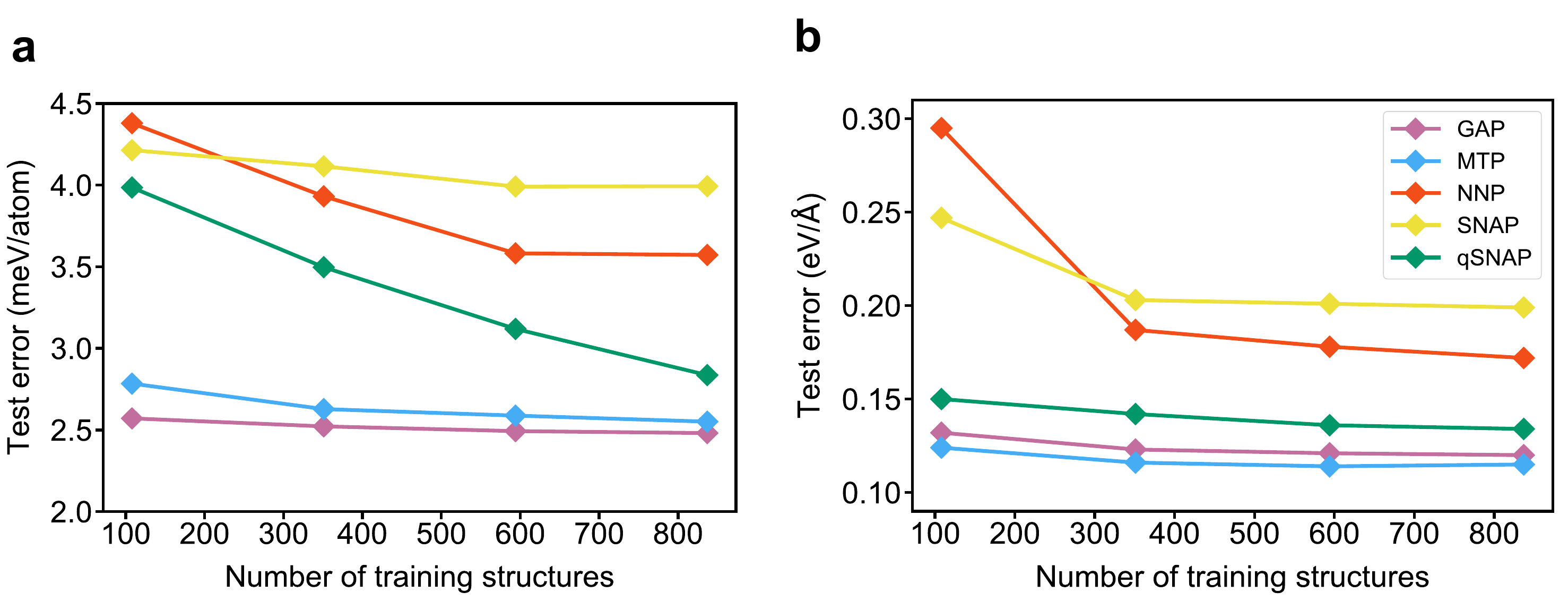}
    \caption{MAE in predicted (a) energies (b) forces of the test set versus the size of the training data for the ML-IAP Mo models.}
    \label{fig:big_data}
\end{figure}

We have also performed a study of the convergence of the ML-IAPs with training data size using Mo as the benchmark system given that it is a bcc metal (for which traditional IAPs tend to perform poorly) with large force distributions. Here, the length of the AIMD simulations were increased four-fold, and more training structures were sampled at the same time interval. The convergence results are shown in Figure \ref{fig:big_data}. While the prediction errors of all models decrease with increase in the number of training structures, the most substantial improvements in accuracy, especially in predicted energies, are observed for the NNP and qSNAP models. The SNAP Mo model appears to have converged in energy and force at a training data size of $\sim 600$ and $\sim 400$ structures, respectively. For the NNP, additional training structures offer modest improvements in force accuracy, but large improvements in energy accuracy. Indeed, it is possible that the NNP and qSNAP Mo models have not been converged with respect to accuracy in energies even at $\sim$ 800 training structures. We have not attempted to further converge these models in view of the computational expense involved.

\subsection{Accuracy in material properties}

The accuracy in predicting basic material properties is critical for evaluating the performance of ML-IAPs. Here, we perform the climbing-image nudged elastic band (CI-NEB) method\cite{neb_method} as well as molecular dynamics (MD) with ML-IAPs to obtain the cubic lattice parameter, elastic constants, migration energies and vacancy formation energies. The comparison of these predicted material properties with respect to the DFT values is provided in Table \ref{table:elastic}.  The performances of all ML-IAPs are generally excellent, with lattice parameters within 0.1-2.0\% of the DFT values and elastic constants that are typically within 10\% of DFT values. It should be noted that the large percentage error in Li for elastic constants is due to the small reference values. The MTP, SNAP and qSNAP models perform well on elastic constants on fcc and bcc systems, but exhibit slightly higher prediction errors in the diamond systems. A possible explanation for the slightly poorer prediction of elastic constants of the NNP model could be the limitation of the size of training data, which restrict the potential of a fully connected neural network. However, it should be noted that despite the slightly higher prediction errors of elastic components for the NNP model, its prediction errors of Voigt-Reuss-Hill approximated bulk modulus\cite{bulkmodulus} across various elemental systems are in good agreement with DFT reference values. 

In terms of diffusion properties, the GAP and MTP models perform well across different chemistries, with most of the prediction errors within 10\% of DFT values, albeit with a moderate underestimate of the migration energy for diamond systems, in line with the previous study\cite{GAP_Si}. While SNAP and qSNAP models show high accuracy in predicting diffusion properties for fcc systems, they considerably underestimate the vacancy formation energy as well as activation barrier for diamond systems. It is noteworthy that all ML-IAPs overestimate the migration energy of Mo system by more than 20\%, which has also been observed in a previous work\cite{SNAP_Mo}.

\begin{center}
\footnotesize
\LTcapwidth=\textwidth
\begin{longtable}{  m{2.3cm} m{0.6cm} m{2.6cm} m{2.6cm} m{2.5cm} m{2.5cm} m{2.5cm}  }
\caption{Calculated cubic lattice parameter $a$, elastic constants ($c_{ij}$), Voigt-Reuss-Hill bulk modulus $B_{\text{VRH}}$, migration energy ($E_{m}$), vacancy formation energy ($E_{m}$) as well as activation barrier for vacancy diffusion ($E_{a} = E_{v} + E_{m}$) with DFT and the four ML-IAPs. Lowest absolute errors with respect to DFT for each property are bolded for ease of reference. Error percentages with respect to DFT values are shown in parentheses.}
\label{table:elastic}
\\
\hline
\hline
\noalign{\smallskip}
            & \multicolumn{1}{c}{\textbf{DFT}} & \multicolumn{1}{c}{\textbf{GAP}}  &  \multicolumn{1}{c}{\textbf{MTP}}    & \multicolumn{1}{c}{\textbf{NNP}}     & \multicolumn{1}{c}{\textbf{SNAP}}  &
            \multicolumn{1}{c}{\textbf{qSNAP}}\\

\hline
\endfirsthead

\multicolumn{7}{c}%
{{\bfseries \tablename\ \thetable{} -- continued from previous page}} \\

\hline
\hline
\noalign{\smallskip}
            & \multicolumn{1}{c}{\textbf{DFT}} & \multicolumn{1}{c}{\textbf{GAP}}  &  \multicolumn{1}{c}{\textbf{MTP}}    & \multicolumn{1}{c}{\textbf{NNP}}     & \multicolumn{1}{c}{\textbf{SNAP}}  &
            \multicolumn{1}{c}{\textbf{qSNAP}}\\

\hline
\endhead

\hline \multicolumn{7}{r}{{Continued on next page}} \\
\endfoot

\hline \hline
\endlastfoot

\hline
\textbf{Ni} &              &                &                   &                 &                  &                \\[1.35ex]

 $a$ ({\AA})  &   3.508      & 3.523 (0.4\%)       & 3.522 (0.4\%)   & 3.523 (0.4\%)   & 3.522 (0.4\%) & \textbf{3.521 (0.4\%)}\\

 $c_{11}$ (GPa) & 276 & 281 (1.8\%)         & 284 (2.9\%)    & \textbf{274 ($\textbf{-0.8\%}$)}    & 283 (2.5\%)   & 267 ($-3.3\%$)\\

 $c_{12}$ (GPa) & 159 & \textbf{159 (0.0\%)}        & 172 (8.2\%)    & 169 (6.3\%)     & 168 (5.7\%)   & 155 ($-2.5\%$)\\

 $c_{44}$ (GPa) & 132 & 126 ($-4.5\%$)        & 127 (-3.8\%)     & 113 ($-14.4\%$)   & \textbf{129 ($\textbf{-2.3\%}$)}   & 125 ($-5.3\%$)\\

 $\textit{B}_{\text{VRH}}$\ (\text{GPa}) & 198 & \textbf{200 (1.0\%)}  & 209 ($5.6\%$)   & 204 (3.0\%)     & 206 (4.0\%)   & 193 ($-2.5\%$)\\
 
 $E_{v}$ (eV) & 1.49    &   1.46 ($-2.0\%$) &   1.43 ($-4.0\%$) &   1.65 (10.7\%) &   \textbf{1.47 ($\textbf{-1.3\%}$)}    &   \textbf{1.47 ($\textbf{-1.3\%}$)}\\

 $E_{m}$ (eV) & 1.12    &   1.14 (1.8\%)    &   1.11 ($-0.9\%$) &   1.14 (1.8\%)    &   \textbf{1.12 (0.0\%)}    &   1.05 ($-6.3\%$)\\
 
 $E_{a}$ (eV) & 2.61    &   \textbf{2.60 ($\textbf{-0.4\%}$)} &   2.54 ($-2.7\%$) &   2.79 (6.9\%)    &   2.59 ($-0.8\%$)    &   2.52 ($-3.4\%$)\\

\hline
\textbf{Cu} &              &                &                   &                 &                  &                \\[1.35ex]

 $a$ ({\AA})  &   3.621      & \textbf{3.634 (0.4\%)}       & 3.636 (0.4\%)   & 3.637 (0.4\%)   & 3.634 (0.4\%) & \textbf{3.636 (0.4\%)}\\

 $c_{11}$ (GPa) & 173 & \textbf{175 (1.2\%)}         & 177 (2.3\%)     & 182 (5.2\%)     & 178 (2.9\%)   & \textbf{178 (2.9\%)}\\

 $c_{12}$ (GPa) & 133 & 120 ($-9.8\%$)        & 120 ($9.8\%$)    & 125 ($-6.0\%$)    & 126 ($-5.3\%$)  & 124 ($-6.8\%$)\\

 $c_{44}$ (GPa) & 88  & 82 ($-6.8\%$)          & 81 ($-8.0\%$)     & 76 ($-13.6\%$)    & \textbf{86 ($\textbf{-2.3\%}$)}   & 82 ($-6.8\%$)\\

 $\textit{B}_{\text{VRH}}$\ (\text{GPa}) & 146 & 138 ($-5.5\%$)  & 139 ($-4.8\%$) & 144 ($-1.4\%$)    & \textbf{143 ($\textbf{-2.1\%}$)}  & 142 ($-2.7\%$)\\
 
 $E_{v}$ (eV) & 1.15    &   1.05 ($-8.7\%$) &   1.10 ($-4.3\%$) &   1.23 (7.0\%)    &   1.19 (3.5\%)   &   \textbf{1.15 (0.0\%)} \\

 $E_{m}$ (eV) & 0.79    &   0.76 ($-3.8\%$) &   \textbf{0.77 ($\textbf{-2.5\%}$)} &   \textbf{0.77 ($\textbf{-2.5\%}$)} &   0.82 (3.8\%)    &   0.74 ($-6.3\%$)\\
 
 $E_{a}$ (eV) & 1.94    &   1.81 ($-6.7\%$) &   1.87 ($-3.6\%$) &   2.00 (3.1\%)    &   2.01 (3.6\%)   &   \textbf{1.89 ($\textbf{-2.6\%}$)}\\

\hline
\textbf{Li} &              &                &                   &                 &                  &                \\[1.35ex]

 $a$ ({\AA})  &   3.427      & 3.450 (0.7\%)       & 3.446 (0.6\%)   & \textbf{3.434 (0.2\%)}   & 3.506 (2.3\%) & 3.469 (1.2\%)\\

 $c_{11}$ (GPa) & 15 & 18 (20.0\%)         & 14 ($-6.7\%$)     & \textbf{17 (13.3\%)}     & 18 (20.0\%)   & 12 ($-20.0\%$)\\

 $c_{12}$ (GPa) & 13 & 14 (7.7\%)          & \textbf{13 (0.0\%)}    & \textbf{12 ($\textbf{-7.7\%}$)}     & 7 ($-46.2\%$)   & 6 ($-53.8\%$)\\

 $c_{44}$ (GPa) & 11 & 12 (9.1\%)         & \textbf{11 (0.0\%)}    & 12 (9.1\%)      & 10 ($-9.1\%$)   & \textbf{11 (0.0\%)}\\

 $\textit{B}_{\text{VRH}}$\ (\text{GPa}) & 14 & \textbf{15 (7.1\%)}  & \textbf{13 ($\textbf{-7.1\%}$)}    & \textbf{13 ($\textbf{-7.1\%}$)}      & 11 ($-21.4\%$)   & 8 ($-42.9\%$)\\
 
 $E_{v}$ (eV)   &   0.62    &   0.56 ($-9.7\%$) &   0.53 ($-14.5\%$)    &   0.50 ($-19.4\%$)    & \textbf{0.63 (1.6\%)}    &   0.58 ($-6.5\%$) \\
 
 $E_{m}$ (eV)   &   0.06    &   \textbf{0.06 (0.0\%)}    &   0.08 (33.3\%)   &   0.05 ($-16.7\%$)    &   0.09 (50.0\%)   &   0.09 (50.0\%) \\
 
 $E_{a}$ (eV)   &   0.68    &   0.62 ($-8.8\%$) &   0.61 ($-10.3\%$)    &   0.55 ($-19.1\%$)    & 0.72 (5.9\%)    &   \textbf{0.67 ($\textbf{-1.5\%}$)}\\

\hline
\textbf{Mo} &              &                &                   &                 &                  &                \\[1.35ex]

 $a$ ({\AA})  &   3.168      & \textbf{3.168 (0.0\%)}       & 3.169 (0.0\%)   & 3.165 ($-0.1\%$)   & 3.169 (0.0\%) & 3.170 (0.1\%)\\

 $c_{11}$ (GPa) & 472 & 481 (1.9\%)         & \textbf{472 (0.0\%)}     & 441 ($-6.6\%$)     & 457 ($-3.2\%$)   & 436 ($-7.6\%$)\\

 $c_{12}$ (GPa) & 158 & 169 (7.0\%)          & \textbf{154 ($\textbf{-2.5\%}$)}    & 192 (21.5\%)     & 158 (0.0\%)   & 166 (5.1\%)\\

 $c_{44}$ (GPa) & 106 & 112 (5.7\%)        & 103 ($-2.8\%$)    & 114 (7.5\%)      & 109 (2.8\%)   & \textbf{104 ($\textbf{-1.9\%}$)} \\

 $\textit{B}_{\text{VRH}}$\ (\text{GPa}) & 263 & 271 (3.8\%)  & \textbf{260 ($\textbf{-1.1\%}$)}    & \textbf{266 (1.1\%)}      & 258 ($-1.9\%$)   & 256 ($-2.7\%$)\\
 
 $E_{v}$ (eV)   &   2.70    &   \textbf{2.68 ($\textbf{-0.7\%}$)} &   2.61 ($-3.3\%$) &   2.94 (8.9\%)    &   \textbf{2.72 (0.7\%)}    &   2.79 (3.3\%) \\

 $E_{m}$ (eV)   &   1.22    &   1.60 (31.1\%)   &   1.51 (23.8\%)   &   1.59 (30.3\%)   &   \textbf{1.49 (22.1\%)}   &   1.50 (23.0\%)   \\
 
 $E_{a}$ (eV) & 3.92    &   4.28 (9.2\%)    &   \textbf{4.12 (5.1\%)}    &   4.53 (15.6\%)   &   4.21 (7.4\%)    &   4.29 (9.4\%) \\

\hline
\textbf{Si} &              &                &                   &                 &                  &                \\[1.35ex]

 $a$ ({\AA})  &   5.469      & 5.458 ($-0.2\%$)       & \textbf{5.465 ($\textbf{-0.1\%}$)}   & 5.459 ($-0.2\%$)   & 5.466 (0.1\%) & 5.464 (-0.1\%)\\

 $c_{11}$ (GPa) & 156 & 168 (7.7\%)           & \textbf{155 ($\textbf{-0.6\%}$)}   & 130 ($-9.7\%$)    & 141 ($-9.6\%$)   & \textbf{155 ($\textbf{-0.6\%}$)}\\

 $c_{12}$ (GPa) & 65  & \textbf{62 ($\textbf{-4.6\%}$)}             & 76  (16.9\%)    & 77 ($18.5\%$)      & 61 ($-6.2\%$)   & 58 ($-10.8\%$)\\

 $c_{44}$ (GPa) & 76  & 69 ($-9.2\%$)             & \textbf{75 ($\textbf{-1.3\%}$)}      & 64 ($-15.8\%$)     & 71 ($-6.6\%$)   & 69 ($-9.2\%$)\\

 $\textit{B}_{\text{VRH}}$\ (\text{GPa}) & 95 & 97 (2.1\%)  & 102 (7.4\%)    & \textbf{95 (0.0\%)}      & 93 ($-2.1\%$)   & 90 ($-5.3\%$)\\
 
 $E_{v}$ (eV)   &   3.25    &   3.04 ($-6.5\%$)     &   \textbf{3.11 ($\textbf{-4.3\%}$)}     &   2.63 ($-19.1\%$)    &   2.71 ($-16.6\%$)    &   2.37 ($-27.1\%$) \\
 
 $E_{m}$ (eV)   &   0.21    &   \textbf{0.21 (0.0\%)}    &   0.16 ($-23.8\%$)    &   0.25 (19.0\%)   &  0.26 (23.8\%)   &   0.20 ($-4.7\%$) \\
 
 $E_{a}$ (eV)   &   3.46    &   3.25 ($-6.1\%$)     &   \textbf{3.27 ($\textbf{-5.5\%}$)}     &   2.88 ($-16.8\%$)    &   2.97 ($-14.2\%$)    &   2.57 ($-25.7\%$) \\

\hline
\textbf{Ge} &              &                &                   &                 &                  &                \\[1.35ex]

 $a$ ({\AA})  &   5.763      & 5.777 (0.2\%)       & \textbf{5.770 (0.1\%)}   & 5.751 ($-0.5\%$)   & 5.773 (0.2\%) & 5.775 (0.2\%)\\

 $c_{11}$ (GPa) & 116 & 127 (9.5\%)           & 106 ($-8.6\%$)     & 101 ($-12.9\%$)     & 101 ($-12.9\%$)   & \textbf{121 (4.3\%)}\\

 $c_{12}$ (GPa) & 48  & 45 ($-6.3\%$)             & 54 (12.5\%)      & \textbf{46 ($\textbf{-4.2\%}$)}      & 41 ($-14.6\%$)   & 43 ($-10.4\%$)\\

 $c_{44}$ (GPa) & 58  & 54 ($-6.9\%$)            & \textbf{55 ($\textbf{-5.2\%}$)}    & 46 ($-20.7\%$)     & 54 ($-6.9\%$)   & 50 ($-13.8\%$)\\

 $\textit{B}_{\text{VRH}}$\ (\text{GPa}) & 71 & 72 (1.4\%)  & \textbf{71 (0.0\%)}    &  69($-2.8\%$)      & 61 ($-14.1\%$)   & 69 ($-2.8\%$) \\
 
 $E_{v}$ (eV)   &   2.19    &   \textbf{2.10 ($\textbf{-4.1\%}$)}     &   1.98 ($-9.6\%$)    &   1.98 ($-9.6\%$)    &   1.77 ($-19.2\%$)    &   1.67 ($-23.7\%$) \\
 
 $E_{m}$ (eV)   &   0.19    &   0.17 ($-10.5\%$)    &   0.17 ($-10.5\%$)    &   \textbf{0.20 (5.3\%)}   &  0.28 (47.4\%)   &   \textbf{0.18 ($\textbf{-5.3\%}$)} \\
 
 $E_{a}$ (eV)   &   2.38    &   \textbf{2.27 ($\textbf{-4.6\%}$)}     &   2.15 ($-9.7\%$)    &   2.18 ($-8.4\%$)    &   2.05 ($-13.9$)  &   1.85 ($-22.3\%$) \\
\end{longtable}
\end{center}

\subsection{Accuracy in equations of state}

To provide an evaluation of the performance of ML-IAPs far from equilibrium, we have computed a pairwise comparison of the equation of state (EOS) curves for all elements studied using the $\Delta_{\text{EOS}}$ gauge of Lejaeghere et al.\cite{Delta_gauge, delta_gauge_assessment1, delta_gauge_assessment2} The $\Delta_{\text{EOS}}$ gauge, which has been used to evaluate accuracy differences between DFT codes, is the root-mean-square difference between two EOS curves over a $\pm 6\%$ interval around the equilibrium volume, defined as follows:
\begin{equation}
    \Delta_{\rm EOS} = \sqrt{\frac{\int_{0.94 V_{0}}^{1.06 V_{0}}[E^{\text{a}}(V)-E^{\text{b}}(V)]^{2} dV}{0.12 V_{0}}}
\end{equation}
where $E^{\text{a}}$ and $E^{\text{b}}$ denote energies computed using methods a and b, respectively. 

\begin{figure}
    \centering
    \subfigure[$\Delta$ gauge comparisons between ML-IAPs with respect to DFT]{\label{fig:gauge}\includegraphics[width=0.6\textwidth]{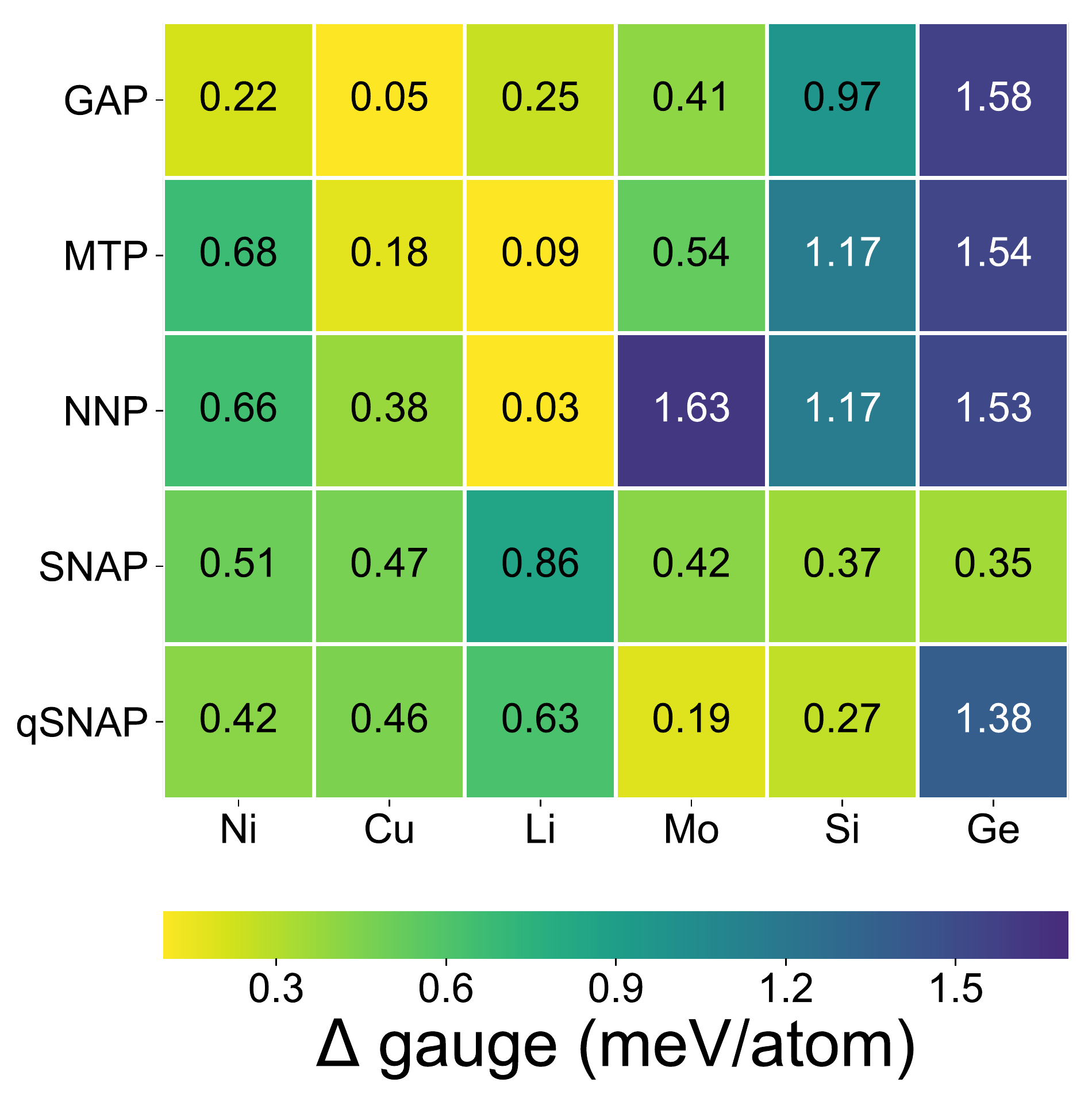}}
    \subfigure[Energy vs volume curves]{\label{fig:eos}\includegraphics[width=0.8\textwidth]{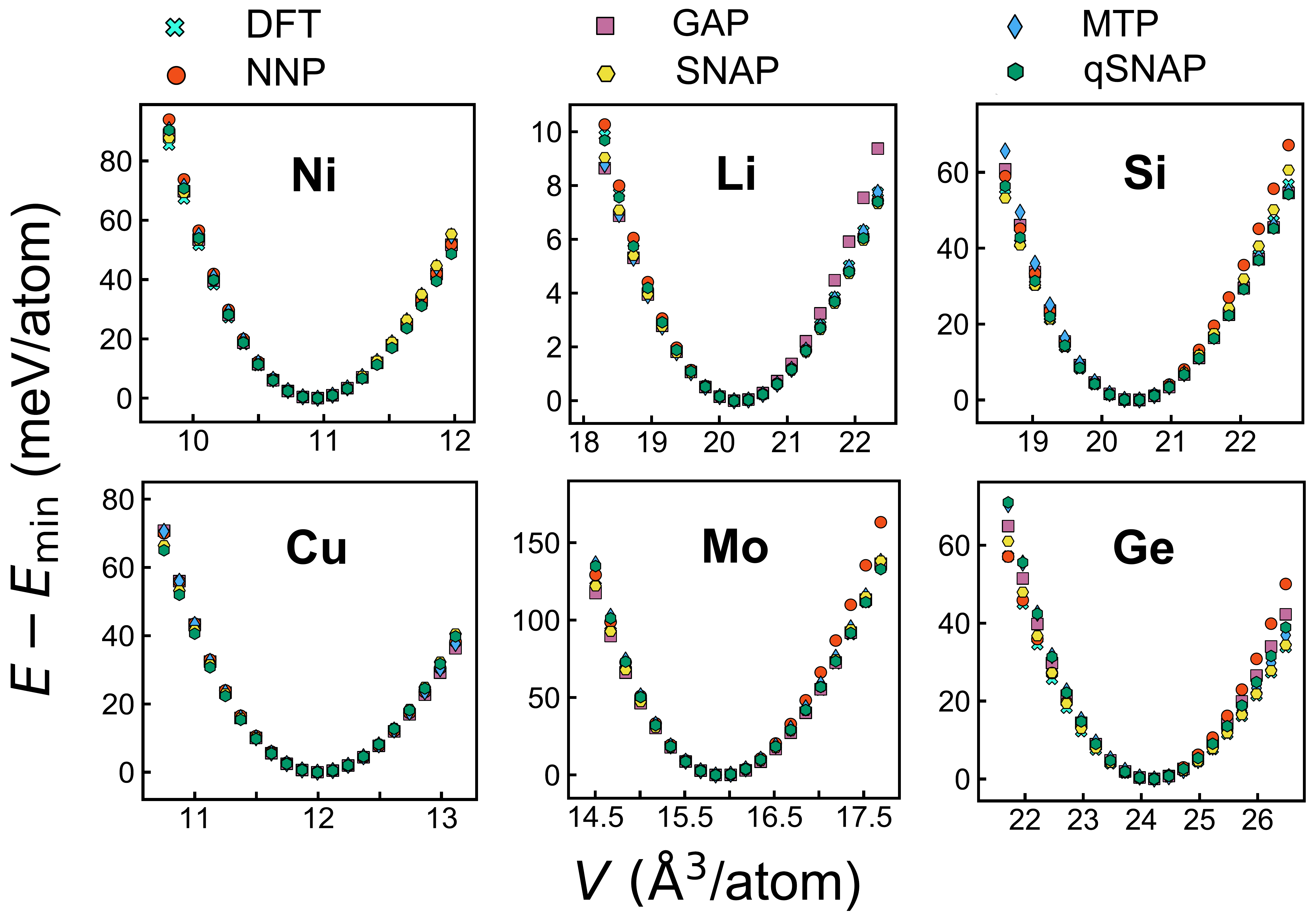}}
    \caption{Assessment of accuracy of ML-IAPs in predicting equation of state. (a) $\Delta$ gauge comparison provides quantitative estimate of deviation between the EOS curve from each ML-IAP with that of DFT. (b) EOS curves for all six elements using DFT and the four ML-IAPs.}
    \label{fig:eos_prediction}
\end{figure}

Figure \ref{fig:eos} shows the $\Delta_{\text{EOS}}$ values of various machine learning models with respect to DFT reference data for different elemental systems as well as the EOS curves of these ML-IAPs. In all cases, the $\Delta_{\text{EOS}}$ for all ML-IAPs for all elements are within 2 meV/atom, which is the threshold for ``indistinguishable EOS'' previously used in evaluating different DFT codes\cite{Science}. It is noteworthy that despite the relatively high prediction errors of SNAP models presented in Figure \ref{fig:energy_error}, they perform considerably better in predicting the EOS curves, with all the $\Delta_{\text{EOS}}$ lower than 1 meV/atom across different chemistries. The NNP models deviate slightly from DFT curves at both tensile and compressive strains for fcc systems, while for diamond systems, the deviation of the NNP models from DFT curve is comparable with those of GAP and MTP models, as evidenced in $\Delta$ gauge comparison. In general, it is more challenging to give highly accurate predictions of EOS in diamond system than in fcc and bcc systems. In addition to the DFT-level accuracy in equations of state prediction, the predicted phonon dispersion curves by all ML-IAPs investigated in this work are in excellent agreement with the DFT reference (see Figure S1--Figure S6 in the Supplementary Information).

\subsection{Accuracy in molecular dynamics (MD) trajectories}

\begin{figure}
    \centering
    \includegraphics[width=1.0\textwidth]{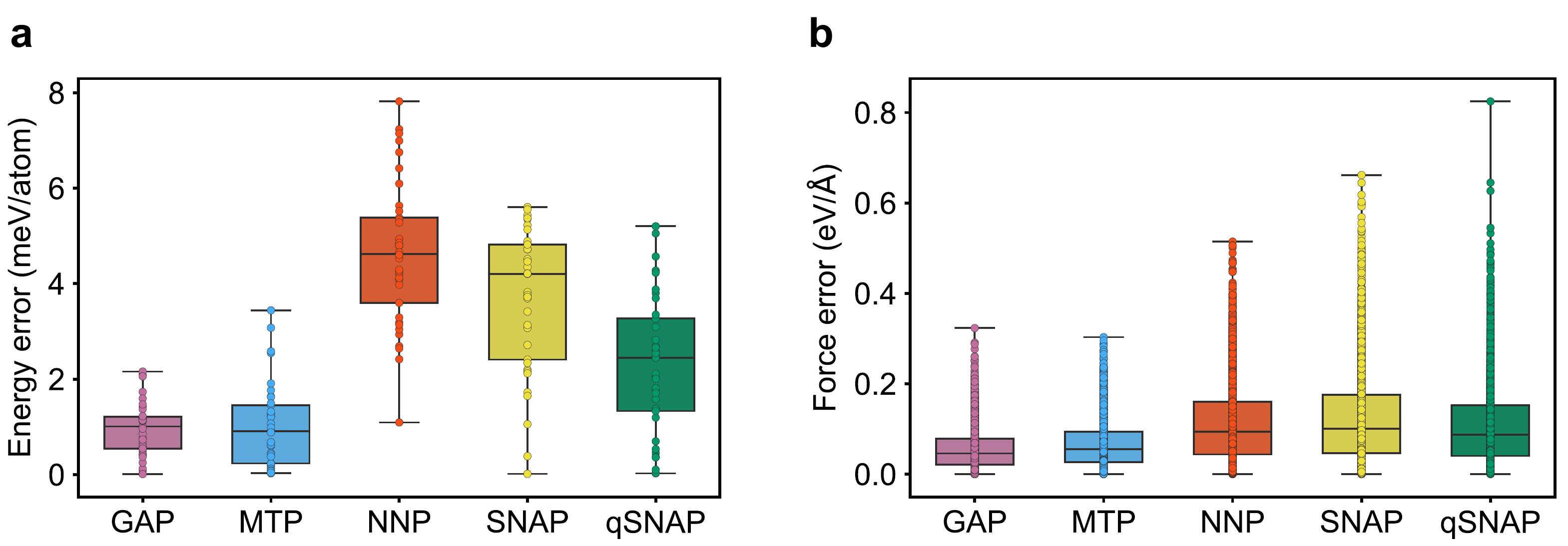}
    \caption{Error distributions in (a) predicted energies (b) predicted forces for sampled structures from MD simulations using each ML-IAP. The rectangular box indicates the interquartile range (IQR), while the line within the box indicates the median.}
    \label{fig:md_error}
\end{figure}

One of the principal applications of ML-IAPs is in molecular dynamics (MD) simulations. To assess the ability of the ML-IAPs to provide stable MD trajectories, we carried out $NVT$ MD simulations at 1,300 K ($0.5\times$ melting point) on a $3 \times 3 \times 3$ 54-atoms supercell of bulk Mo for 0.25 ns using LAMMPS with the different ML-IAPs. A total of 40 snapshots at an interval of 2.5 ps were then sampled from each MD trajectory, and DFT static calculations were performed on these snapshots. Figure \ref{fig:md_error} shows the distribution of the errors in the energies and forces of sampled structures. In line with the previous results, the GAP and MTP models generally exhibit smaller errors in the energies and forces than the NNP, SNAP and qSNAP models. The GAP model not only have the lowest median, but also the smallest interquartile range (IQR) in the errors in energies and forces. Somewhat interestingly, the NNP model has higher energy errors, but smaller force errors than SNAP and qSNAP. For consistency of comparison, all models shown here are the ``optimal'' models based on $\sim100$ training structures. It is likely that a larger training set would improve the performance of the NNP and qSNAP models. (Figure \ref{fig:big_data}).

\subsection{Accuracy in polymorphic energy differences}

\begin{figure}
    \centering
    \includegraphics[width=0.8\textwidth]{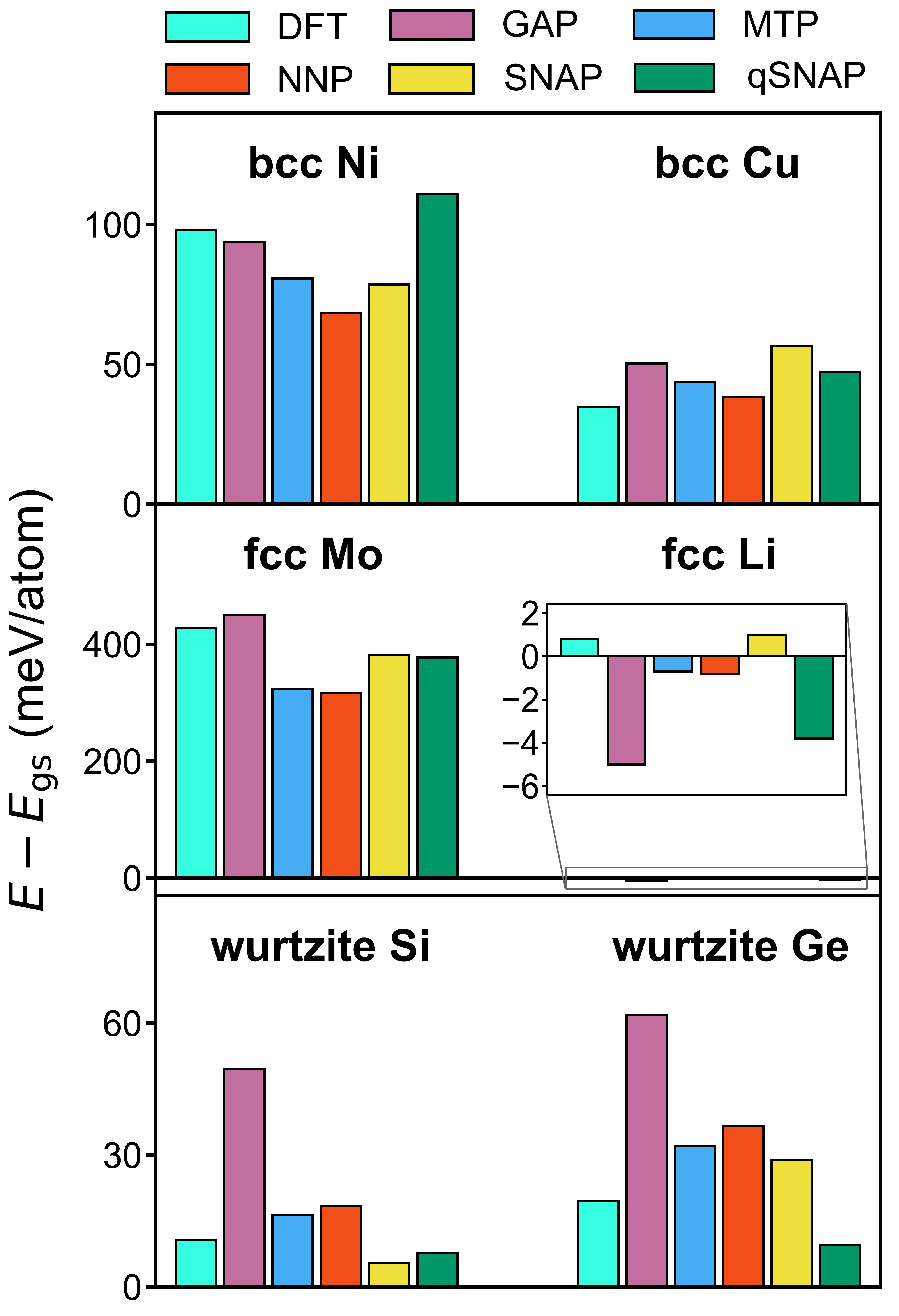}
    \caption{Calculated energetic differences between the typical low energy polymorph and ground-state polymorph of each elemental system. The inset shows the magnified bar chart for Li system due to its relatively small range. The typical low energy polymorph is indicated with the label above each bar chart.}
    \label{fig:transferability}
\end{figure}

To evaluate the ability of the ML-IAPs to extrapolate to unseen data, we have computed the energy differences between the DFT ground state polymorph and a low-energy polymorph for each element, presented in Figure \ref{fig:transferability}. The low-energy polymorphs correspond to the bcc, fcc, and wurtzite (hexagonal diamond) structures for the fcc, bcc, and diamond systems, respectively. It should be noted that only the ground state structures were used in training the ML-IAPs, and these low-energy polymorphs were not present in the training structures. Except for Li which has an extremely small energy difference between the fcc and bcc structures in DFT, all ML-IAPs are able to reproduce qualitatively the energy difference between polymorphs. For most systems, the ML-IAPs are able to reproduce energy differences between the polymorphs to within 10-20 meV/atom; the main exception is Mo, which exhibits a large energy difference between the fcc and bcc structures. One notable observation is that the GAP model shows the largest error in predicting the energy difference between the wurtzite and diamond structures in Si and Ge compared to the other ML-IAPs, despite having relatively low MAE in predicted energies in these systems (see Figure \ref{fig:energy_error}). We believe that this may be due to fact that the GAP may be more sensitive to missing reference configurations, while the other IAPs are able to extrapolate the interactions to this unseen configuration more effectively. Somewhat surprisingly, the linear SNAP model exhibits among the best performance in reproducing the polymorphic energy differences across all systems, outperforming even the GAP and MTP for Mo, Si and Ge, despite having substantially larger MAEs in energies and forces.

\section{Conclusions}

We have performed a comprehensive unbiased evaluation of the GAP, MTP, NNP, SNAP, and qSNAP ML-IAP models using consistently-generated DFT data on six elemental systems spanning different crystal structures (fcc, bcc, and diamond), chemistries (main group metals, transition metals and semiconductors) and bonding (metallic and covalent). This evaluation is carried out across three key metrics that are of critical importance for any potential user of these ML-IAPs:
\begin{enumerate}
    \item Accuracy in predicted energies, forces and properties for both seen and unseen structures;
    \item Training data requirements, which influence the number of expensive DFT computations that have to performed to train a ML-IAP to a given accuracy; and
    \item Computational cost, which influence the size of the systems on which computations can be performed for a given computing budget.
\end{enumerate}

These three metrics are inextricably linked - for all the four ML-IAPs, an increase in number of degrees of freedom (with increase in computational cost) and increase in training structures generally leads to higher accuracy. We demonstrate the application of the Pareto frontier as a means to identify the optimal trade-offs between these metrics. For all ML-IAPs, we find that there is an ``optimal'' configuration at which further expansion of the number of degrees of freedom yield little improvement in accuracy with increases in computational cost.

We find that all ML-IAPs are able to achieve near-DFT accuracy in predicting energies, forces and material properties, substantially outperforming traditional IAPs. The GAP and MTP models exhibit the smallest MAEs in energies and forces. However, the GAP models are among the most computationally expensive for a given accuracy (based on current implementations) and show poor extrapolability to higher energy polymorphs in the diamond systems. Indeed, the simple linear SNAP model, which has among the highest MAEs in predicted energies and forces, show the best extrapolability to higher energy polymorphs as well as reproducing the equation of state for the diamond systems. The NNP and qSNAP models show relatively high MAEs in energies with small data sizes, but these can be mitigated with increases in training data.

Another somewhat surprising conclusion is also that even with relatively small training data sets of $\sim$100-200 structures, the GAP, MTP and SNAP models appear to be reasonably well-converged to meV atom$^{-1}$ accuracy in energies and 0.01 eV \AA$^{-1}$ accuracy in forces. The NNP and qSNAP models can be further improved with larger training data sets, but the MAEs even at $\sim 100$ structures are not excessively high. We attribute this performance to the training data generation procedure, which is aimed at sampling a diversity of structures from both ground state and multi-temperature AIMD simulations. In other words, diversity of training data is arguably a more important consideration than quantity.

Finally, we note that one limitation of this study is that we have not attempted to combine the different local environment descriptors (symmetry functions, SOAP, bispectrum, moment tensors) with different ML frameworks (e.g., linear regression versus gaussian process regression versus neural network). The choice of descriptor affects how efficiently diverse local environments can be encoded, while the choice of ML framework determines the functional flexibility in mapping the relationship between descriptors and energies/forces. Different descriptor-model choices may yield better tradeoffs between accuracy and cost for a particular application.

\section{Acknowledgments}

This work was primarily supported by the Office of Naval Research (ONR) Young Investigator Program (YIP) under Award No. N00014-16-1-2621. The authors also acknowledge computational resources provided by Triton Shared Computing Cluster (TSCC) at the University of California, San Diego, the National Energy Research Scientific Computing Center (NERSC), and the Extreme Science and Engineering Discovery Environment (XSEDE) supported by the National Science Foundation under Grant No. ACI-1053575. J. Behler thanks the Deutsche Forschungsgemeinschaft for a Heisenberg professorship (Be3264/11-2, project number 329898176). A. V. Shapeev was supported by the Russian Science Foundation under Grant No. 18-13-00479.

A. Thompson and M. Wood are employees of Sandia National Laboratories, a multimission laboratory managed and operated by National Technology and Engineering Solutions of Sandia, LLC, a wholly owned subsidiary of Honeywell International Inc., for the U.S. Department of Energy’s National Nuclear Security Administration under contract DE-NA0003525.  This paper describes objective technical results and analysis. Any subjective views or opinions that might be expressed in the paper do not necessarily represent the views of the U.S. Department of Energy or the United States Government.

\bibliography{new_collection}% Produces the bibliography via BibTeX.

\end{document}

\title[Supplementary Information --- A Performance and Cost Assessment of Machine Learning Interatomic Potentials]{Supplementary Information\\A Performance and Cost Assessment of Machine Learning Interatomic Potentials}

\author{Yunxing Zuo}
\affiliation{Department of NanoEngineering, University of California San Diego, 9500 Gilman Dr, Mail Code 0448, La Jolla, CA 92093-0448, United States}
\author{Chi Chen}
\affiliation{Department of NanoEngineering, University of California San Diego, 9500 Gilman Dr, Mail Code 0448, La Jolla, CA 92093-0448, United States}
\author{Xiangguo Li}
\affiliation{Department of NanoEngineering, University of California San Diego, 9500 Gilman Dr, Mail Code 0448, La Jolla, CA 92093-0448, United States}
\author{Zhi Deng}
\affiliation{Department of NanoEngineering, University of California San Diego, 9500 Gilman Dr, Mail Code 0448, La Jolla, CA 92093-0448, United States}
\author{Yiming Chen}
\affiliation{Department of NanoEngineering, University of California San Diego, 9500 Gilman Dr, Mail Code 0448, La Jolla, CA 92093-0448, United States}
\author{J{\"o}rg Behler}
\affiliation{Universit{\"a}t G{\"o}ttingen, Institut f{\"u}r Physikalische Chemie, Theoretische Chemie, Tammannstra{\ss}e 6, 37077 G{\"o}ttingen, Germany}
\author{G{\'a}bor Cs{\'a}nyi}
\affiliation{Department of Engineering, University of Cambridge, Trumpington Street, Cambridge, CB2 1PZ, United Kingdom}
\author{Alexander V. Shapeev}
\affiliation{Skolkovo Institute of Science and Technology, Skolkovo Innovation Center, Building 3, Moscow, 143026, Russia}
\author{Aidan P. Thompson}
\affiliation{Center for Computing Research, Sandia National Laboratories, Albuquerque, New Mexico 87185, United States}
\author{Mitchell A. Wood}
\affiliation{Center for Computing Research, Sandia National Laboratories, Albuquerque, New Mexico 87185, United States}
\author{Shyue Ping Ong}
\email{ongsp@eng.ucsd.edu}
\affiliation{Department of NanoEngineering, University of California San Diego, 9500 Gilman Dr, Mail Code 0448, La Jolla, CA 92093-0448, United States}

\maketitle

\pagebreak

\section{Detailed formalism of ML-IAPs and Optimized parameters}

\subsection{Behler-Parrinello neural network}

The Behler-Parrinello neural network potentials (NNPs)\cite{NNP_Si, NNP_review_2011, NNP_review_2015} express the potential energy surface (PES) in terms of the atom-centered symmetry functions (ACSF)\cite{Atom-centered-symmetry-functions} using fully-connected feed-forward neural networks.

The ACSF converts the local atomic environment to numeric vectors that fulfill rotational and translational invariance.\cite{Atom-centered-symmetry-functions, NNP_review_2014}. For atom \textit{i} in the structure, the radial ACSF $G_{i}^{\rm atom, \rm rad}$, which provides information about pair correlations between the atoms, is given as follows:
\begin{equation}
    G_{i}^{\rm atom, \rm rad} = \sum_{j\neq i}^{N_{\rm atom}} e^{-\eta(R_{ij}-R_{s})^{2}} \cdot f_{c}(R_{ij}),
\end{equation}
where $\eta$ determines the width of the Gaussian basis, and $R_{s}$ is the position shift over all neighboring atoms within the cutoff radius $R_{c}$. The cutoff function $f_{c}$, which ensures a smooth decay in value and slope at cutoff radius $R_{c}$, is given as follows:
\begin{equation}
    f_{c}(R_{ij}) = 
  \begin{cases} 
   \ 0.5 \cdot [\cos{(\frac{\pi R_{ij}}{R_{c}})} + 1], & \text{for } R_{ij} \leq R_{c} \\
   \ 0.0,                                     & \text{for } R_{ij} > R_{c}.
  \end{cases}
\end{equation}
The angular ACSF takes the following form:
\begin{equation}
\begin{split}
    G_{i}^{\rm atom, \rm ang}&=2^{1-\zeta}\sum_{j,k \neq i}^{N_{\rm atom}} (1 + \lambda \cos \theta_{ijk})^{\zeta} \cdot e^{-\eta^{\prime}(R_{ij}^{2}+R_{ik}^{2}+R_{jk}^{2})}
    \cdot f_{c}(R_{ij}) \cdot f_{c}(R_{ik}) \cdot f_{c}(R_{jk}),
\end{split}
\end{equation}
where the summation loops over neighbors $j$ and $k$, $\theta_{ijk}$ is the angle, and $\zeta$, $\eta^{\prime}$, and $\lambda$ are three parameters that determine the shape of angular symmetry functions.

Symmetry functions determined by an appropriate choice of hyperparameters ($\eta$, $R_s$, etc.) should reflect the effective relationship between atomic representations and corresponding properties (\textit{e.g.}, energy and force components). A Kalman filter algorithm was used to optimize the neural network weights, and a grid search was conducted to optimize the hyperparameters so that the fitted model can reproduce not only the basic energy and force values but also the elastic constants. The optimized parameters for all elemental systems studied in this work are listed in \textbf{Table \ref{table:NNP_parameters_for_Ni}} - \textbf{Table \ref{table:NNP_parameters_for_Ge}}.

\pagebreak

\begin{center}
\LTcapwidth=\textwidth
\begin{longtable}{M{2.0cm} M{2.0cm} M{0.8cm} M{1.2cm} | M{2.0cm} M{2.0cm} M{2.0cm} M{0.8cm} M{0.8cm} M{1.2cm}}
\caption{\label{table:NNP_parameters_for_Ni} Symmetry function parameters for optimized Ni NNP. The units of $R_{c}$, $R_{s}$, $\eta$ and $\eta^{\prime}$ are {\AA}, {\AA}, \AA$^{-2}$ and \AA$^{-2}$, respectively.}
\\
\hline
\hline
\multicolumn{4}{c}{$R_{c}$} & \multicolumn{6}{c}{3.9}\\
\hline
\multicolumn{4}{c}{$G_{i}^{\rm atom, \rm rad}$} & \multicolumn{6}{c}{$G_{i}^{\rm atom, \rm ang}$}\\
\hline
 Reference & Neighbor & $R_{s}$ &  $\eta$ & Reference & Neighbor 1 & Neighbor 2 & $\lambda$ & $\zeta$ & $\eta^{\prime}$\\
\hline
\endfirsthead

\multicolumn{10}{c}%
{{\bfseries \tablename\ \thetable{} -- continued from previous page}} \\

\hline
\multicolumn{4}{c}{$G_{i}^{\rm atom, \rm rad}$} & \multicolumn{6}{c}{$G_{i}^{\rm atom, \rm ang}$}\\
\hline
 Reference & Neighbor & $R_{s}$ &  $\eta$  & Reference & Neighbor 1 & Neighbor 2 & $\lambda$ & $\zeta$ & $\eta^{\prime}$\\
\hline
\endhead

\hline \multicolumn{10}{r}{{Continued on next page}} \\
\endfoot

\hline \hline
\endlastfoot

  Ni    &   Ni  &   0.0 &  0.036   &    Ni  &   Ni  &   Ni  &   1   &   1   &   0.036   \\
  Ni    &   Ni  &   0.0 &  0.071   &    Ni  &   Ni  &   Ni  &   1   &   2   &   0.036   \\
  Ni    &   Ni  &   0.0 &  0.179   &    Ni  &   Ni  &   Ni  &   1   &   4   &   0.036   \\
        &       &          &       &    Ni  &   Ni  &   Ni  &   1   &   16  &   0.036   \\
        &       &          &       &    Ni  &   Ni  &   Ni  &   -1   &   1  &   0.036   \\
        &       &          &       &    Ni  &   Ni  &   Ni  &   -1   &   2  &   0.036   \\
        &       &          &       &    Ni  &   Ni  &   Ni  &   -1   &   4  &   0.036   \\
        &       &          &       &    Ni  &   Ni  &   Ni  &   -1   &   16  &  0.036   \\
        &       &          &       &    Ni  &   Ni  &   Ni  &   1   &   1  &   0.071   \\
        &       &          &       &    Ni  &   Ni  &   Ni  &   1   &   2  &   0.071   \\
        &       &          &       &    Ni  &   Ni  &   Ni  &   1   &   4  &   0.071   \\
        &       &          &       &    Ni  &   Ni  &   Ni  &   1   &   16  &   0.071   \\
        &       &          &       &    Ni  &   Ni  &   Ni  &   -1   &   1  &   0.071   \\
        &       &          &       &    Ni  &   Ni  &   Ni  &   -1   &   2  &   0.071   \\
        &       &          &       &    Ni  &   Ni  &   Ni  &   -1   &   4  &   0.071   \\
        &       &          &       &    Ni  &   Ni  &   Ni  &   -1   &   16  &  0.071   \\
        &       &          &       &    Ni  &   Ni  &   Ni  &   1   &   1  &   0.179   \\
        &       &          &       &    Ni  &   Ni  &   Ni  &   1   &   2  &   0.179   \\
        &       &          &       &    Ni  &   Ni  &   Ni  &   1   &   4  &   0.179   \\
        &       &          &       &    Ni  &   Ni  &   Ni  &   1   &   16  &   0.179   \\
        &       &          &       &    Ni  &   Ni  &   Ni  &   -1   &   1  &   0.179   \\
        &       &          &       &    Ni  &   Ni  &   Ni  &   -1   &   2  &   0.179   \\
        &       &          &       &    Ni  &   Ni  &   Ni  &   -1   &   4  &   0.179   \\
        &       &          &       &    Ni  &   Ni  &   Ni  &   -1   &   16  &  0.179   \\
\end{longtable}
\end{center}

\pagebreak

\begin{center}
\LTcapwidth=\textwidth
\begin{longtable}{M{2.0cm} M{2.0cm} M{0.8cm} M{1.2cm} | M{2.0cm} M{2.0cm} M{2.0cm} M{0.8cm} M{0.8cm} M{1.2cm}}
\caption{\label{table:NNP_parameters_for_Cu} Symmetry function parameters for optimized Cu NNP. The units of $R_{c}$, $R_{s}$, $\eta$ and $\eta^{\prime}$ are {\AA}, {\AA}, \AA$^{-2}$ and \AA$^{-2}$, respectively.}
\\
\hline
\hline
\multicolumn{4}{c}{$R_{c}$} & \multicolumn{6}{c}{4.1}\\
\hline
\multicolumn{4}{c}{$G_{i}^{\rm atom, \rm rad}$} & \multicolumn{6}{c}{$G_{i}^{\rm atom, \rm ang}$}\\
\hline
 Reference & Neighbor & $R_{s}$ &  $\eta$ & Reference & Neighbor 1 & Neighbor 2 & $\lambda$ & $\zeta$ & $\eta^{\prime}$\\
\hline
\endfirsthead

\multicolumn{10}{c}%
{{\bfseries \tablename\ \thetable{} -- continued from previous page}} \\

\hline
\multicolumn{4}{c}{$G_{i}^{\rm atom, \rm rad}$} & \multicolumn{6}{c}{$G_{i}^{\rm atom, \rm ang}$}\\
\hline
 Reference & Neighbor & $R_{s}$ &  $\eta$  & Reference & Neighbor 1 & Neighbor 2 & $\lambda$ & $\zeta$ & $\eta^{\prime}$\\
\hline
\endhead

\hline \multicolumn{10}{r}{{Continued on next page}} \\
\endfoot

\hline \hline
\endlastfoot

  Cu    &   Cu  &   3.0 &  0.036   &    Cu  &   Cu  &   Cu  &   1   &   1   &   0.036   \\
  Cu    &   Cu  &   3.0 &  0.071   &    Cu  &   Cu  &   Cu  &   1   &   2   &   0.036   \\
  Cu    &   Cu  &   3.0 &  0.179   &    Cu  &   Cu  &   Cu  &   1   &   4   &   0.036   \\
        &       &          &       &    Cu  &   Cu  &   Cu  &   1   &   16  &   0.036   \\
        &       &          &       &    Cu  &   Cu  &   Cu  &   -1   &   1  &   0.036   \\
        &       &          &       &    Cu  &   Cu  &   Cu  &   -1   &   2  &   0.036   \\
        &       &          &       &    Cu  &   Cu  &   Cu  &   -1   &   4  &   0.036   \\
        &       &          &       &    Cu  &   Cu  &   Cu  &   -1   &   16  &  0.036   \\
        &       &          &       &    Cu  &   Cu  &   Cu  &   1   &   1  &   0.071   \\
        &       &          &       &    Cu  &   Cu  &   Cu  &   1   &   2  &   0.071   \\
        &       &          &       &    Cu  &   Cu  &   Cu  &   1   &   4  &   0.071   \\
        &       &          &       &    Cu  &   Cu  &   Cu  &   1   &   16  &   0.071   \\
        &       &          &       &    Cu  &   Cu  &   Cu  &   -1   &   1  &   0.071   \\
        &       &          &       &    Cu  &   Cu  &   Cu  &   -1   &   2  &   0.071   \\
        &       &          &       &    Cu  &   Cu  &   Cu  &   -1   &   4  &   0.071   \\
        &       &          &       &    Cu  &   Cu  &   Cu  &   -1   &   16  &  0.071   \\
        &       &          &       &    Cu  &   Cu  &   Cu  &   1   &   1  &   0.179   \\
        &       &          &       &    Cu  &   Cu  &   Cu  &   1   &   2  &   0.179   \\
        &       &          &       &    Cu  &   Cu  &   Cu  &   1   &   4  &   0.179   \\
        &       &          &       &    Cu  &   Cu  &   Cu  &   1   &   16  &   0.179   \\
        &       &          &       &    Cu  &   Cu  &   Cu  &   -1   &   1  &   0.179   \\
        &       &          &       &    Cu  &   Cu  &   Cu  &   -1   &   2  &   0.179   \\
        &       &          &       &    Cu  &   Cu  &   Cu  &   -1   &   4  &   0.179   \\
        &       &          &       &    Cu  &   Cu  &   Cu  &   -1   &   16  &  0.179   \\
\end{longtable}
\end{center}

\pagebreak

\begin{center}
\LTcapwidth=\textwidth
\begin{longtable}{M{2.0cm} M{2.0cm} M{0.8cm} M{1.2cm} | M{2.0cm} M{2.0cm} M{2.0cm} M{0.8cm} M{0.8cm} M{1.2cm}}
\caption{\label{table:NNP_parameters_for_Li} Symmetry function parameters for optimized Li NNP. The units of $R_{c}$, $R_{s}$, $\eta$ and $\eta^{\prime}$ are {\AA}, {\AA}, \AA$^{-2}$ and \AA$^{-2}$, respectively.}
\\
\hline
\hline
\multicolumn{4}{c}{$R_{c}$} & \multicolumn{6}{c}{5.2}\\
\hline
\multicolumn{4}{c}{$G_{i}^{\rm atom, \rm rad}$} & \multicolumn{6}{c}{$G_{i}^{\rm atom, \rm ang}$}\\
\hline
 Reference & Neighbor & $R_{s}$ &  $\eta$ & Reference & Neighbor 1 & Neighbor 2 & $\lambda$ & $\zeta$ & $\eta^{\prime}$\\
\hline
\endfirsthead

\multicolumn{10}{c}%
{{\bfseries \tablename\ \thetable{} -- continued from previous page}} \\

\hline
\multicolumn{4}{c}{$G_{i}^{\rm atom, \rm rad}$} & \multicolumn{6}{c}{$G_{i}^{\rm atom, \rm ang}$}\\
\hline
 Reference & Neighbor & $R_{s}$ &  $\eta$  & Reference & Neighbor 1 & Neighbor 2 & $\lambda$ & $\zeta$ & $\eta^{\prime}$\\
\hline
\endhead

\hline \multicolumn{10}{r}{{Continued on next page}} \\
\endfoot

\hline \hline
\endlastfoot

  Li    &   Li  &   0.0 &  0.036   &    Li  &   Li  &   Li  &   1   &   1   &   0.036   \\
  Li    &   Li  &   1.0 &  0.036   &    Li  &   Li  &   Li  &   1   &   2   &   0.036   \\
  Li    &   Li  &   2.0 &  0.036   &    Li  &   Li  &   Li  &   1   &   4   &   0.036   \\
  Li    &   Li  &   3.0 &  0.036   &    Li  &   Li  &   Li  &   1   &   16  &   0.036   \\
  Li    &   Li  &   4.0 &  0.036   &    Li  &   Li  &   Li  &   -1   &   1  &   0.036   \\
  Li    &   Li  &   0.0 &  0.071   &    Li  &   Li  &   Li  &   -1   &   2  &   0.036   \\
  Li    &   Li  &   1.0 &  0.071   &    Li  &   Li  &   Li  &   -1   &   4  &   0.036   \\
  Li    &   Li  &   2.0 &  0.071   &    Li  &   Li  &   Li  &   -1   &   16  &  0.036   \\
  Li    &   Li  &   3.0 &  0.071  &    Li  &   Li  &   Li  &   1   &   1  &   0.071   \\
  Li    &   Li  &   4.0 &  0.071   &    Li  &   Li  &   Li  &   1   &   2   &   0.071   \\
  Li    &   Li  &   0.0 &  0.179   &    Li  &   Li  &   Li  &   1   &   4   &   0.071   \\
  Li    &   Li  &   1.0 &  0.179   &    Li  &   Li  &   Li  &   1   &   16  &   0.071   \\
  Li    &   Li  &   2.0 &  0.179   &    Li  &   Li  &   Li  &   -1   &   1  &   0.071   \\
  Li    &   Li  &   3.0 &  0.179   &    Li  &   Li  &   Li  &   -1   &   2  &   0.071   \\
  Li    &   Li  &   4.0 &  0.179   &    Li  &   Li  &   Li  &   -1   &   4  &   0.071   \\
  Li    &   Li  &   0.0 &  0.357   &    Li  &   Li  &   Li  &   -1   &   16  &  0.071   \\
  Li    &   Li  &   1.0 &  0.357   &    Li  &   Li  &   Li  &   1   &   1   &   0.179   \\
  Li    &   Li  &   2.0 &  0.357  &    Li  &   Li  &   Li  &   1   &   2   &   0.179   \\
  Li    &   Li  &   3.0 &  0.357   &    Li  &   Li  &   Li  &   1   &   4   &   0.179   \\
  Li    &   Li  &   4.0 &  0.357   &    Li  &   Li  &   Li  &   1   &   16  &   0.179   \\
  Li    &   Li  &   0.0 &  0.714   &    Li  &   Li  &   Li  &   -1   &   1  &   0.179   \\
  Li    &   Li  &   1.0 &  0.714   &    Li  &   Li  &   Li  &   -1   &   2  &   0.179   \\
  Li    &   Li  &   2.0 &  0.714   &    Li  &   Li  &   Li  &   -1   &   4  &   0.179   \\
  Li    &   Li  &   3.0 &  0.714   &    Li  &   Li  &   Li  &   -1   &   16  &  0.179   \\
  Li    &   Li  &   4.0 &  0.714   &    Li  &   Li  &   Li  &   1   &   1  &   0.357   \\
  Li    &   Li  &   0.0 &  1.786   &    Li  &   Li  &   Li  &   1   &   2   &   0.357   \\
  Li    &   Li  &   1.0 &  1.786  &    Li  &   Li  &   Li  &   1   &   4   &   0.357   \\
  Li    &   Li  &   2.0 &  1.786   &    Li  &   Li  &   Li  &   1   &   16  &   0.357   \\
  Li    &   Li  &   3.0 &  1.786   &    Li  &   Li  &   Li  &   -1   &   1  &   0.357   \\
  Li    &   Li  &   4.0 &  1.786   &    Li  &   Li  &   Li  &   -1   &   2  &   0.357   \\
  Li    &   Li  &   0.0 &  3.571   &    Li  &   Li  &   Li  &   -1   &   4  &   0.357   \\
  Li    &   Li  &   1.0 &  3.571   &    Li  &   Li  &   Li  &   -1   &   16  &  0.357   \\
  Li    &   Li  &   2.0 &  3.571   &    Li  &   Li  &   Li  &   1   &   1  &   0.714   \\
  Li    &   Li  &   3.0 &  3.571   &    Li  &   Li  &   Li  &   1   &   2   &   0.714   \\
  Li    &   Li  &   4.0 &  3.571   &    Li  &   Li  &   Li  &   1   &   4   &   0.714   \\
  Li    &   Li  &   0.0 &  7.142  &    Li  &   Li  &   Li  &   1   &   16  &   0.714   \\
  Li    &   Li  &   1.0 &  7.142   &    Li  &   Li  &   Li  &   -1   &   1  &   0.714   \\
  Li    &   Li  &   2.0 &  7.142   &    Li  &   Li  &   Li  &   -1   &   2  &   0.714   \\
  Li    &   Li  &   3.0 &  7.142   &    Li  &   Li  &   Li  &   -1   &   4  &   0.714   \\
  Li    &   Li  &   4.0 &  7.142   &    Li  &   Li  &   Li  &   -1   &   16  &  0.714   \\
  Li    &   Li  &   0.0 &  17.855   &    Li  &   Li  &   Li  &   1   &   1  &   1.786   \\
  Li    &   Li  &   1.0 &  17.855   &    Li  &   Li  &   Li  &   1   &   2  &   1.786   \\
  Li    &   Li  &   2.0 &  17.855   &    Li  &   Li  &   Li  &   1   &   4  &   1.786   \\
  Li    &   Li  &   3.0 &  17.855   &    Li  &   Li  &   Li  &   1   &   16  &  1.786   \\
  Li    &   Li  &   4.0 &  17.855  &    Li  &   Li  &   Li  &   -1   &   1  &   1.786   \\
        &       &          &       &    Li  &   Li  &   Li  &   -1   &   2  &   1.786   \\
        &       &          &       &    Li  &   Li  &   Li  &   -1   &   4  &   1.786   \\
        &       &          &       &    Li  &   Li  &   Li  &   -1   &   16  &   1.786   \\
        &       &          &       &    Li  &   Li  &   Li  &   1   &   1  &   3.571   \\
        &       &          &       &    Li  &   Li  &   Li  &   1   &   2  &   3.571   \\
        &       &          &       &    Li  &   Li  &   Li  &   1   &   4  &   3.571   \\
        &       &          &       &    Li  &   Li  &   Li  &   1   &   16  &  3.571   \\
        &       &          &       &    Li  &   Li  &   Li  &   -1   &   1  &   3.571   \\
        &       &          &       &    Li  &   Li  &   Li  &   -1   &   2  &   3.571   \\
        &       &          &       &    Li  &   Li  &   Li  &   -1   &   4  &   3.571   \\
        &       &          &       &    Li  &   Li  &   Li  &   -1   &   16  &   3.571   \\
        &       &          &       &    Li  &   Li  &   Li  &   1   &   1  &   7.142   \\
        &       &          &       &    Li  &   Li  &   Li  &   1   &   2  &   7.142   \\
        &       &          &       &    Li  &   Li  &   Li  &   1   &   4  &   7.142   \\
        &       &          &       &    Li  &   Li  &   Li  &   1   &   16  &  7.142   \\
        &       &          &       &    Li  &   Li  &   Li  &   -1   &   1  &   7.142   \\
        &       &          &       &    Li  &   Li  &   Li  &   -1   &   2  &   7.142   \\
        &       &          &       &    Li  &   Li  &   Li  &   -1   &   4  &   7.142   \\
        &       &          &       &    Li  &   Li  &   Li  &   -1   &   16  &   7.142   \\
        &       &          &       &    Li  &   Li  &   Li  &   1   &   1  &   17.855   \\
        &       &          &       &    Li  &   Li  &   Li  &   1   &   2  &   17.855   \\
        &       &          &       &    Li  &   Li  &   Li  &   1   &   4  &   17.855   \\
        &       &          &       &    Li  &   Li  &   Li  &   1   &   16  &  17.855   \\
        &       &          &       &    Li  &   Li  &   Li  &   -1   &   1  &   17.855   \\
        &       &          &       &    Li  &   Li  &   Li  &   -1   &   2  &   17.855   \\
        &       &          &       &    Li  &   Li  &   Li  &   -1   &   4  &   17.855   \\
        &       &          &       &    Li  &   Li  &   Li  &   -1   &   16  &   17.855   \\
\end{longtable}
\end{center}

\pagebreak

\begin{center}
\LTcapwidth=\textwidth
\begin{longtable}{M{2.0cm} M{2.0cm} M{0.8cm} M{1.2cm} | M{2.0cm} M{2.0cm} M{2.0cm} M{0.8cm} M{0.8cm} M{1.2cm}}
\caption{\label{table:NNP_parameters_for_Mo} Symmetry function parameters for optimized Mo NNP. The units of $R_{c}$, $R_{s}$, $\eta$ and $\eta^{\prime}$ are {\AA}, {\AA}, \AA$^{-2}$ and \AA$^{-2}$, respectively.}
\\
\hline
\hline
\multicolumn{4}{c}{$R_{c}$} & \multicolumn{6}{c}{5.2}\\
\hline
\multicolumn{4}{c}{$G_{i}^{\rm atom, \rm rad}$} & \multicolumn{6}{c}{$G_{i}^{\rm atom, \rm ang}$}\\
\hline
 Reference & Neighbor & $R_{s}$ &  $\eta$ & Reference & Neighbor 1 & Neighbor 2 & $\lambda$ & $\zeta$ & $\eta^{\prime}$\\
\hline
\endfirsthead

\multicolumn{10}{c}%
{{\bfseries \tablename\ \thetable{} -- continued from previous page}} \\

\hline
\multicolumn{4}{c}{$G_{i}^{\rm atom, \rm rad}$} & \multicolumn{6}{c}{$G_{i}^{\rm atom, \rm ang}$}\\
\hline
 Reference & Neighbor & $R_{s}$ &  $\eta$  & Reference & Neighbor 1 & Neighbor 2 & $\lambda$ & $\zeta$ & $\eta^{\prime}$\\
\hline
\endhead

\hline \multicolumn{10}{r}{{Continued on next page}} \\
\endfoot

\hline \hline
\endlastfoot

  Mo    &   Mo  &   0.0 &  0.036   &    Mo  &   Mo  &   Mo  &   1   &   1   &   0.036   \\
  Mo    &   Mo  &   1.0 &  0.036   &    Mo  &   Mo  &   Mo  &   1   &   2   &   0.036   \\
  Mo    &   Mo  &   2.0 &  0.036   &    Mo  &   Mo  &   Mo  &   1   &   4   &   0.036   \\
  Mo    &   Mo  &   3.0 &  0.036   &    Mo  &   Mo  &   Mo  &   1   &   16  &   0.036   \\
  Mo    &   Mo  &   4.0 &  0.036   &    Mo  &   Mo  &   Mo  &   -1   &   1   &   0.036   \\
  Mo    &   Mo  &   0.0 &  0.071   &    Mo  &   Mo  &   Mo  &   -1   &   2   &   0.036   \\
  Mo    &   Mo  &   1.0 &  0.071   &    Mo  &   Mo  &   Mo  &   -1   &   4   &   0.036   \\
  Mo    &   Mo  &   2.0 &  0.071   &    Mo  &   Mo  &   Mo  &   -1   &   16  &   0.036   \\
  Mo    &   Mo  &   3.0 &  0.071   &    Mo  &   Mo  &   Mo  &   1   &   1  &   0.071   \\
  Mo    &   Mo  &   4.0 &  0.071   &    Mo  &   Mo  &   Mo  &   1   &   2  &   0.071   \\
  Mo    &   Mo  &   0.0 &  0.179   &    Mo  &   Mo  &   Mo  &   1   &   4  &   0.071   \\
  Mo    &   Mo  &   1.0 &  0.179   &    Mo  &   Mo  &   Mo  &   1   &   16  &   0.071   \\
  Mo    &   Mo  &   2.0 &  0.179   &    Mo  &   Mo  &   Mo  &   -1   &   1  &   0.071   \\
  Mo    &   Mo  &   3.0 &  0.179   &    Mo  &   Mo  &   Mo  &   -1   &   2  &   0.071   \\
  Mo    &   Mo  &   4.0 &  0.179   &    Mo  &   Mo  &   Mo  &   -1   &   4  &   0.071   \\
        &       &       &          &    Mo  &   Mo  &   Mo  &   -1   &   16  &  0.071   \\
        &       &       &          &    Mo  &   Mo  &   Mo  &   1   &   1  &   0.179   \\
        &       &       &          &    Mo  &   Mo  &   Mo  &   1   &   2  &   0.179   \\
        &       &       &          &    Mo  &   Mo  &   Mo  &   1   &   4  &   0.179   \\
        &       &       &          &    Mo  &   Mo  &   Mo  &   1   &   16  &   0.179   \\
        &       &       &          &    Mo  &   Mo  &   Mo  &   -1   &   1  &   0.179   \\
        &       &       &          &    Mo  &   Mo  &   Mo  &   -1   &   2  &   0.179   \\
        &       &       &          &    Mo  &   Mo  &   Mo  &   -1   &   4  &   0.179   \\
        &       &       &          &    Mo  &   Mo  &   Mo  &   -1   &   16  &  0.179   \\
\end{longtable}
\end{center}

\pagebreak

\begin{center}
\LTcapwidth=\textwidth
\begin{longtable}{M{2.0cm} M{2.0cm} M{0.8cm} M{1.2cm} | M{2.0cm} M{2.0cm} M{2.0cm} M{0.8cm} M{0.8cm} M{1.2cm}}
\caption{\label{table:NNP_parameters_for_Si} Symmetry function parameters for optimized Si NNP. The units of $R_{c}$, $R_{s}$, $\eta$ and $\eta^{\prime}$ are {\AA}, {\AA}, \AA$^{-2}$ and \AA$^{-2}$, respectively.}
\\
\hline
\hline
\multicolumn{4}{c}{$R_{c}$} & \multicolumn{6}{c}{5.2}\\
\hline
\multicolumn{4}{c}{$G_{i}^{\rm atom, \rm rad}$} & \multicolumn{6}{c}{$G_{i}^{\rm atom, \rm ang}$}\\
\hline
 Reference & Neighbor & $R_{s}$ &  $\eta$ & Reference & Neighbor 1 & Neighbor 2 & $\lambda$ & $\zeta$ & $\eta^{\prime}$\\
\hline
\endfirsthead

\multicolumn{10}{c}%
{{\bfseries \tablename\ \thetable{} -- continued from previous page}} \\

\hline
\multicolumn{4}{c}{$G_{i}^{\rm atom, \rm rad}$} & \multicolumn{6}{c}{$G_{i}^{\rm atom, \rm ang}$}\\
\hline
 Reference & Neighbor & $R_{s}$ &  $\eta$  & Reference & Neighbor 1 & Neighbor 2 & $\lambda$ & $\zeta$ & $\eta^{\prime}$\\
\hline
\endhead

\hline \multicolumn{10}{r}{{Continued on next page}} \\
\endfoot

\hline \hline
\endlastfoot

  Si    &   Si  &   0.0 &  0.036   &    Si  &   Si  &   Si  &   1   &   1   &   0.036   \\
  Si    &   Si  &   0.0 &  0.071   &    Si  &   Si  &   Si  &   1   &   1   &   0.071   \\
  Si    &   Si  &   0.0 &  0.179   &    Si  &   Si  &   Si  &   1   &   1   &   0.179   \\
  Si    &   Si  &   0.0 &  0.357   &    Si  &   Si  &   Si  &   1   &   1  &   0.357   \\
  Si    &   Si  &   0.0 &  0.714   &    Si  &   Si  &   Si  &   1   &   1   &   0.714   \\
  Si    &   Si  &   0.0 &  1.786   &    Si  &   Si  &   Si  &   1   &   1   &   1.786   \\
  Si    &   Si  &   0.0 &  3.571   &    Si  &   Si  &   Si  &   1   &   1  &   3.571   \\
  Si    &   Si  &   0.0 &  7.142   &    Si  &   Si  &   Si  &   1   &   1  &   7.142   \\
  Si    &   Si  &   0.0 &  17.855  &    Si  &   Si  &   Si  &   1   &   1  &   17.855   \\
        &       &       &          &    Si  &   Si  &   Si  &   -1   &   1  &   0.036   \\
        &       &       &          &    Si  &   Si  &   Si  &   -1   &   1  &   0.071   \\
        &       &       &          &    Si  &   Si  &   Si  &   -1   &   1  &   0.179   \\
        &       &       &          &    Si  &   Si  &   Si  &   -1   &   1  &   0.357   \\
        &       &       &          &    Si  &   Si  &   Si  &   -1   &   1  &   0.714   \\
        &       &       &          &    Si  &   Si  &   Si  &   -1   &   1  &   1.786   \\
        &       &       &          &    Si  &   Si  &   Si  &   -1   &   1  &   3.571   \\
        &       &       &          &    Si  &   Si  &   Si  &   -1   &   1  &   7.142   \\
        &       &       &          &    Si  &   Si  &   Si  &   -1   &   1  &   17.855   \\
\end{longtable}
\end{center}

\pagebreak

\begin{center}
\LTcapwidth=\textwidth
\begin{longtable}{M{2.0cm} M{2.0cm} M{0.8cm} M{1.2cm} | M{2.0cm} M{2.0cm} M{2.0cm} M{0.8cm} M{0.8cm} M{1.2cm}}
\caption{\label{table:NNP_parameters_for_Ge} Symmetry function parameters for optimized Ge NNP. The units of $R_{c}$, $R_{s}$, $\eta$ and $\eta^{\prime}$ are {\AA}, {\AA}, \AA$^{-2}$ and \AA$^{-2}$, respectively.}
\\
\hline
\hline
\multicolumn{4}{c}{$R_{c}$} & \multicolumn{6}{c}{5.1}\\
\hline
\multicolumn{4}{c}{$G_{i}^{\rm atom, \rm rad}$} & \multicolumn{6}{c}{$G_{i}^{\rm atom, \rm ang}$}\\
\hline
 Reference & Neighbor & $R_{s}$ &  $\eta$ & Reference & Neighbor 1 & Neighbor 2 & $\lambda$ & $\zeta$ & $\eta^{\prime}$\\
\hline
\endfirsthead

\multicolumn{10}{c}%
{{\bfseries \tablename\ \thetable{} -- continued from previous page}} \\

\hline
\multicolumn{4}{c}{$G_{i}^{\rm atom, \rm rad}$} & \multicolumn{6}{c}{$G_{i}^{\rm atom, \rm ang}$}\\
\hline
 Reference & Neighbor & $R_{s}$ &  $\eta$  & Reference & Neighbor 1 & Neighbor 2 & $\lambda$ & $\zeta$ & $\eta^{\prime}$\\
\hline
\endhead

\hline \multicolumn{10}{r}{{Continued on next page}} \\
\endfoot

\hline \hline
\endlastfoot

  Ge    &   Ge  &   0.0 &  0.036   &    Ge  &   Ge  &   Ge  &   1   &   1   &   0.036   \\
  Ge    &   Ge  &   1.0 &  0.036   &    Ge  &   Ge  &   Ge  &   -1   &   1  &   0.036   \\
  Ge    &   Ge  &   2.0 &  0.036   &    Ge  &   Ge  &   Ge  &   1   &   1  &   0.071   \\
  Ge    &   Ge  &   3.0 &  0.036   &    Ge  &   Ge  &   Ge  &   -1   &   1  &   0.071   \\
  Ge    &   Ge  &   4.0 &  0.036   &    Ge  &   Ge  &   Ge  &   1   &   1   &   0.179   \\
  Ge    &   Ge  &   0.0 &  0.071   &    Ge  &   Ge  &   Ge  &   -1   &   1  &   0.179   \\
  Ge    &   Ge  &   1.0 &  0.071   &    Ge  &   Ge  &   Ge  &   1   &   1  &   0.357   \\
  Ge    &   Ge  &   2.0 &  0.071   &    Ge  &   Ge  &   Ge  &   -1   &   1  &   0.357   \\
  Ge    &   Ge  &   3.0 &  0.071  &    Ge  &   Ge  &   Ge  &   1   &   1  &   0.714   \\
  Ge    &   Ge  &   4.0 &  0.071   &    Ge  &   Ge  &   Ge  &   -1   &   1  &   0.714   \\
  Ge    &   Ge  &   0.0 &  0.179   &    Ge  &   Ge  &   Ge  &   1   &   1  &   1.786   \\
  Ge    &   Ge  &   1.0 &  0.179   &    Ge  &   Ge  &   Ge  &   -1   &   1  &   1.786   \\
  Ge    &   Ge  &   2.0 &  0.179   &    Ge  &   Ge  &   Ge  &   1   &   1  &   3.571   \\
  Ge    &   Ge  &   3.0 &  0.179   &    Ge  &   Ge  &   Ge  &   -1   &   1  &   3.571   \\
  Ge    &   Ge  &   4.0 &  0.179   &    Ge  &   Ge  &   Ge  &   1   &   1  &   7.142   \\
  Ge    &   Ge  &   0.0 &  0.357   &    Ge  &   Ge  &   Ge  &   -1   &   1  &   7.142   \\
  Ge    &   Ge  &   1.0 &  0.357   &    Ge  &   Ge  &   Ge  &   1   &   1  &   17.855   \\
  Ge    &   Ge  &   2.0 &  0.357  &    Ge  &   Ge  &   Ge  &   -1   &   1  &   17.855   \\
  Ge    &   Ge  &   3.0 &  0.357   &        &       &       &       &       &      \\
  Ge    &   Ge  &   4.0 &  0.357   &        &       &       &       &      &      \\
  Ge    &   Ge  &   0.0 &  0.714   &        &       &       &       &      &      \\
  Ge    &   Ge  &   1.0 &  0.714   &        &       &       &       &      &      \\
  Ge    &   Ge  &   2.0 &  0.714   &        &       &       &       &      &      \\
  Ge    &   Ge  &   3.0 &  0.714   &        &       &       &       &      &     \\
  Ge    &   Ge  &   4.0 &  0.714   &        &       &       &       &      &      \\
  Ge    &   Ge  &   0.0 &  1.786   &        &       &       &       &       &      \\
  Ge    &   Ge  &   1.0 &  1.786  &        &       &       &       &       &      \\
  Ge    &   Ge  &   2.0 &  1.786   &        &       &       &       &      &      \\
  Ge    &   Ge  &   3.0 &  1.786   &        &       &       &       &      &      \\
  Ge    &   Ge  &   4.0 &  1.786   &        &       &       &       &      &      \\
  Ge    &   Ge  &   0.0 &  3.571   &        &       &       &       &      &      \\
  Ge    &   Ge  &   1.0 &  3.571   &        &       &       &       &      &     \\
  Ge    &   Ge  &   2.0 &  3.571   &        &       &       &       &      &      \\
  Ge    &   Ge  &   3.0 &  3.571   &        &       &       &       &       &      \\
  Ge    &   Ge  &   4.0 &  3.571   &        &       &       &       &       &      \\
  Ge    &   Ge  &   0.0 &  7.142  &        &       &       &       &       &      \\
  Ge    &   Ge  &   1.0 &  7.142   &        &       &       &       &      &      \\
  Ge    &   Ge  &   2.0 &  7.142   &        &       &       &       &      &      \\
  Ge    &   Ge  &   3.0 &  7.142   &        &       &       &       &      &      \\
  Ge    &   Ge  &   4.0 &  7.142   &        &       &       &       &      &     \\
  Ge    &   Ge  &   0.0 &  17.855   &        &       &       &       &      &      \\
  Ge    &   Ge  &   1.0 &  17.855   &        &       &       &       &      &      \\
  Ge    &   Ge  &   2.0 &  17.855   &        &       &       &       &      &      \\
  Ge    &   Ge  &   3.0 &  17.855   &        &       &       &       &      &     \\
  Ge    &   Ge  &   4.0 &  17.855  &        &       &       &        &      &     \\
\end{longtable}
\end{center}

\pagebreak

\subsection{Gaussian approximation potentials}

The Gaussian Approximation Potentials (GAPs)\cite{GAP_review_2010, GAP_SOAP, GAP_review_2015}, implemented in the QUIP program package\cite{libatoms}, applies Gaussian process regression (GPR) to interpolate the atomic energy based on the spatial distribution of the neighboring atoms. The atomic energy function is given as follows:
\begin{equation}
    \epsilon(\boldsymbol{R}) = \sum_{k} b_{k} \boldsymbol{K}(\boldsymbol{R}, \boldsymbol{R}_{k}),
\end{equation}
where $\boldsymbol{R}$ represents the geometry of neighboring atoms within cutoff radius $R_{c}$, and $k$ indexes a set of reference data points ${\boldsymbol{R}_{k}}$ that serve as a basis on which the atomic energy function is expanded. $\boldsymbol{K}$ is a kernel function that captures the variation of the energy function in terms of changing neighboring configurations.

In this work, the kernel function used is the ``Smooth Overlap of Atomic Positions'' (SOAP)\cite{GAP_SOAP, GAP_molecule_bulk, GAP_Boron, GAP_Si}, which represents the rotationally integrated overlap of neighbor densities. The atomic configuration of a central atom \textit{i} is represented by its neighbor density as a sum of Gaussians centered over all neighboring atoms \textit{j} within a cutoff distance $R_c$,
\begin{equation}
    \rho_{i}(\boldsymbol{R}) = \sum_{j} f_{c}(R_{ij}) \cdot \exp(-\frac{|\boldsymbol{R} - \boldsymbol{R}_{ij}|^{2}}{2\sigma_{\rm atom}^{2}}),
\end{equation}
where $f_{c}$ is a cutoff function ensures smooth decay in value and slope at cutoff radius $R_{c}$, and $\sigma_{atom}$ is a smearing parameter. This neighbor density is then expanded in terms of a basis of spherical harmonics $Y_{lm}(\hat{\boldsymbol{R}})$, and radial functions $g_{n}(R)$:
\begin{equation}
    \rho_{i}(\boldsymbol{R}) = \sum_{nlm} c_{nlm} \ g_{n}(R) Y_{lm}(\hat{\boldsymbol{R}}),
\end{equation}
and the rotationally invariant spherical power spectrum of atom \textit{i} is then expressed in terms of expansion coefficients $c_{nlm}$ as follows:
\begin{equation}
    p_{n_{1}n_{2}l} (\boldsymbol{R}_{i}) = \sum_{m = -l}^{l} c_{n_{1}lm}^{\ast} c_{n_{2}lm},
\end{equation}
Finally, the SOAP kernel is written as a dot product of power spectrums and raised to a small integer power as follows:
\begin{equation}
    \boldsymbol{K}(\boldsymbol{R}, \boldsymbol{R^{\prime}}) = \sum_{n_{1}n_{2}l} (p_{n_{1}n_{2}l}(\boldsymbol{R}) p_{n_{1}n_{2}l}(\boldsymbol{R^{\prime}}))^{\zeta},
\end{equation}
where $\zeta = 4$ in the present work. Normalization is then carried out to ensure that the kernel of each atomic environment with itself is unity.

The choice of hyperparameters such as cutoff radius and regularization parameters in the linear system has a significant influence on the performance of GAP models\cite{GAP_tungsten, GAP_C}. The optimized parameters for each elemental system, determined via a grid search, are listed in \textbf{Table \ref{table:GAP_hyperparameters}}.

\begin{center}
\begin{table}
\begin{tabular}{  M{4cm}  M{1.5cm}  M{1.5cm}  M{1.5cm}  M{1.5cm}  M{1.5cm}  M{1.5cm} }
\hline
\hline
\noalign{\smallskip}
    &   Ni  &   Cu  &   Li  &   Mo  &   Si  &   Ge  \\
\noalign{\smallskip}
\hline
\noalign{\smallskip}
cutoff radius ({\AA})    &   3.9 &   4.1 &   4.8    &   5.2 &   5.1 &   5.4   \\
\noalign{\smallskip}
\hline
\noalign{\smallskip}
$\sigma_{\text{energy}}$ (eV/atom)  &   0.01    &   0.01    &   0.01    &   0.01    &   0.01    &   0.01    \\
\noalign{\smallskip}
$\sigma_{\text{force}}$ (eV/{\AA})  &   0.05    &   0.05    &   0.05    &   0.05    &   0.05    &   0.05    \\
\noalign{\smallskip}
$(n_{\text{max}}, l_{\text{max}})$                &   (8, 8)  &   (8, 8)  &   (8, 8)  &   (8, 8)  &   (8, 8)    &   (8, 8)  \\
\noalign{\smallskip}
\hline
\end{tabular}
\caption{\label{table:GAP_hyperparameters} Key parameters of the GAP model optimized for each elemental system}
\end{table}
\end{center}

\pagebreak

\subsection{Spectral neighbor analysis potential}

The bispectrum of the neighbor density and spectral neighbor analysis potential (SNAP) formalism has been extensively studied in the previous works\cite{GAP_SOAP, SNAP}. The basic concept of the bispectrum formalism is to project the 3D local atomic neighbor density into a set of coefficients that satisfy the invariant properties by expansion on a spherical harmonics basis.

In SNAP, the atomic neighbor density around a central atom \textit{i} at position $\boldsymbol{R}$ is defined as follows:
\begin{equation}
    \rho_{i}(\textbf{R}) = \delta(\textbf{R}) + \sum_{R_{ij} < R_{c}} f_{c}(R_{ij}) \cdot \omega_{j} \cdot \delta(\textbf{R} - \textbf{R}_{ij}),
\end{equation}
where $R_{ij}$ is the position of neighbor atom \textit{j} relative to central atom \textit{i}, and $\omega_{j}$ is the dimensionless weight to discriminate atom types. The cutoff function $f_{c}$ ensures that the neighbor atomic density decreases smoothly to zero at the cutoff radius $R_{c}$.

The angular distribution of neighbor density function can be projected onto spherical harmonic functions $Y_{m}^{l}(\theta, \phi)$. In the bispectrum approach, the radial distribution is converted into an additional polar angle $\theta_{0}$ defined by $\theta_{0}=\theta_{0}^{\rm max}\frac{r}{R_{c}}$. Thus the density function can be represented in the 3-sphere $(\theta, \phi, \theta_{0})$ coordinates instead of $(\theta, \phi, r)$. The density function on 3-sphere can then be expanded with 4-dimensional hyperspherical harmonics $U_{m, m\prime}^{j}$, as follows:
\begin{equation}
    \rho_{i}(\textbf{R}) = \sum_{j=0}^{\infty} \sum_{m, m^{\prime}=-j}^{j} u_{m, m^{\prime}}^{j} U_{m, m^{\prime}}^{j},
\end{equation}
where the coefficients $u_{m, m^{\prime}}^{j}$ are obtained as the inner product of the neighbor density function with the basic function given by the following:
\begin{equation}
    u_{m, m^{\prime}}^{j} = U_{m, m^{\prime}}^{j}(0, 0, 0) + \sum_{R_{ij} < R_{c}} f_{c}(R_{ij}) \cdot \omega_{j} \cdot U_{m, m^{\prime}}^{j}(\theta, \phi, \theta_{0}).
\end{equation}
The bispectrum components $B_{j_{1}, j_{2}, j}$ can then obtained via following:
\begin{equation}
\begin{split}
    B_{j_{1}, j_{2}, j} &= \sum_{m_{1}, m_{1}^{\prime}=-j_{1}}^{j_{1}} 
                       \sum_{m_{2}, m_{2}^{\prime}=-j_{2}}^{j_{2}}
                       \sum_{m, m^{\prime}=-j}^{j} (u_{m, m^{\prime}}^{j})^{\ast}
                    \\&
    \times H\substack{j m m\prime \\ j_{1} m_{1} m_{1}^{\prime} \\ j_{2} m_{2} m_{2}^{\prime}} u_{m_{1}, m_{1}^{\prime}}^{j_{1}} u_{m_{2}, m_{2}^{\prime}}^{j_{2}},
\end{split}
\end{equation}
where constants $H\substack{j m m\prime \\ j_{1} m_{1} m_{1}^{\prime} \\ j_{2} m_{2} m_{2}^{\prime}}$ are coupling coefficients satisfying the conditions $\|j_{1} - j_{2}\| \leq j \leq \|j_{1} + j_{2}\|$.

In the SNAP formalism, the energy $E_{\rm SNAP}$, force $\boldsymbol{F}_{\rm SNAP}^{j}$, and stresses $\boldsymbol{\sigma}^{j}_{\rm SNAP}$ are expressed as linear functions of the bispectrum components\cite{SNAP, SNAP_tungsten} as follows:
\begin{equation}
    E_{\rm SNAP} = \beta_{0}N + \boldsymbol{\beta} \cdot \sum_{i=1}^{N}\boldsymbol{B}^{i},
\end{equation}
\begin{equation}
    \boldsymbol{F}_{\rm SNAP}^{j} = -\boldsymbol{\beta} \cdot \sum_{i=1}^{N} \frac{\partial \boldsymbol{B}^{i}}{\partial \boldsymbol{r}_{j}},
\end{equation}
\begin{equation}
    \boldsymbol{\sigma}_{\rm SNAP}^{j} = -\boldsymbol{\beta} \cdot \sum_{i=1}^{N}\boldsymbol{r}_{j} \otimes \sum_{i=1}^{N} \frac{\partial \boldsymbol{B}^{i}}{\partial \boldsymbol{r}_{j}},
\end{equation}
where $\beta_{0}$ and the vector $\boldsymbol{\beta}$ are the coefficients derived from fitting with the database of quantum electronic structure calculations.

In our previous works\cite{SNAP_Mo, SNAP_Nickel_Copper, eSNAP_Li3N}, we have developed a robust alternating two-step fitting process to obtain the optimal hyperparameters for SNAP. In each inner iteration, the fitting of a linear model is performed. In the outer loop, the model fitted in the inner iteration is then used to predict materials properties (e.g., elastic tensors), and the differences between predicted and reference values are then used to optimize hyperparameters via the differential evolution optimization algorithm implemented in Scipy package\cite{scipy}. The optimized SNAP model hyperparameters and coefficients for all elemental systems are provided in \textbf{Table \ref{table:SNAP_parameters}}.

\begin{center}
\LTcapwidth=\textwidth
% \begin{longtable}{  r r r r r r r r r r }
\begin{longtable}{ M{0.8cm} M{0.8cm} M{0.8cm} M{0.8cm} M{1.6cm} M{1.6cm} M{1.6cm} M{1.6cm} M{1.6cm} M{1.6cm} }
\caption{\label{table:SNAP_parameters} Key parameters of the SNAP model optimized for each elemental system}
\\
\hline
\hline
    &   &   &   &   \multicolumn{1}{c}{Ni}  &   \multicolumn{1}{c}{Cu}  &   \multicolumn{1}{c}{Li}  &   \multicolumn{1}{c}{Mo}  &   \multicolumn{1}{c}{Si}  &   \multicolumn{1}{c}{Ge}  \\
\hline
\multicolumn{4}{c}{cutoff radius ({\AA})}    &  \multicolumn{1}{c}{3.9}  &  \multicolumn{1}{c}{3.8}  &  \multicolumn{1}{c}{5.1}  &  \multicolumn{1}{c}{5.2}  &  \multicolumn{1}{c}{4.9} &   \multicolumn{1}{c}{5.0}   \\
\hline

$k$   &   $2j_{1}$    &   $2j_{2}$    &   $2j$    &   \multicolumn{1}{c}{$\beta_{k}^{\rm Ni}$}    &   \multicolumn{1}{c}{$\beta_{k}^{\rm Cu}$}     &   \multicolumn{1}{c}{$\beta_{k}^{\rm Li}$}     &   \multicolumn{1}{c}{$\beta_{k}^{\rm Mo}$}     &   \multicolumn{1}{c}{$\beta_{k}^{\rm Si}$}     &   \multicolumn{1}{c}{$\beta_{k}^{\rm Ge}$}  \\

\hline
\endfirsthead

\multicolumn{10}{c}%
{{\bfseries \tablename\ \thetable{} -- continued from previous page}} \\
\hline
    &   &   &   &   \multicolumn{1}{c}{Ni}  &   \multicolumn{1}{c}{Cu}  &   \multicolumn{1}{c}{Li}  &   \multicolumn{1}{c}{Mo}  &   \multicolumn{1}{c}{Si}  &   \multicolumn{1}{c}{Ge}  \\
\hline
\multicolumn{4}{c}{cutoff radius ({\AA})}    &  \multicolumn{1}{c}{3.9}  &  \multicolumn{1}{c}{4.1}  &  \multicolumn{1}{c}{5.1}  &  \multicolumn{1}{c}{4.6}  &  \multicolumn{1}{c}{4.9} &   \multicolumn{1}{c}{5.5}   \\
\hline

$k$   &   $2j_{1}$    &   $2j_{2}$    &   $2j$    &   \multicolumn{1}{c}{$\beta_{k}^{\rm Ni}$}    &   \multicolumn{1}{c}{$\beta_{k}^{\rm Cu}$}     &   \multicolumn{1}{c}{$\beta_{k}^{\rm Li}$}     &   \multicolumn{1}{c}{$\beta_{k}^{\rm Mo}$}     &   \multicolumn{1}{c}{$\beta_{k}^{\rm Si}$}     &   \multicolumn{1}{c}{$\beta_{k}^{\rm Ge}$}  \\

\hline
\endhead

\hline \multicolumn{10}{r}{{Continued on next page}} \\
\endfoot

\hline \hline
\endlastfoot

0   &   &   &   &   $-13.981$   &   $-12.560$   &   $-3.443$    &   $-17.762$  &    $-5.6640$    &   $-8.0321$    \\

1   & 0 & 0 & 0 &   $0.0145$     &   $0.0086$     &   $0.0029$     &   $0.0218$    &    $0.0000$     &   $0.0036$     \\
2   & 1 & 0 & 1 &   $0.0101$     &   $0.0137$     &   $0.0041$     &   $0.0034$    &    $0.0181$     &   $-0.0013$     \\
3   & 1 & 1 & 2 &   $0.1205$     &   $0.0639$     &   $0.0184$     &   $0.3091$    &    $0.0589$     &   $-0.0746$     \\
4   & 2 & 0 & 2 &   $-0.0069$     &   $-0.0094$     &   $0.0063$     &   $0.0290$    &    $-0.0160$     &   $-0.0148$     \\
5   & 2 & 1 & 3 &   $0.2776$     &   $0.1906$     &   $0.0562$     &   $0.6210$    &    $0.2494$     &   $-0.0045$     \\
6   & 2 & 2 & 2 &   $0.1109$     &   $0.0975$     &   $0.0065$     &   $0.2611$    &    $0.0307$     &   $0.0249$     \\
7   & 2 & 2 & 4 &   $0.1105$     &   $0.0810$     &   $0.0218$     &   $0.1901$    &    $0.0790$     &   $0.0239$     \\
8   & 3 & 0 & 3 &   $0.0103$     &   $0.0071$     &   $-0.0019$     &   $0.0173$    &    $0.0378$     &   $0.0116$     \\
9   & 3 & 1 & 4 &   $0.1234$     &   $0.0929$     &   $0.0281$     &   $0.2779$    &    $0.0801$     &   $0.0727$     \\
10  & 3 & 2 & 3 &   $0.1304$     &   $0.0880$     &   $0.0177$     &   $0.2888$    &    $0.1145$     &   $0.1424$     \\
11  & 3 & 2 & 5 &   $0.1353$     &   $0.1192$     &   $0.0350$     &   $0.1009$    &    $0.0877$     &   $0.0367$     \\
12  & 3 & 3 & 4 &   $0.0614$     &   $0.0653$     &   $0.0178$     &   $0.0940$    &    $-0.0145$     &   $0.0890$     \\
13  & 3 & 3 & 6 &   $0.0248$     &   $0.0314$     &   $0.0118$     &   $0.0116$    &    $-0.0168$     &   $0.0046$     \\
14  & 4 & 0 & 4 &   $0.0051$     &   $0.0028$     &   $0.0010$     &   $0.0181$    &    $0.0111$     &   $0.0046$     \\
15  & 4 & 1 & 5 &   $0.0962$     &   $0.0919$     &   $0.0181$     &   $0.1319$    &    $0.0156$     &   $-0.0040$     \\
16  & 4 & 2 & 4 &   $0.0488$     &   $0.0532$     &   $0.0036$     &   $0.0947$    &    $0.0268$     &   $0.0720$     \\
17  & 4 & 2 & 6 &   $0.0992$     &   $0.0988$     &   $0.0273$     &   $0.0970$    &    $0.0487$     &   $0.0229$     \\
18  & 4 & 3 & 5 &   $0.0818$     &   $0.0837$     &   $0.0204$     &   $0.0680$    &    $-0.0649$     &   $0.0370$     \\
19  & 4 & 3 & 7 &   $0.0200$     &   $0.0190$     &   $0.0094$     &                &    $0.0029$     &   $-0.0024$     \\
20  & 4 & 4 & 4 &   $0.0227$     &   $0.0284$     &   $0.0030$     &   $0.0091$    &    $-0.0199$     &   $-0.0060$     \\
21  & 4 & 4 & 6 &   $0.0316$     &   $0.0387$     &   $0.0037$     &   $-0.0173$    &    $-0.0179$     &   $0.0150$     \\
22  & 4 & 4 & 8 &   $0.0015$     &   $0.0043$     &   $0.0030$     &                &    $0.0031$     &   $0.0006$     \\
23  & 5 & 0 & 5 &   $-0.0004$     &   $0.0018$     &   $-0.0004$     &   $0.0033$    &    $0.0120$     &   $0.0063$     \\
24  & 5 & 1 & 6 &   $0.0584$     &   $0.0542$     &   $0.0102$     &   $0.0688$    &    $0.0037$     &   $0.0208$     \\
25  & 5 & 2 & 5 &   $0.0419$     &   $0.0486$     &   $0.0034$     &   $0.0628$    &    $0.0301$     &   $0.0532$     \\
26  & 5 & 2 & 7 &   $0.0565$     &   $0.0508$     &   $0.0098$     &                &    $0.0213$     &   $0.0115$     \\
27  & 5 & 3 & 6 &   $0.0560$     &   $0.0667$     &   $0.0115$     &   $-0.0080$    &    $-0.0364$     &   $-0.0140$     \\
28  & 5 & 3 & 8 &   $0.0004$     &   $0.0044$     &   $0.0030$     &               &    $-0.0083$     &   $0.0056$     \\
29  & 5 & 4 & 5 &   $0.0328$     &   $0.0447$     &   $0.0065$     &   $0.0162$    &    $-0.0499$     &   $-0.0343$     \\
30  & 5 & 4 & 7 &   $0.0289$     &   $0.0253$     &   $0.0043$     &                &    $-0.0011$     &   $0.0178$     \\
31  & 5 & 5 & 6 &   $0.0137$     &   $0.0222$     &   $0.0008$     &   $-0.0412$    &    $-0.0319$     &   $-0.0118$     \\
32  & 5 & 5 & 8 &   $0.0049$     &   $0.0039$     &   $-0.0005$     &               &    $-0.0031$     &   $0.0092$     \\
33  & 6 & 0 & 6 &   $-0.0070$     &   $-0.0059$     &   $-0.0015$     &   $-0.0033$    &    $0.0026$     &   $0.0028$     \\
34  & 6 & 1 & 7 &   $0.0316$     &   $0.0337$     &   $0.0049$     &                &    $0.0044$     &   $0.0151$     \\
35  & 6 & 2 & 6 &   $0.0367$     &   $0.0392$     &   $0.0002$     &   $0.0584$    &    $0.0325$     &   $0.0359$     \\
36  & 6 & 2 & 8 &   $0.0231$     &   $0.0191$     &   $0.0013$     &                &    $0.0058$     &   $0.0081$     \\
37  & 6 & 3 & 7 &   $0.0317$     &   $0.0304$     &   $0.0060$     &                &    $-0.0049$     &   $-0.0205$     \\
38  & 6 & 4 & 6 &   $0.0170$     &   $0.0250$     &   $0.0024$     &   $-0.0017$    &    $-0.0164$     &   $-0.0007$     \\
39  & 6 & 4 & 8 &   $0.0079$     &   $0.0044$     &   $0.0001$     &              &    $0.0021$     &   $0.0083$     \\
40  & 6 & 5 & 7 &   $0.0137$     &   $0.0175$     &   $-0.0009$     &               &    $-0.0272$     &   $-0.0241$     \\
41  & 6 & 6 & 6 &   $0.0048$     &   $0.0086$     &   $0.0007$     &   $0.0008$    &    $-0.0083$     &   $-0.0002$     \\
42  & 6 & 6 & 8 &   $0.0057$     &   $0.0011$     &   $0.0000$     &               &    $-0.0062$     &   $-0.0035$     \\
43  & 7 & 0 & 7 &   $-0.0037$     &   $0.0016$     &   $-0.0011$     &              &    $0.0006$     &   $0.0008$     \\
44  & 7 & 1 & 8 &   $0.0072$     &   $-0.0095$     &   $0.0029$     &              &    $0.0055$     &   $0.0024$     \\
45  & 7 & 2 & 7 &   $0.0168$     &   $0.0077$     &   $-0.0015$     &              &    $0.0211$     &   $0.0436$     \\
46  & 7 & 3 & 8 &   $0.0086$     &   $0.0053$     &   $0.0029$     &                &    $-0.0073$     &   $-0.0048$     \\
47  & 7 & 4 & 7 &   $0.0025$     &   $0.0038$     &   $0.0005$     &                &    $-0.0029$     &   $-0.0015$     \\
48  & 7 & 5 & 8 &   $0.0007$     &   $0.0009$     &   $0.0000$     &               &    $-0.0113$     &   $-0.0047$     \\
49  & 7 & 6 & 7 &   $0.0043$     &   $0.0017$     &   $0.0006$     &               &    $-0.0088$     &   $-0.0075$     \\
50  & 7 & 7 & 8 &   $0.0008$     &   $-0.0033$     &   $0.0000$     &              &    $-0.0069$     &   $0.0055$     \\
51  & 8 & 0 & 8 &   $-0.0008$     &   $0.0015$     &   $-0.0004$     &               &    $0.0012$     &   $-0.0009$     \\
52  & 8 & 2 & 8 &   $0.0025$     &   $-0.0081$     &   $-0.0005$     &             &    $0.0071$     &   $0.0170$     \\
53  & 8 & 4 & 8 &   $-0.0008$     &   $0.0019$     &   $0.0005$     &               &    $0.0010$     &   $-0.0124$     \\
54  & 8 & 6 & 8 &   $-0.0009$     &   $-0.0023$     &   $-0.0003$     &             &    $0.0069$     &   $0.0038$     \\
55  & 8 & 8 & 8 &   $0.0008$     &   $0.0020$     &   $0.0003$     &                &    $-0.0034$     &   $-0.0014$     \\

\end{longtable}
\end{center}

\pagebreak

\subsection{Quadratic Spectral neighbor analysis potential}

In the initial formulation of SNAP, the linear form of the PES to bispectrum relationship limits the maximum complexity of energy functions to a four-body effect, which may in turn have an impact on its predictive power. Recently \citet{qSNAP} proposed a quadratic extension of SNAP (qSNAP) approach. The quadratic contributions to the energy can be viewed as a kind of embedding energy in analogy with the embedded atom method (EAM)\cite{EAM_fcc, EAM_transition_metals}. In qSNAP, the SNAP potential is extended via the addition of an embedding energy term as follows:
\begin{equation}
    E_{\rm SNAP}^{i} = \boldsymbol{\beta} \cdot \boldsymbol{B}^{i} + F(\rho_{i}),
\end{equation}
where $F(\rho_{i})$ represents the energy of embedding atom \textit{i} into the electron density contributed by its neighboring atoms. The ``host'' electron density of embedding atom \textit{i} can be expressed as a linear function of the bispectrum components as follows:
\begin{equation}
    \rho_{i} = \boldsymbol{a} \cdot \boldsymbol{B}^{i},
\end{equation}
and the embedding energy can be expressed as a Taylor expansion based on a reference structure with density $\rho_{0}$ as follows:
\begin{equation}
    F(\rho) = F_{0} + (\rho - \rho_{0})F^{\prime} + \frac{1}{2}(\rho - \rho_{0})^{2}F^{\prime\prime} + \cdots.
\end{equation}

Thus, the modified SNAP energy is then given by the following expression:
\begin{equation}
\begin{split}
    E_{\rm SNAP}^{i} &= \boldsymbol{\beta} \cdot \boldsymbol{B}^{i} + \frac{1}{2} F^{\prime\prime}(\boldsymbol{a} \cdot \boldsymbol{B}^{i})^{2} \\
    &= \boldsymbol{\beta} \cdot \boldsymbol{B}^{i} + \frac{1}{2}(\boldsymbol{B}^{i})^{T} \cdot \boldsymbol{\alpha} \cdot \boldsymbol{B}^{i},
\end{split}
\end{equation}
where $\boldsymbol{\alpha} = F^{\prime\prime}\boldsymbol{a}\otimes\boldsymbol{a}$ is a symmetric $K \times K$ matrix. Essentially the quadratic extension carries all distinct pairwise products of bispectrum components, and expands the maximum complexity of energy functions to seven-body effects\cite{qSNAP}.

In this work, the same two-step fitting approach is used for both the SNAP and qSNAP models. The optimized qSNAP hyperparameters for all elemental systems are listed in \textbf{Table \ref{table:qSNAP_hyperparameters}}.

\begin{center}
\begin{table}
\begin{tabular}{  M{4cm}  M{1.5cm}  M{1.5cm}  M{1.5cm}  M{1.5cm}  M{1.5cm}  M{1.5cm} }
\hline
\hline
\noalign{\smallskip}
    &   Ni  &   Cu  &   Li  &   Mo  &   Si  &   Ge  \\
\noalign{\smallskip}
\hline
\noalign{\smallskip}
cutoff radius ({\AA})    &   3.8 &   3.9 &   5.1    &   5.2 &   4.8 &   4.9   \\
\noalign{\smallskip}
\hline
\noalign{\smallskip}
$J_{\text{max}}$   &   6   &   6   &   6   &   6   &   8     &   6 \\
\noalign{\smallskip}
\hline
\end{tabular}
\caption{Key parameters of the qSNAP model optimized for each elemental system}
\label{table:qSNAP_hyperparameters}
\end{table}
\end{center}

\pagebreak

\subsection{Moment tensor potentials}

The Moment Tensor Potential (MTP) model and its formalism have been studied in previous works\cite{MTP, MTP_activelearning, MTP_diffusion}. The fundamental idea of MTP is to construct a contracted rotationally invariant representation of the local atomic environment in a tensorial sense and build a linear correlation between potential energy and atomic representation based on the assumption that the total energy can be partitioned into individual atomic environment contributions.

As is denoted previously\cite{MTP, MTP_activelearning}, the potential energy of the atomic environment of central atom \textit{i} can be linearly expanded on a set of basis functions $B(\boldsymbol{R})$,
\begin{equation}
    V_{i}(\boldsymbol{R}) = \sum_{l} \beta_{l} B(\boldsymbol{R}),
\end{equation}
The basis functions $B(\boldsymbol{R})$, in turn, depend on a series of moment tensor descriptors over all neighbor atoms \textit{j},
\begin{equation}
    M_{\mu, \nu}(\boldsymbol{R}) = \sum_{j} f_{\mu}(R_{ij})\ \underbrace{\boldsymbol{R}_{ij} \otimes \dots \otimes \boldsymbol{R}_{ij}}_{\nu\ \rm times},
\end{equation}
where the functions $f_{\mu}$ are the radial distribution of atomic configuration and the terms $\boldsymbol{R}_{ij} \otimes \dots \otimes \boldsymbol{R}_{ij}$ are tensors of rank $\nu$ entailing angular information about the atomic configuration. These descriptors are rotationally invariant in a tensorial sense (or, to be more precise, rotationally covariant).

The choice of $\mu$ and $\nu$ provides the balance between computational complexity and computational efficiency of MTP\cite{MTP}. The basis functions $B(\boldsymbol{R})$ are formulated by different ways of contracting the moment tensors $M_{\mu, \nu}$ to a scalar. Each contraction can be encoded by a symmetric $m\times m$ matrix $\alpha$ where diagonal elements $\alpha_{ij}$ demonstrate dimensions of the \textit{i}th and \textit{j}th tensors being contracted. In this work we have used the implementation described in previous work\cite{MTP_alloy_search} which for a single component reduces to learning the optimal radial functions $f_{\mu}$ instead of fixing them to be a universal radial basis.

We have carefully examined the choice of hyperparameters including the cutoff radius, the number of polynomial powers as well as the number of free parameters by evaluating the performance on reproducing the basic properties, e.g., elastic constants and phonon spectrum, while the weighted parameters are less relevant to the predicted properties. A full set of optimized hyperparameters for each elemental system has been presented in \textbf{Table \ref{table:MTP_hyperparameters}}. The maximum number of iterations has been set to 500, 1000 and 2000 to ensure the convergence of the fitting process for 16 or lower polynomial powers, 20 polynomial powers and 24 polynomial powers, respectively.

\begin{center}
\begin{table}
\begin{tabular}{  M{4.5cm}  M{1.5cm}  M{1.5cm}  M{1.5cm}  M{1.5cm}  M{1.5cm}  M{1.5cm} }
\hline
\hline
\noalign{\smallskip}
    &   Ni  &   Cu  &   Li  &   Mo  &   Si  &   Ge  \\
\noalign{\smallskip}
\hline
\noalign{\smallskip}
cutoff radius ({\AA})    &   4.0 &   3.9 &   5.1    &   5.2 &   4.7 &   5.1   \\
\noalign{\smallskip}
\hline
\noalign{\smallskip}
\# of polynomial powers   &   20    &   20    &   16    &   20    &   24    &   24    \\
\noalign{\smallskip}
\# of free parameters  &   329    &   329    &   125    &   329    &   913    &       913 \\
\noalign{\smallskip}
Energy weight           &   1.00     &   1.00     &   1.00     &   1.00     &   1.00     &     1.00  \\
\noalign{\smallskip}
Force weight            &   0.01       &   0.01       &   0.01       &   0.01       &   0.01       &     0.01 \\
\noalign{\smallskip}
\hline
\end{tabular}
\caption{Key parameters of the MTP model optimized for each elemental system}
\label{table:MTP_hyperparameters}
\end{table}
\end{center}

\section{Phonon dispersion curves computed DFT, GAP, MTP, NNP, SNAP and qSNAP}

\clearpage

\begin{figure}
    \centering
    \subfigure[DFT]{\label{fig:dft_Mo_phonon}\includegraphics[width=0.37\textwidth]{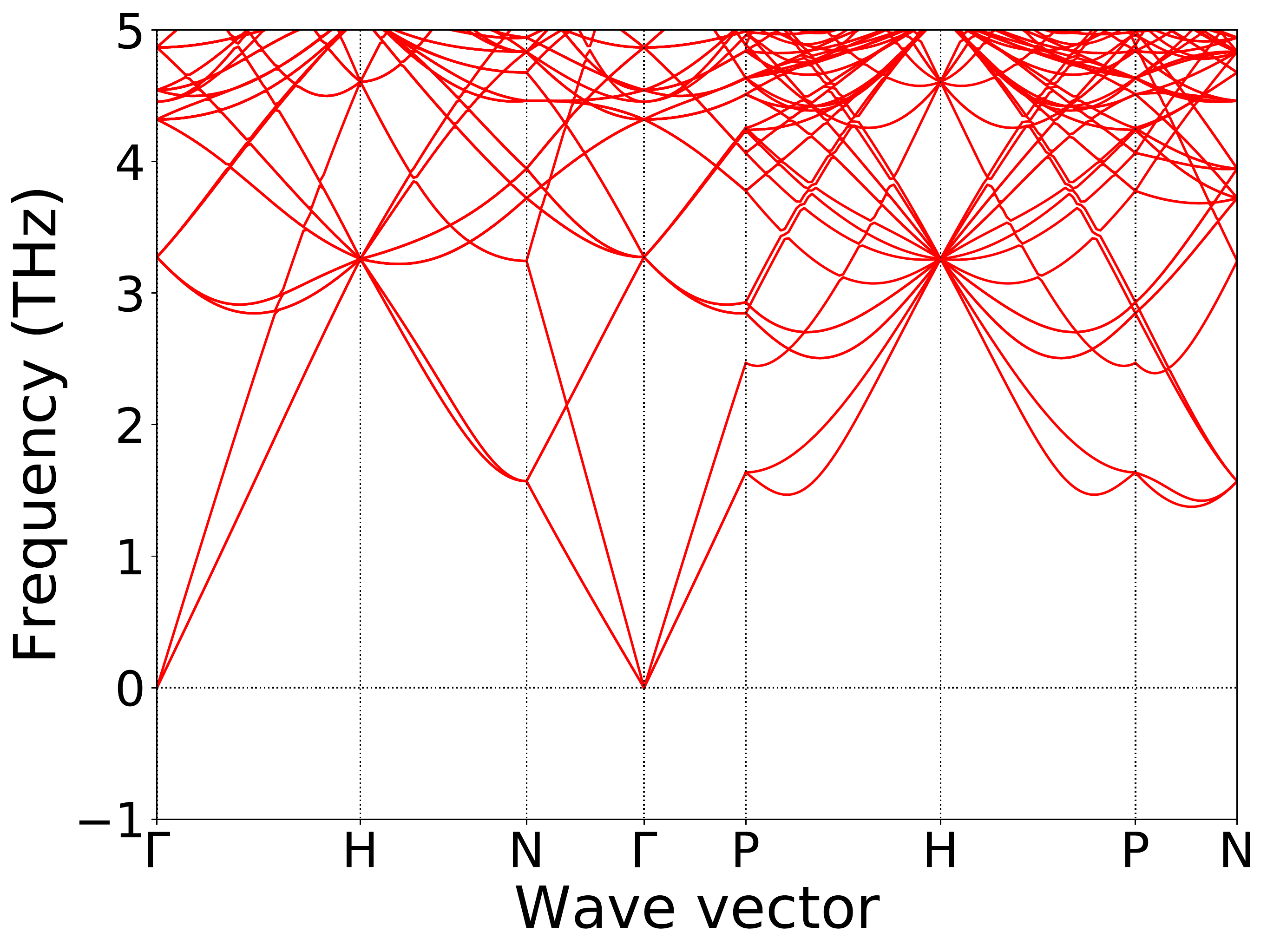}}\hspace{1cm}
    \subfigure[GAP]{\label{fig:gap_Mo_phonon}\includegraphics[width=0.37\textwidth]{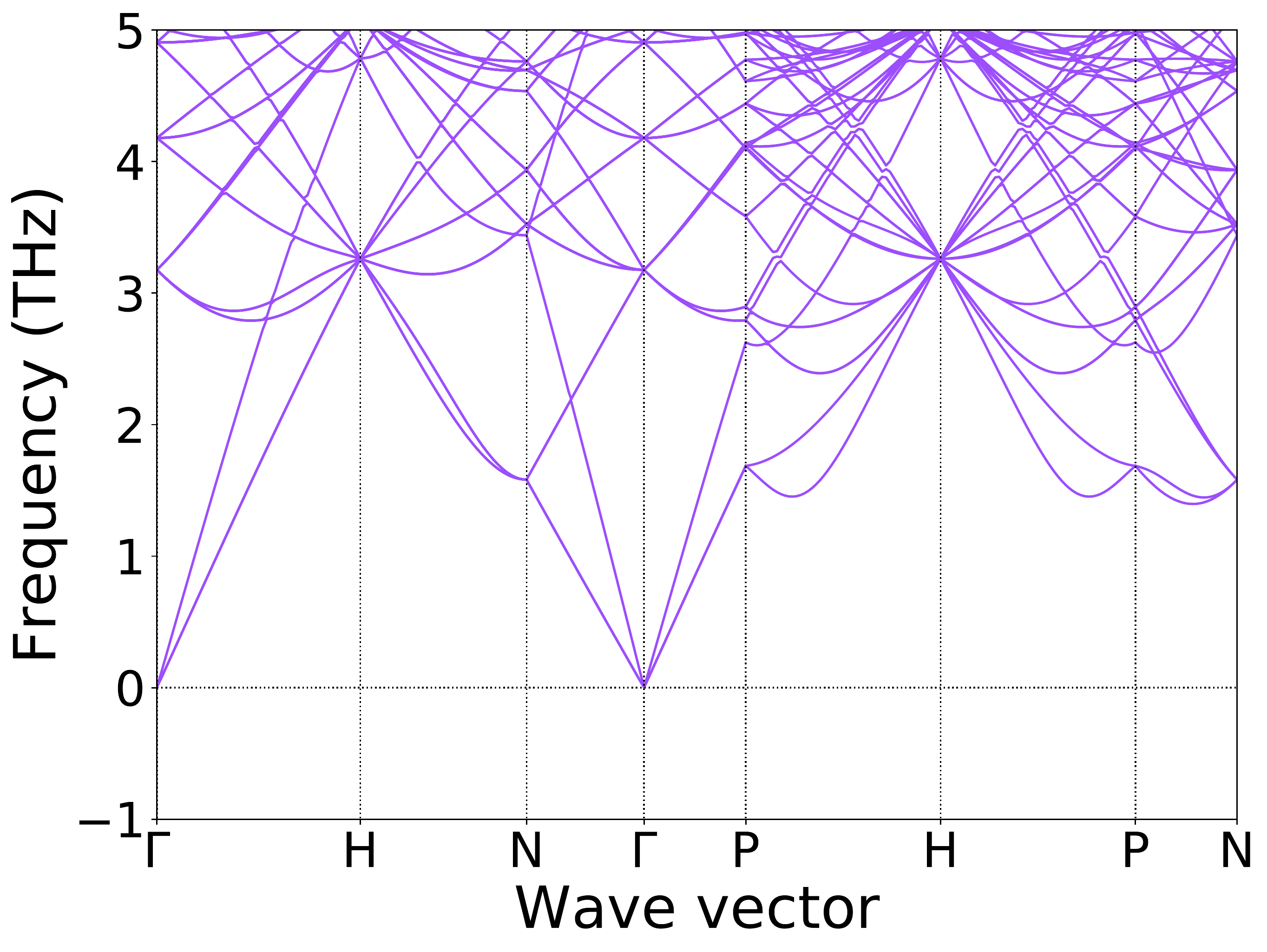}}
    \subfigure[MTP]{\label{fig:mtp_Mo_phonon}\includegraphics[width=0.37\textwidth]{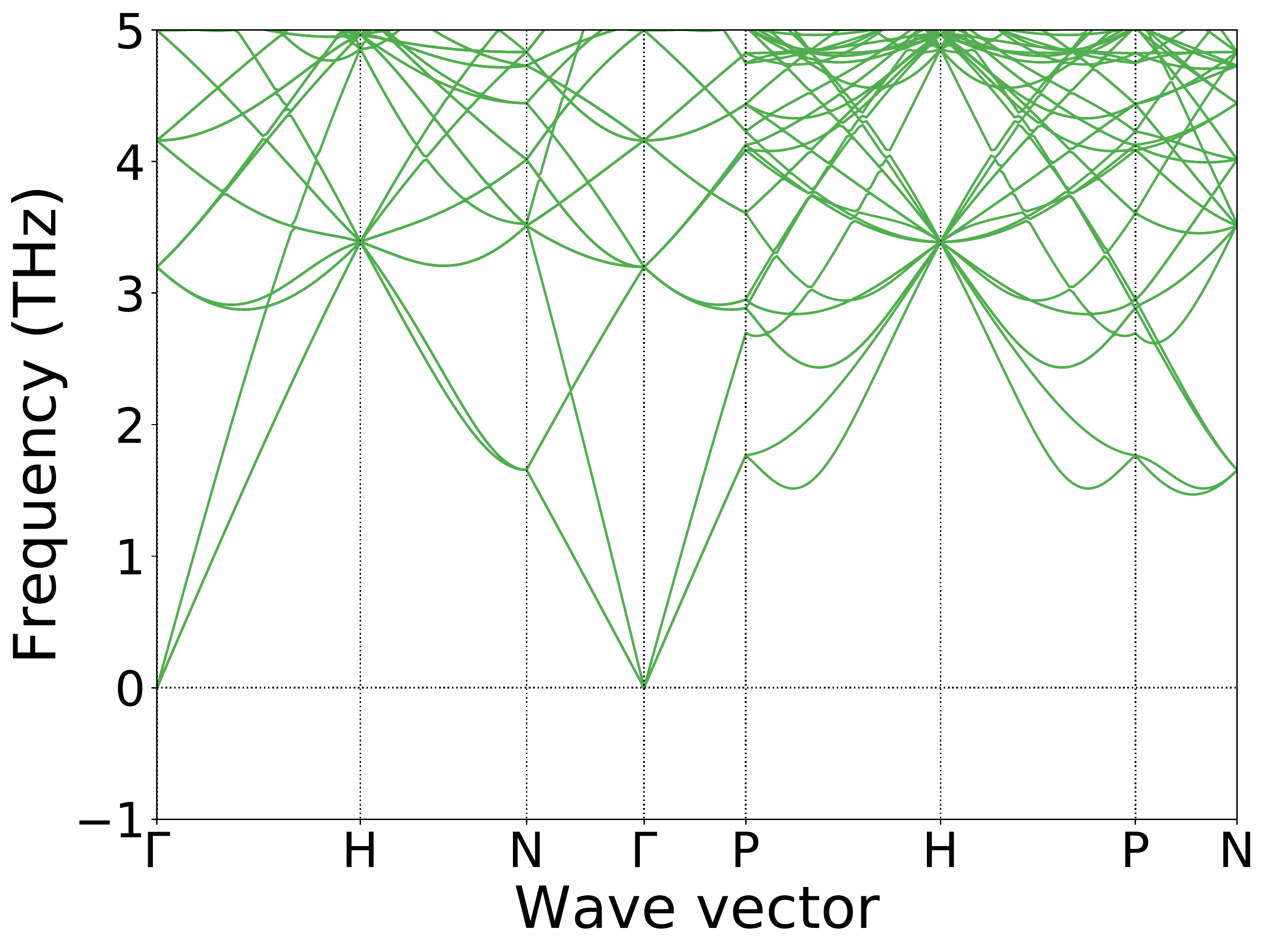}}\hspace{1cm}
    \subfigure[NNP]{\label{fig:nnp_Mo_phonon}\includegraphics[width=0.37\textwidth]{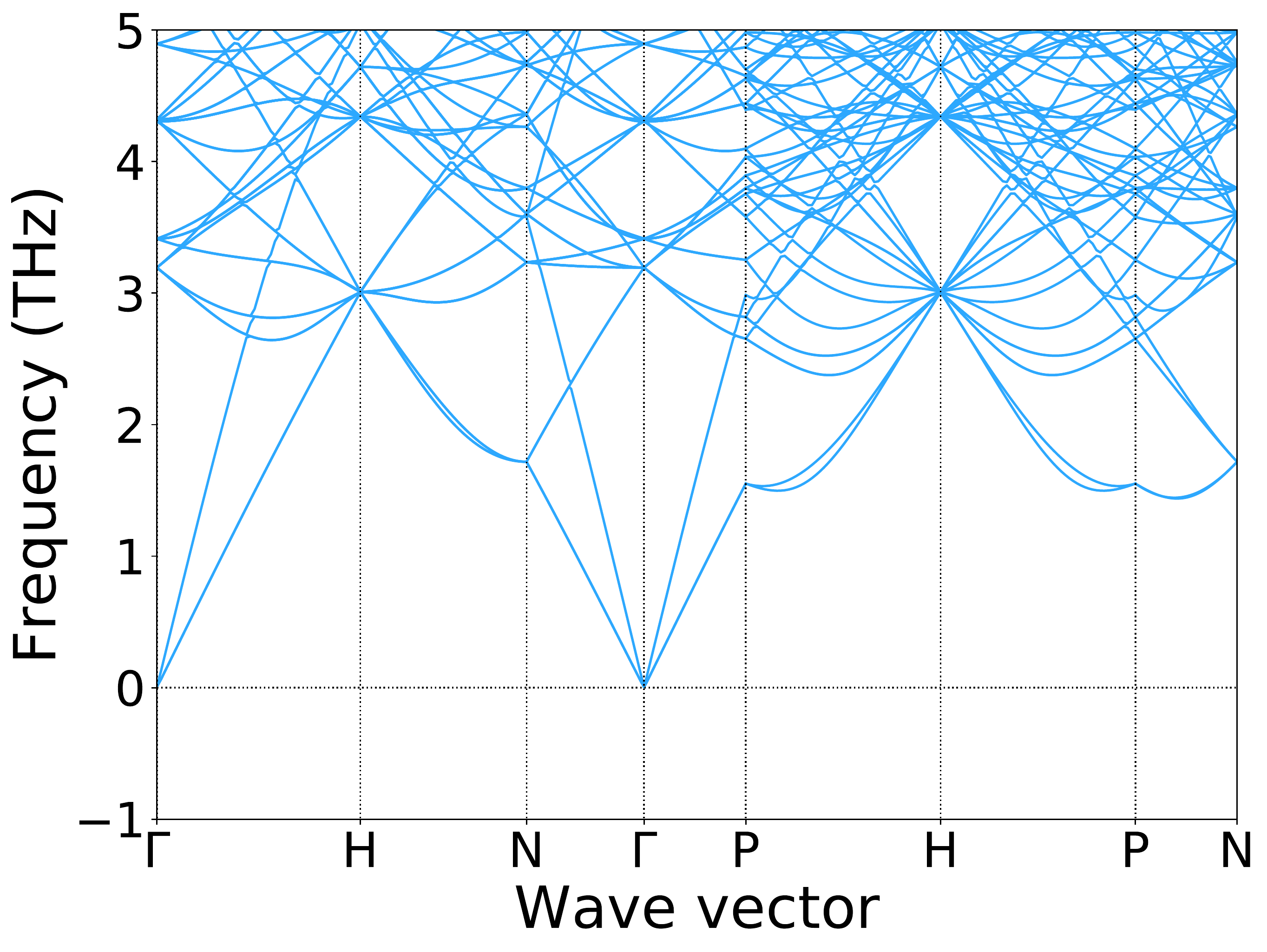}}
    \subfigure[SNAP]{\label{fig:snap_Mo_phonon}\includegraphics[width=0.37\textwidth]{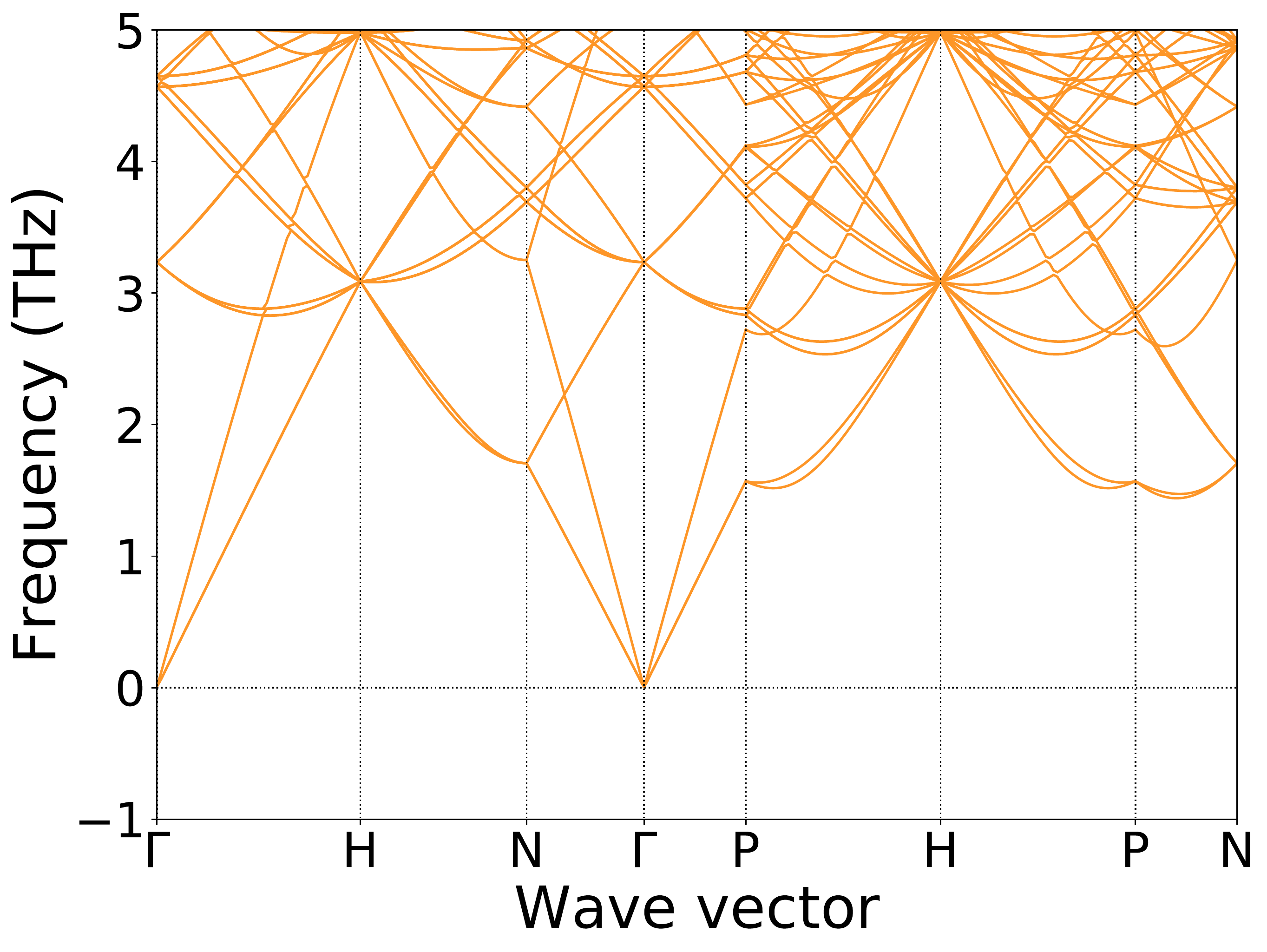}}\hspace{1cm}
    \subfigure[qSNAP]{\label{fig:q_snap_Mo_phonon}\includegraphics[width=0.37\textwidth]{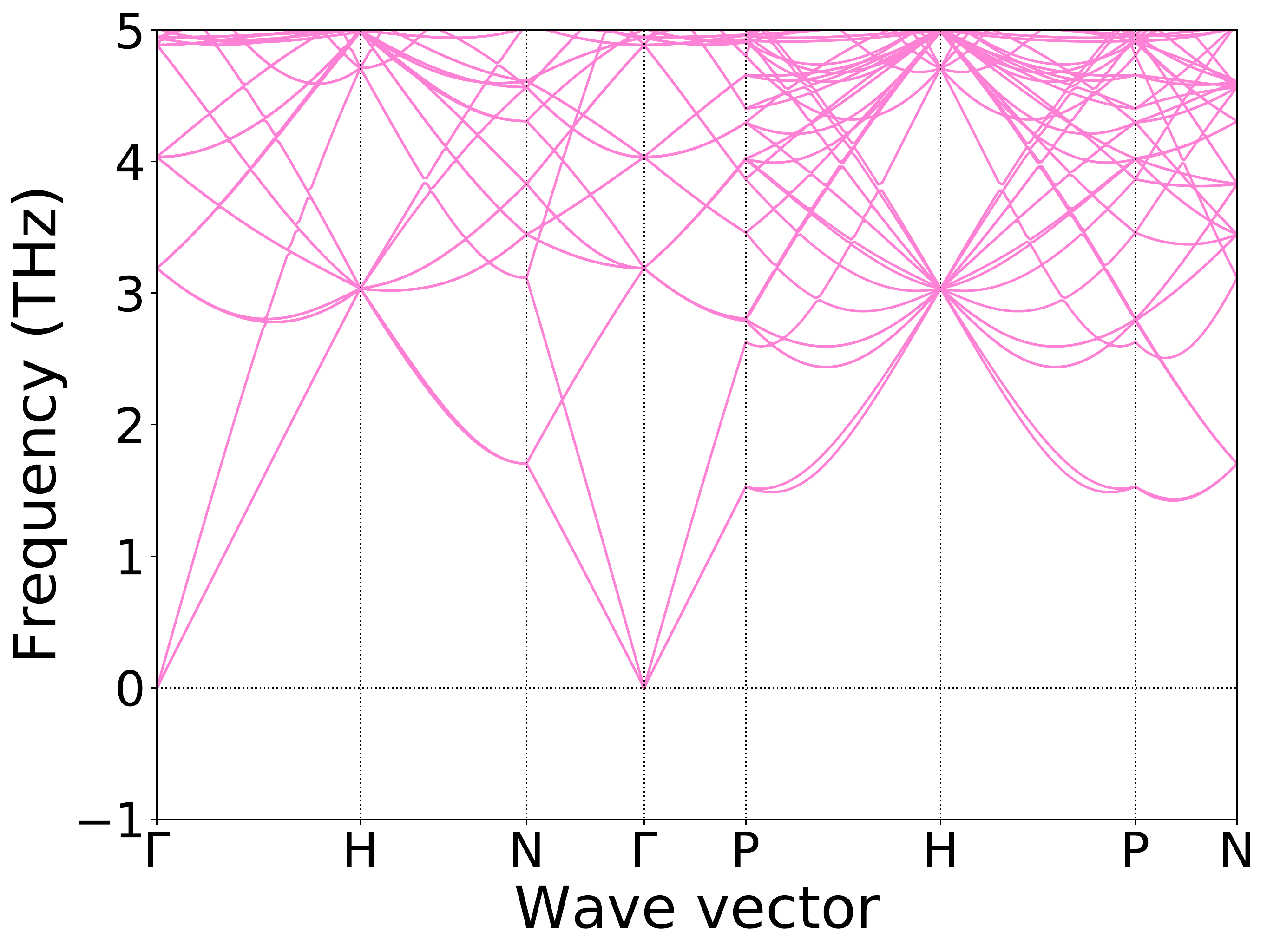}}
    \caption{Phonon dispersion curves for 54-atom bcc Mo supercell via (a) DFT (b) GAP (c) MTP (d) NNP (e) SNAP (f) qSNAP.}
    \label{fig:Mo_phonon}
\end{figure}

\clearpage

\begin{figure}
    \centering
    \subfigure[DFT]{\label{fig:dft_Li_phonon}\includegraphics[width=0.37\textwidth]{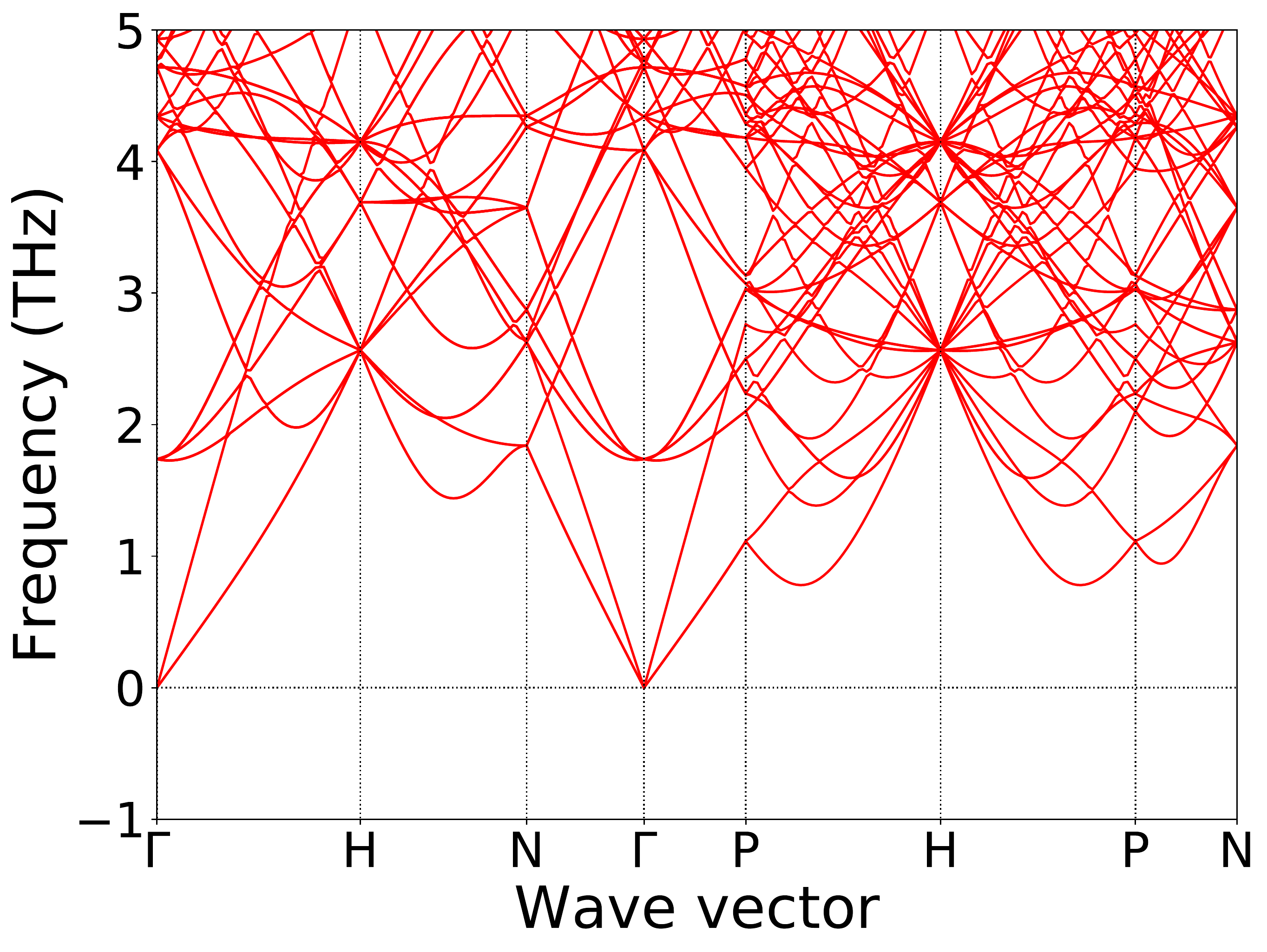}}\hspace{1cm}
    \subfigure[GAP]{\label{fig:gap_Li_phonon}\includegraphics[width=0.37\textwidth]{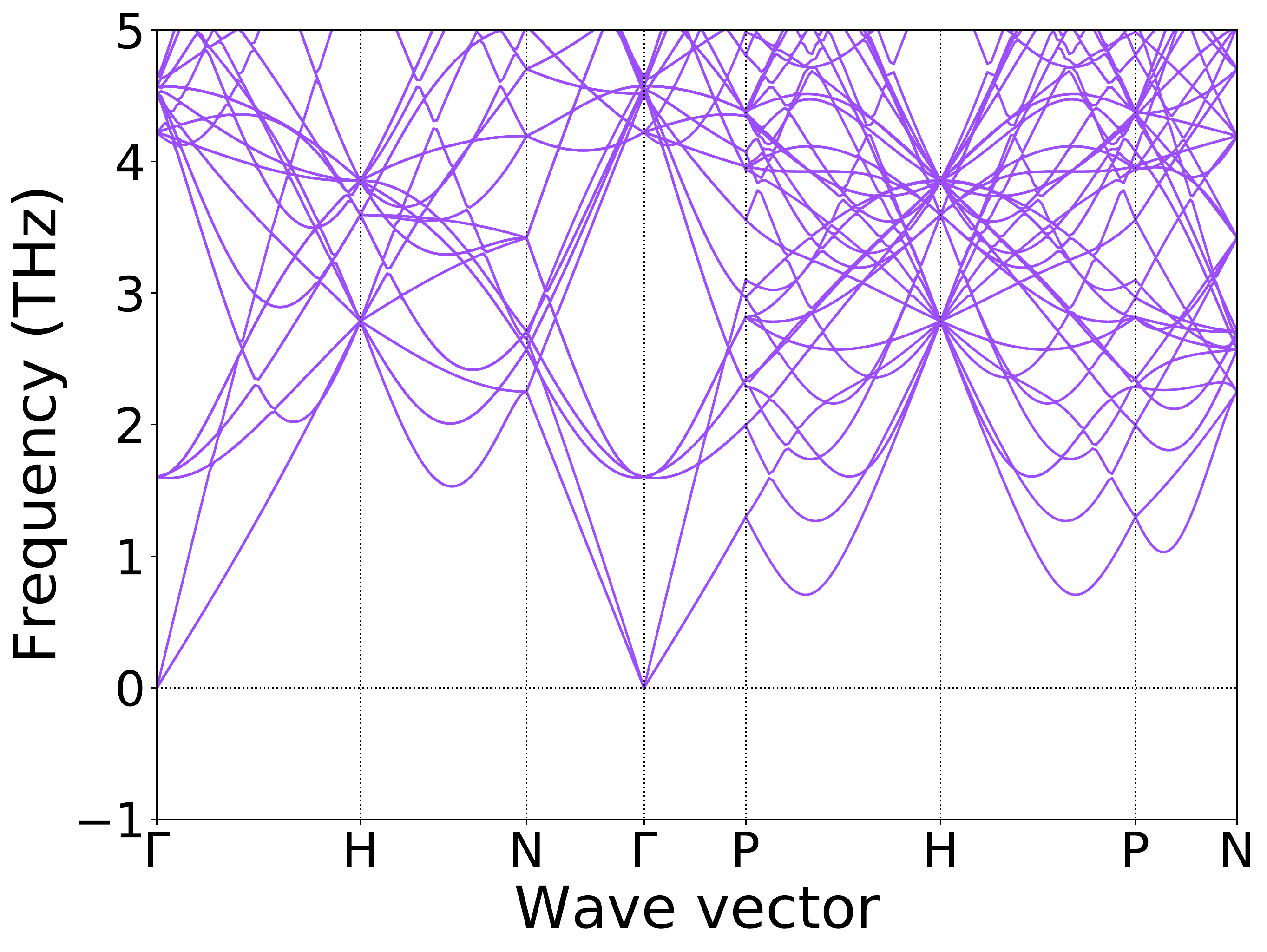}}
    \subfigure[MTP]{\label{fig:mtp_Li_phonon}\includegraphics[width=0.37\textwidth]{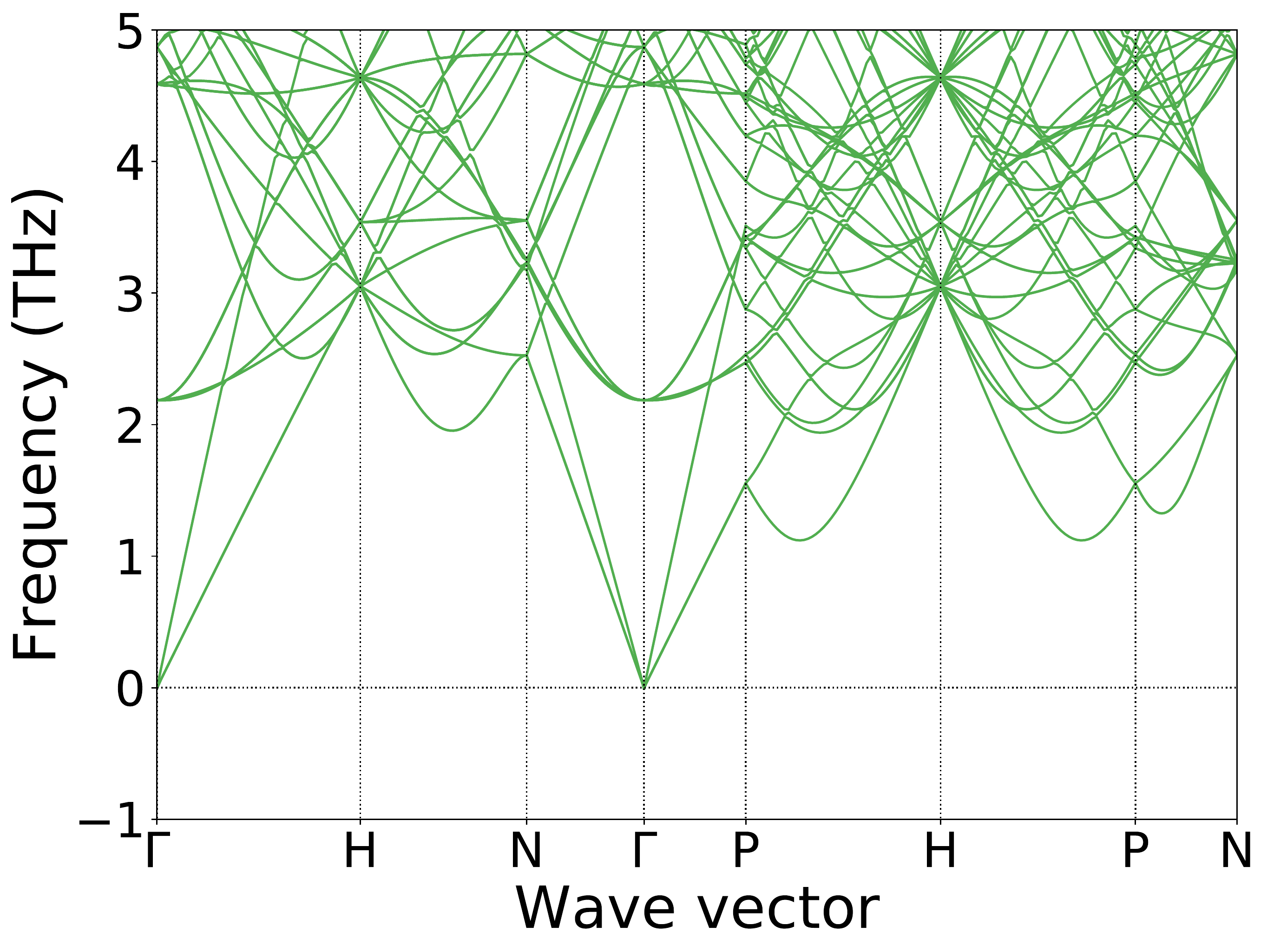}}\hspace{1cm}
    \subfigure[NNP]{\label{fig:nnp_Li_phonon}\includegraphics[width=0.37\textwidth]{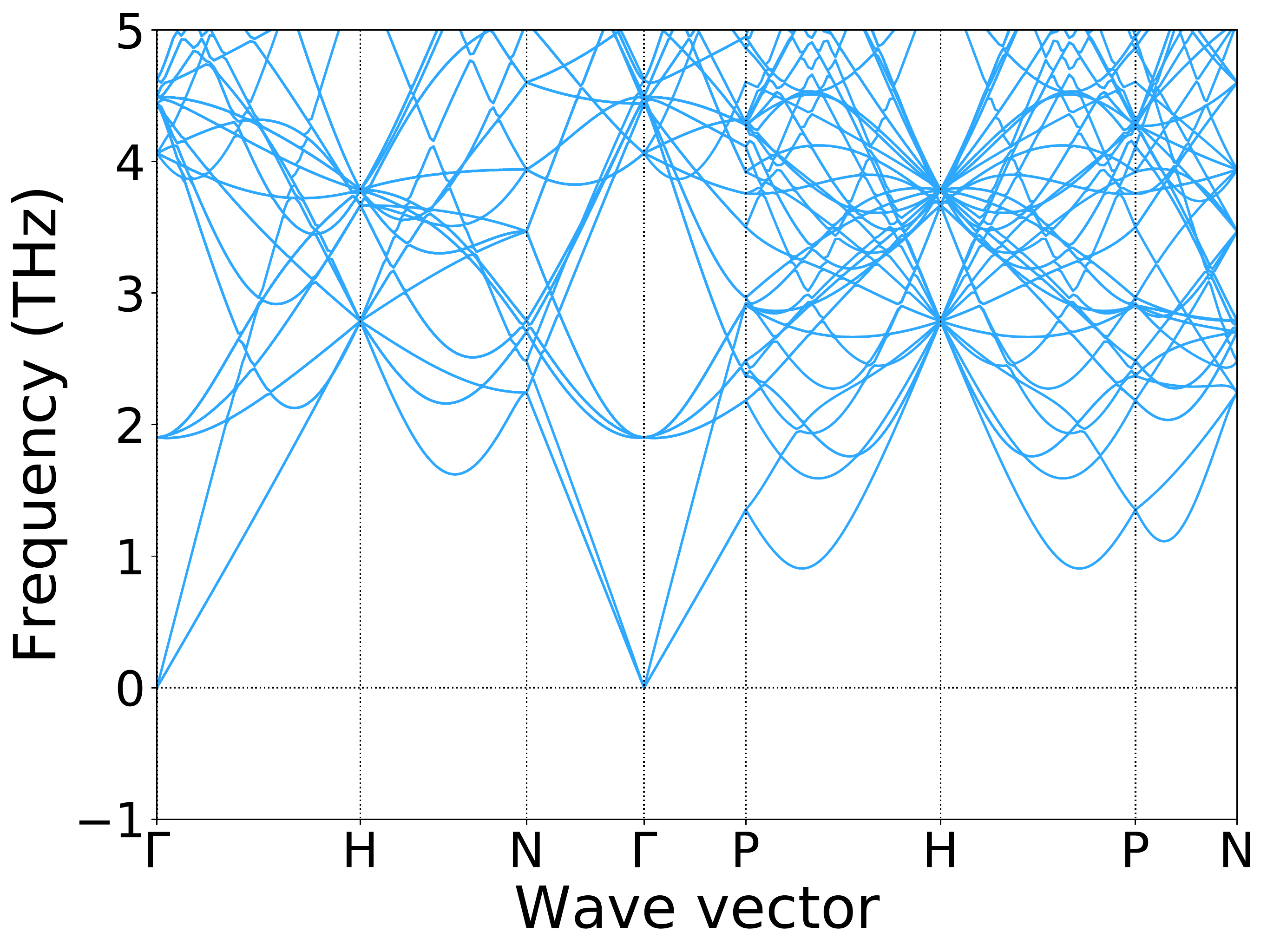}}
    \subfigure[SNAP]{\label{fig:snap_Li_phonon}\includegraphics[width=0.37\textwidth]{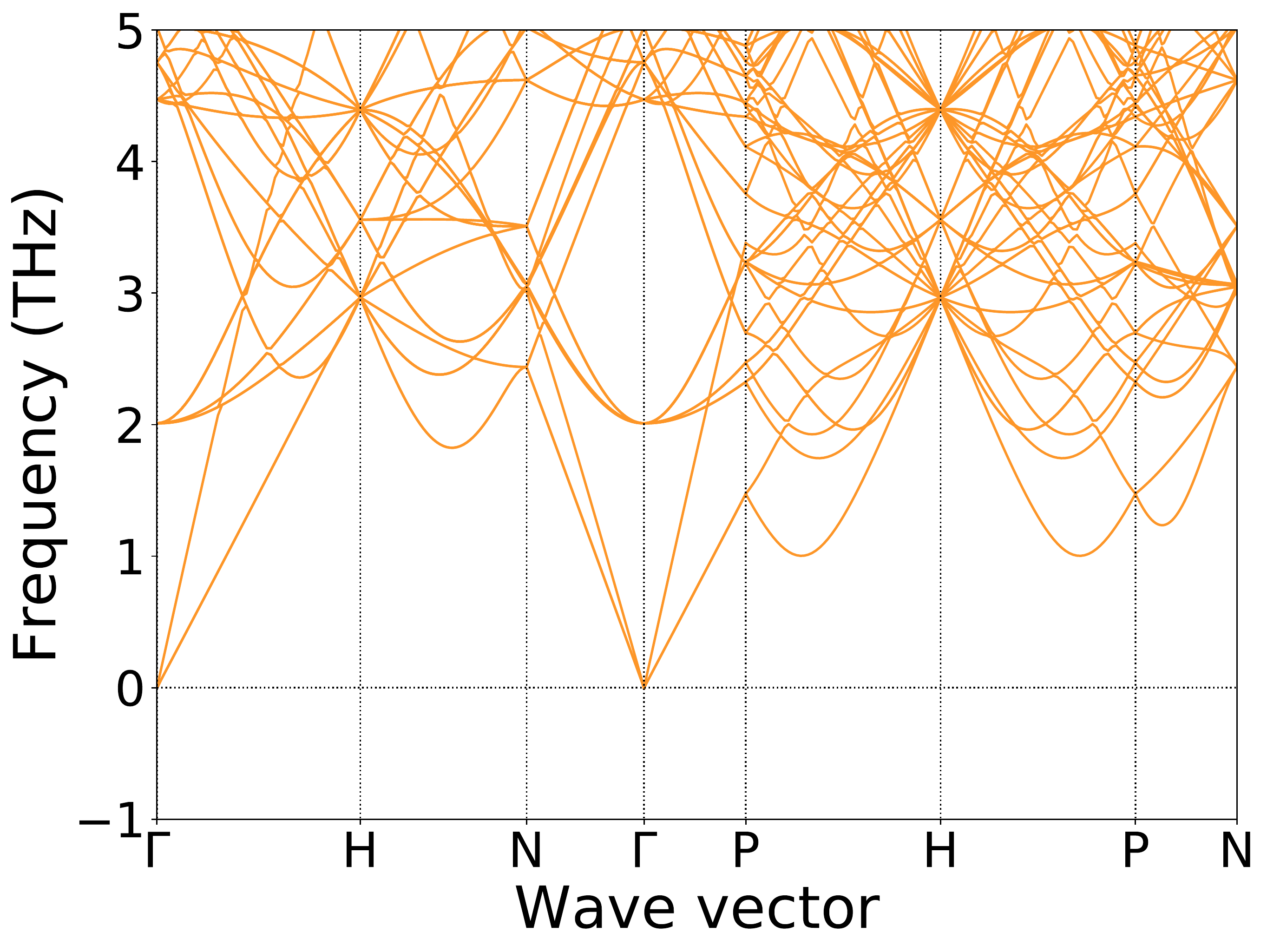}}\hspace{1cm}
    \subfigure[qSNAP]{\label{fig:q_snap_Li_phonon}\includegraphics[width=0.37\textwidth]{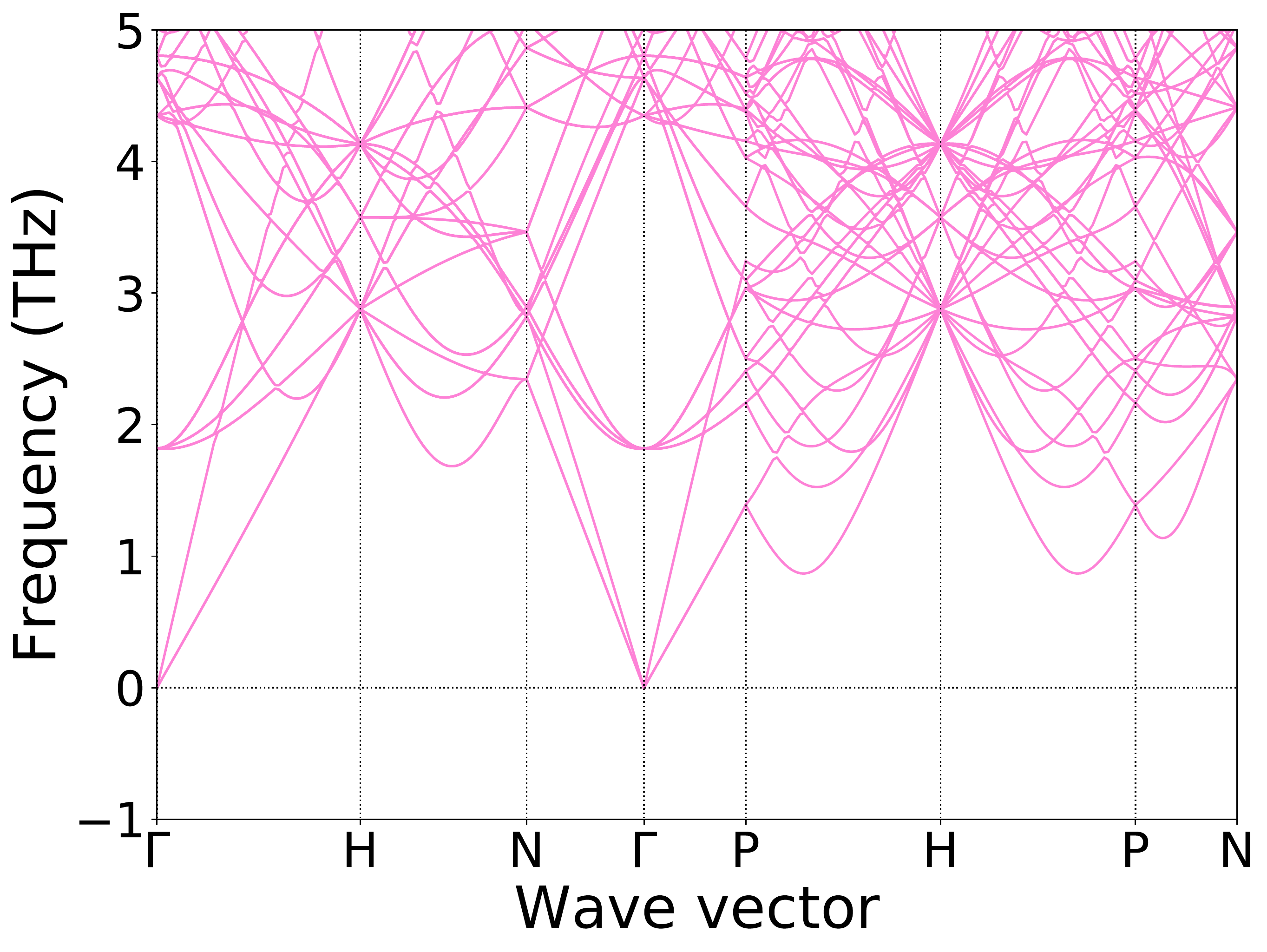}}
    \caption{Phonon dispersion curves for 54-atom bcc Li supercell via (a) DFT (b) GAP (c) MTP (d) NNP (e) SNAP (f) qSNAP.}
    \label{fig:Li_phonon}
\end{figure}

\clearpage

\begin{figure}
    \centering
    \subfigure[DFT]{\label{fig:dft_Ni_phonon}\includegraphics[width=0.37\textwidth]{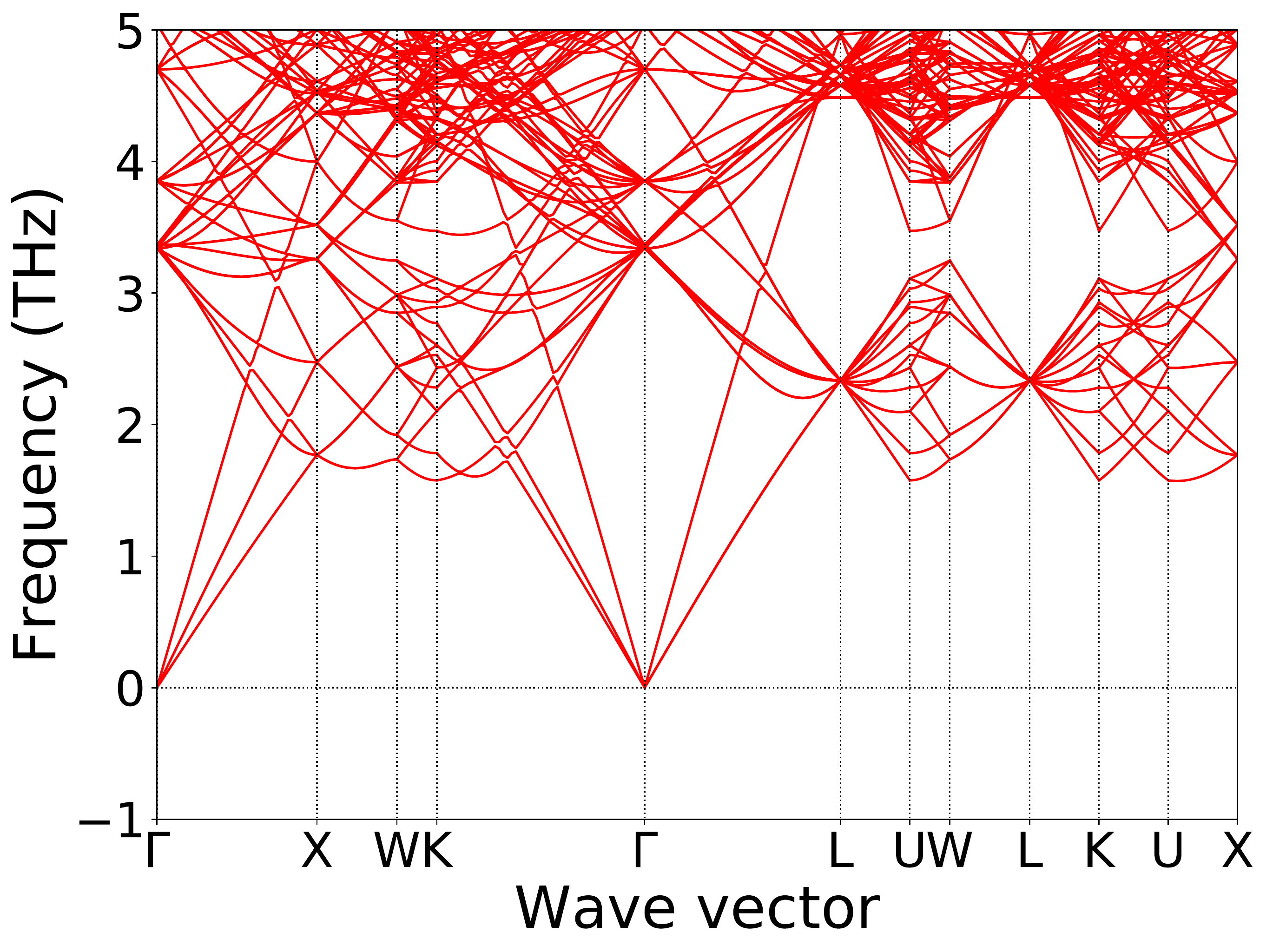}}\hspace{1cm}
    \subfigure[GAP]{\label{fig:gap_Ni_phonon}\includegraphics[width=0.37\textwidth]{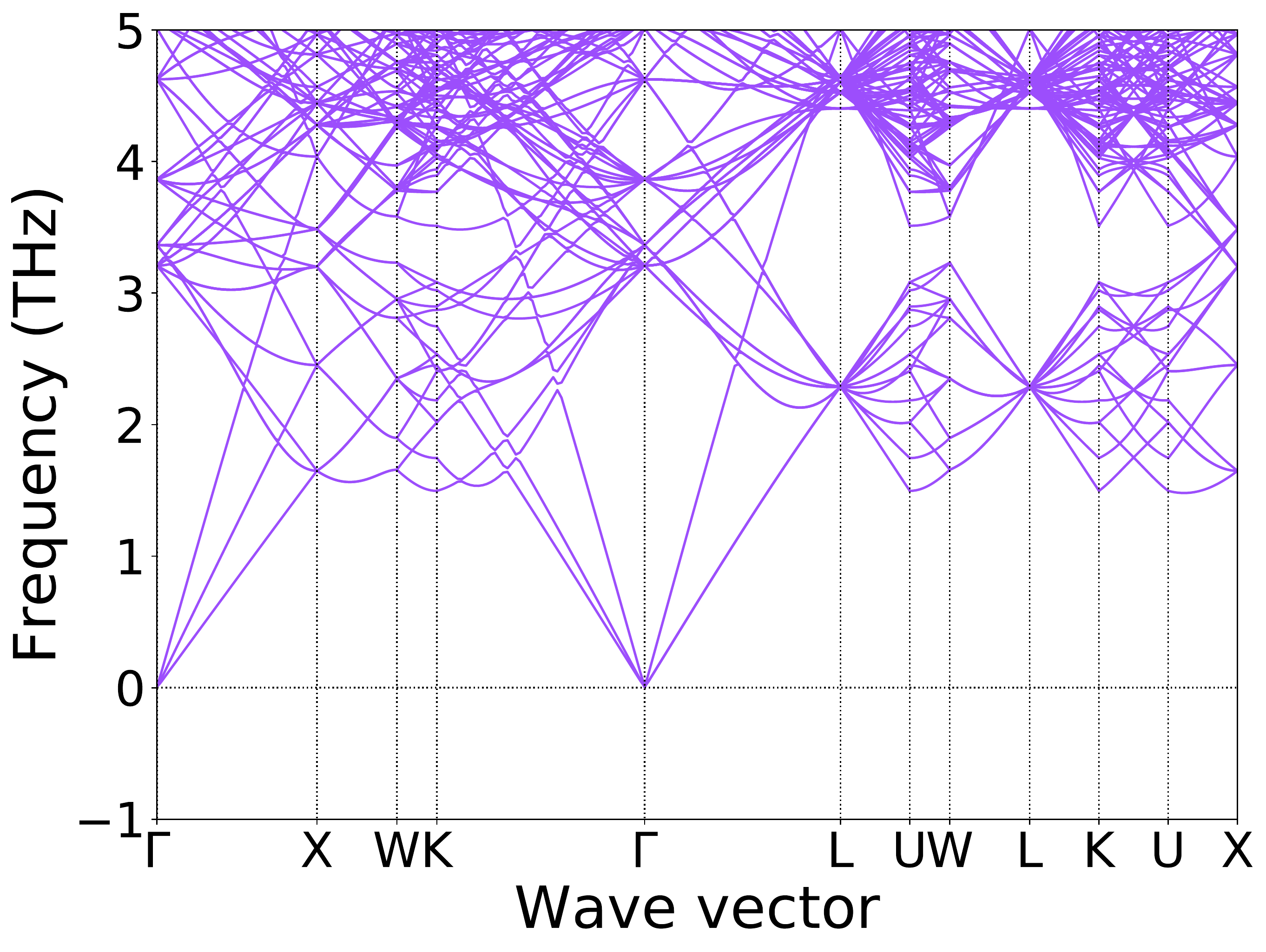}}
    \subfigure[MTP]{\label{fig:mtp_Ni_phonon}\includegraphics[width=0.37\textwidth]{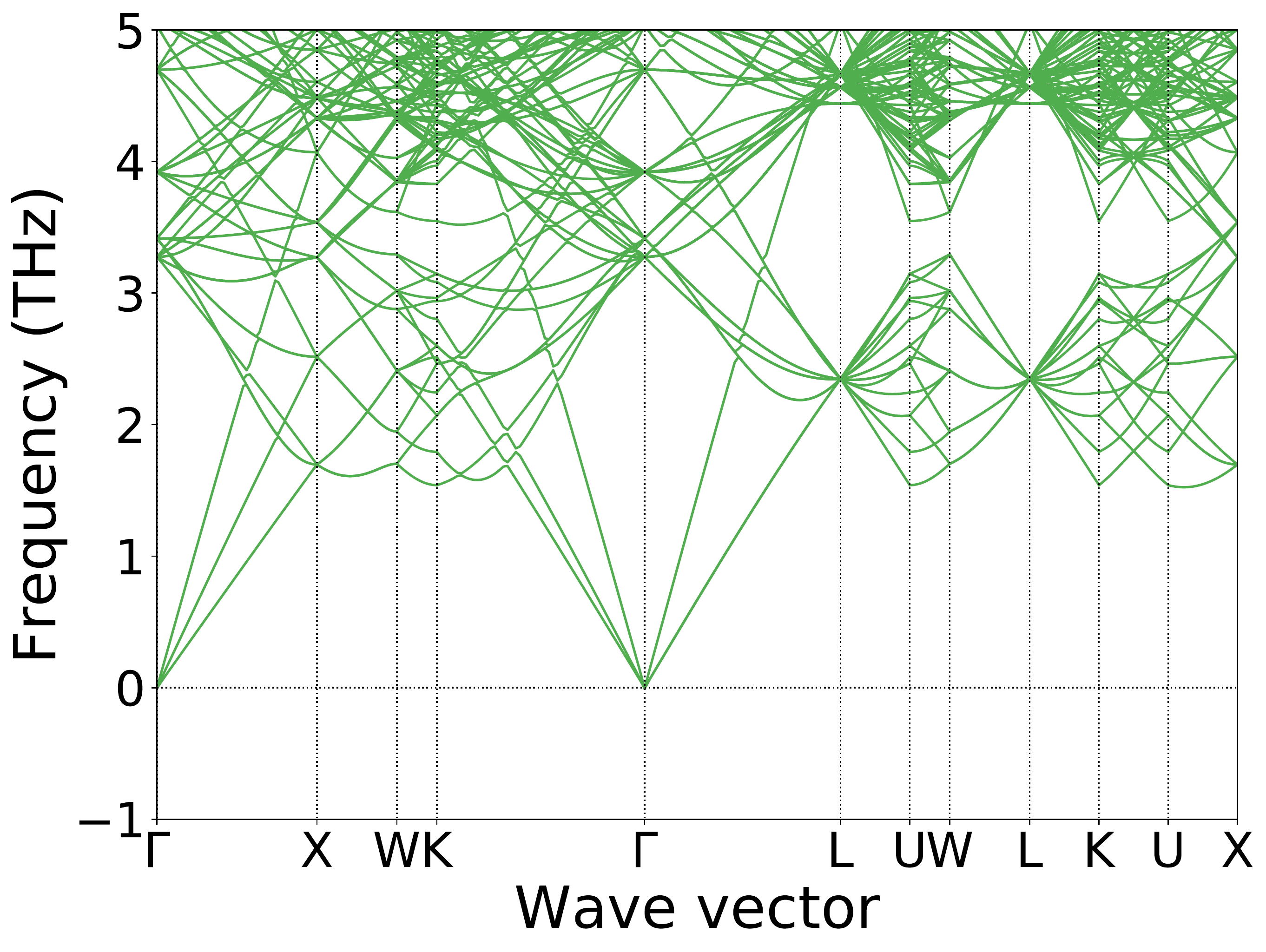}}\hspace{1cm}
    \subfigure[NNP]{\label{fig:nnp_Ni_phonon}\includegraphics[width=0.37\textwidth]{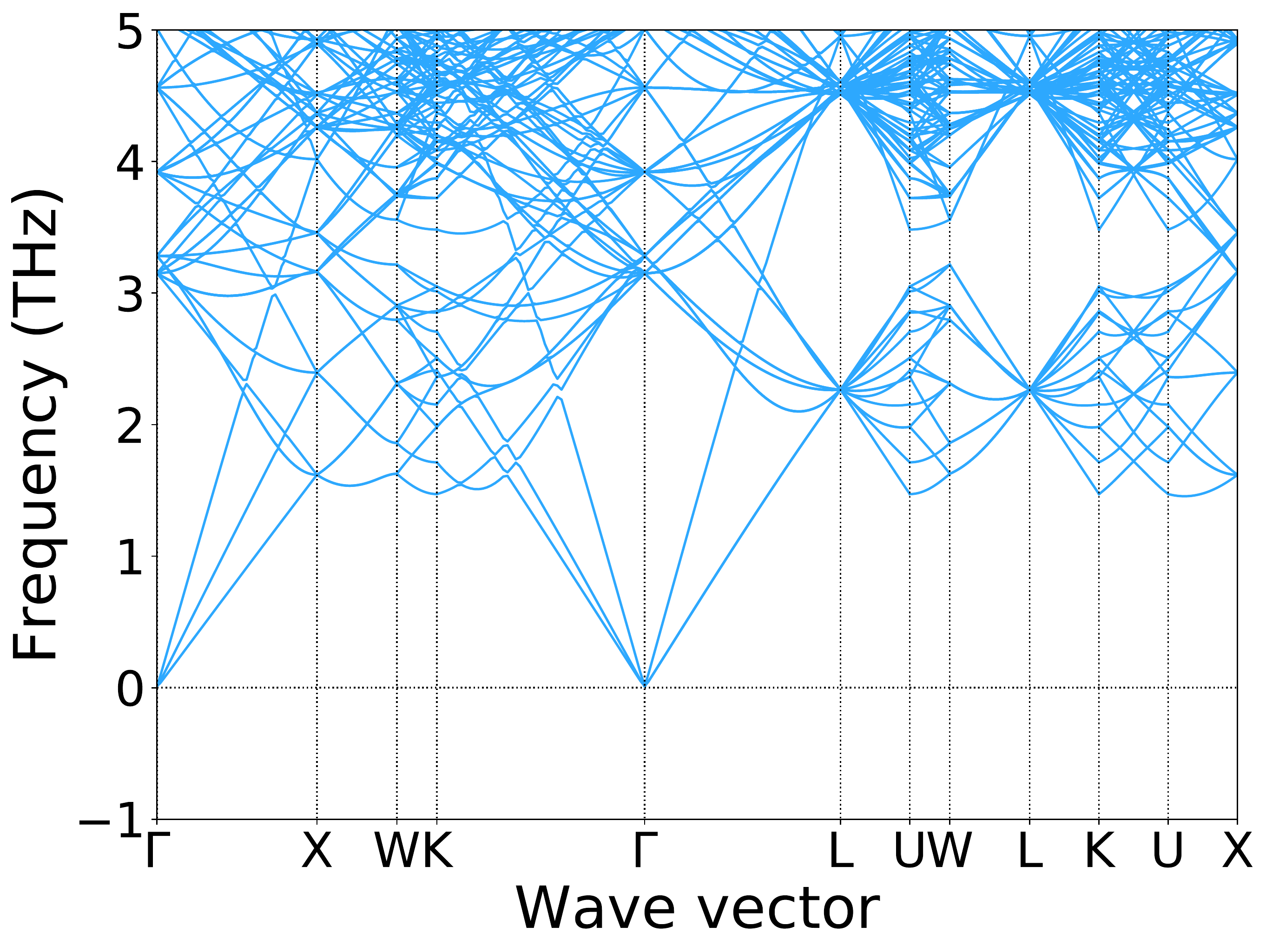}}
    \subfigure[SNAP]{\label{fig:snap_Ni_phonon}\includegraphics[width=0.37\textwidth]{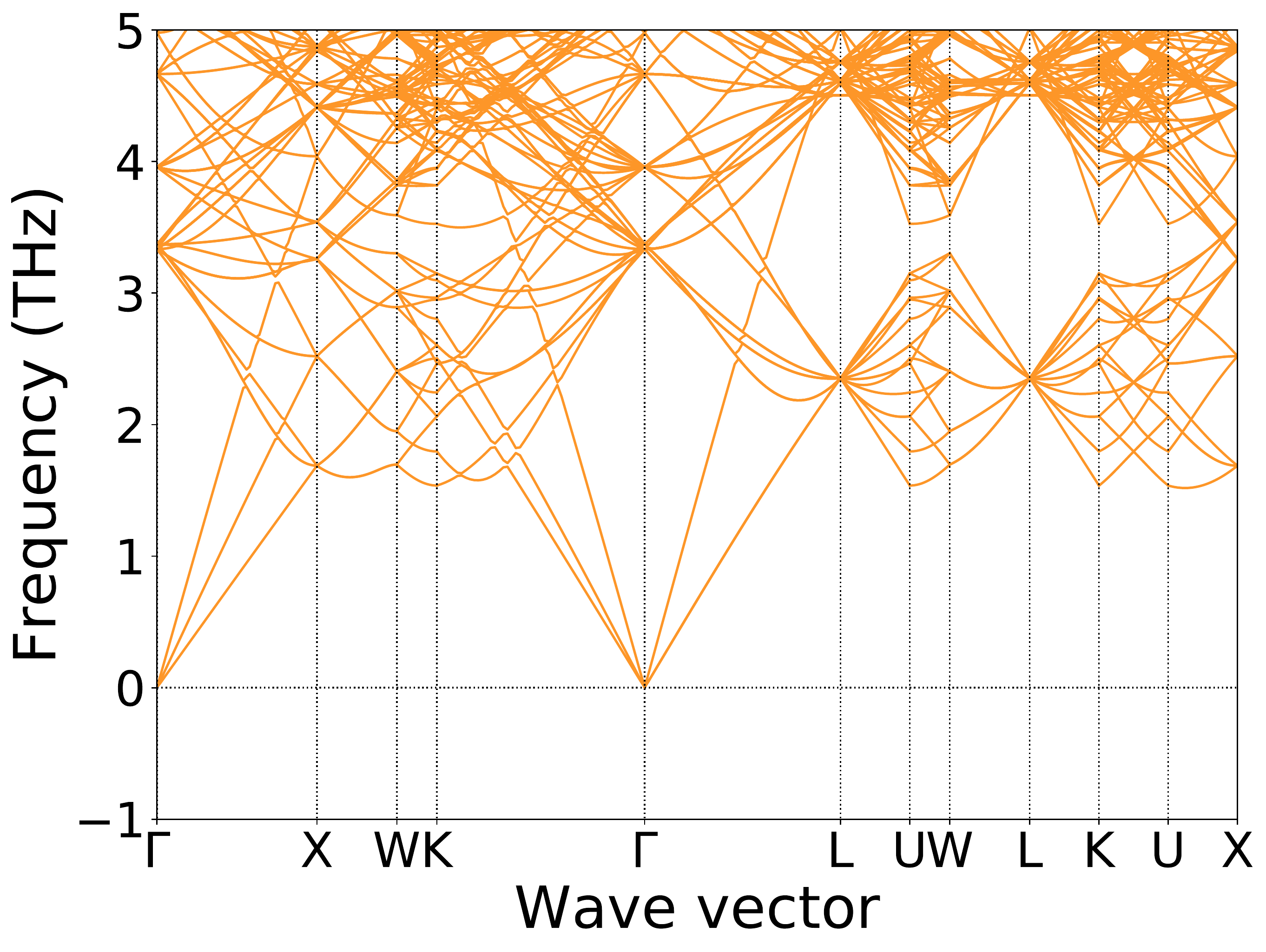}}\hspace{1cm}
    \subfigure[qSNAP]{\label{fig:q_snap_Ni_phonon}\includegraphics[width=0.37\textwidth]{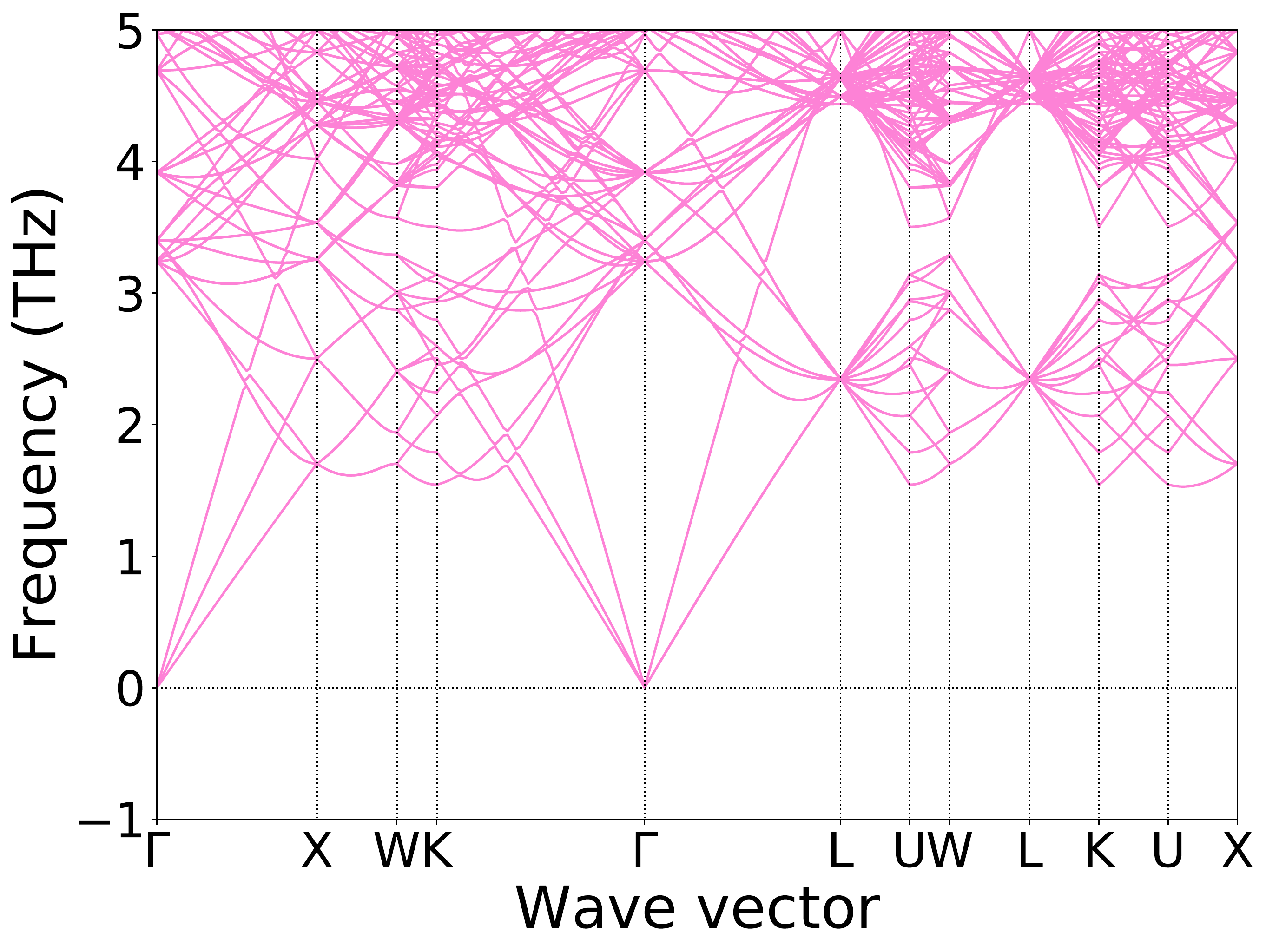}}
    \caption{Phonon dispersion curves for 108-atom fcc Ni supercell via (a) DFT (b) GAP (c) MTP (d) NNP (e) SNAP (f) qSNAP.}
    \label{fig:Ni_phonon}
\end{figure}

\clearpage

\begin{figure}
    \centering
    \subfigure[DFT]{\label{fig:dft_Cu_phonon}\includegraphics[width=0.37\textwidth]{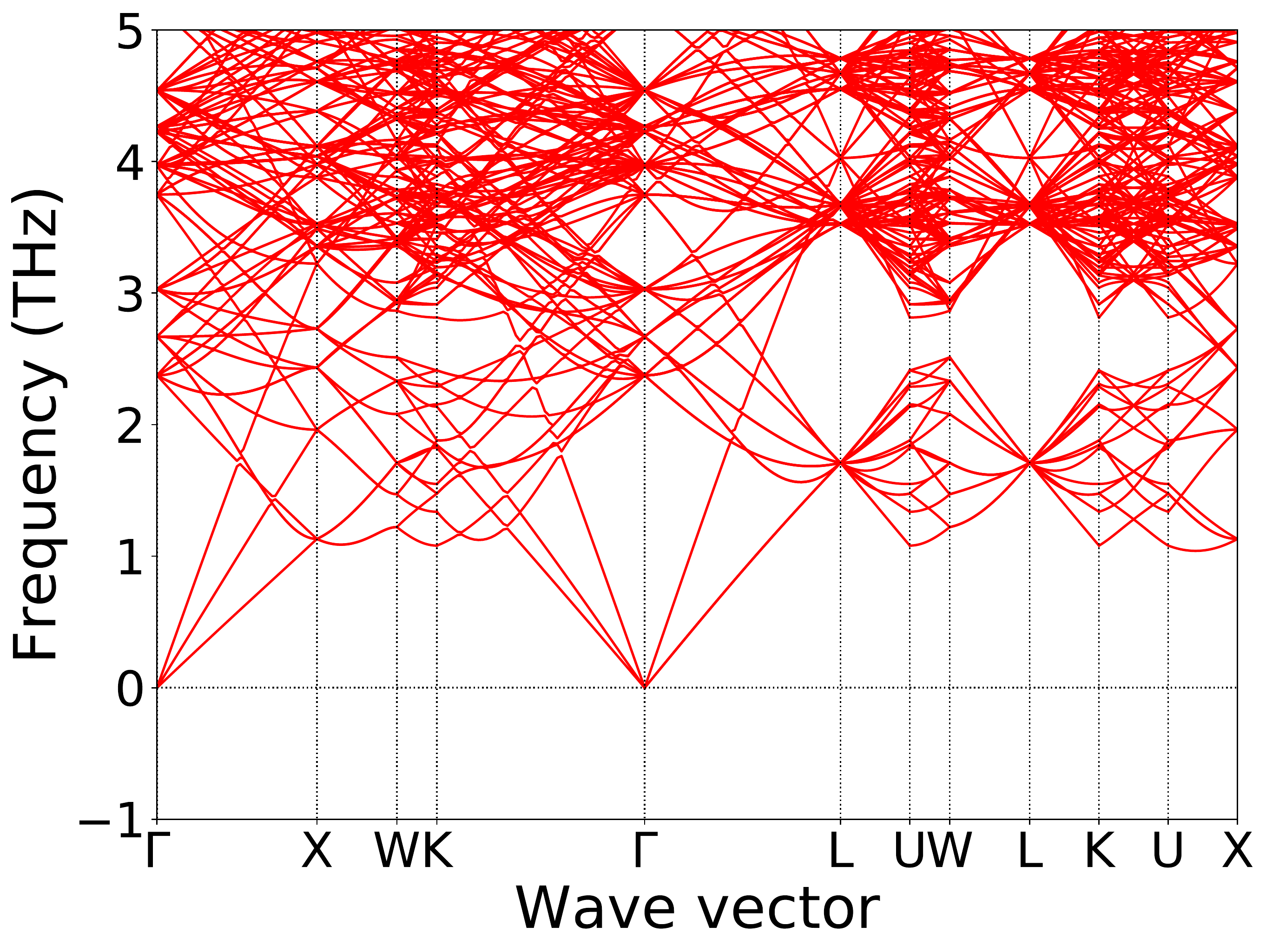}}\hspace{1cm}
    \subfigure[GAP]{\label{fig:gap_Cu_phonon}\includegraphics[width=0.37\textwidth]{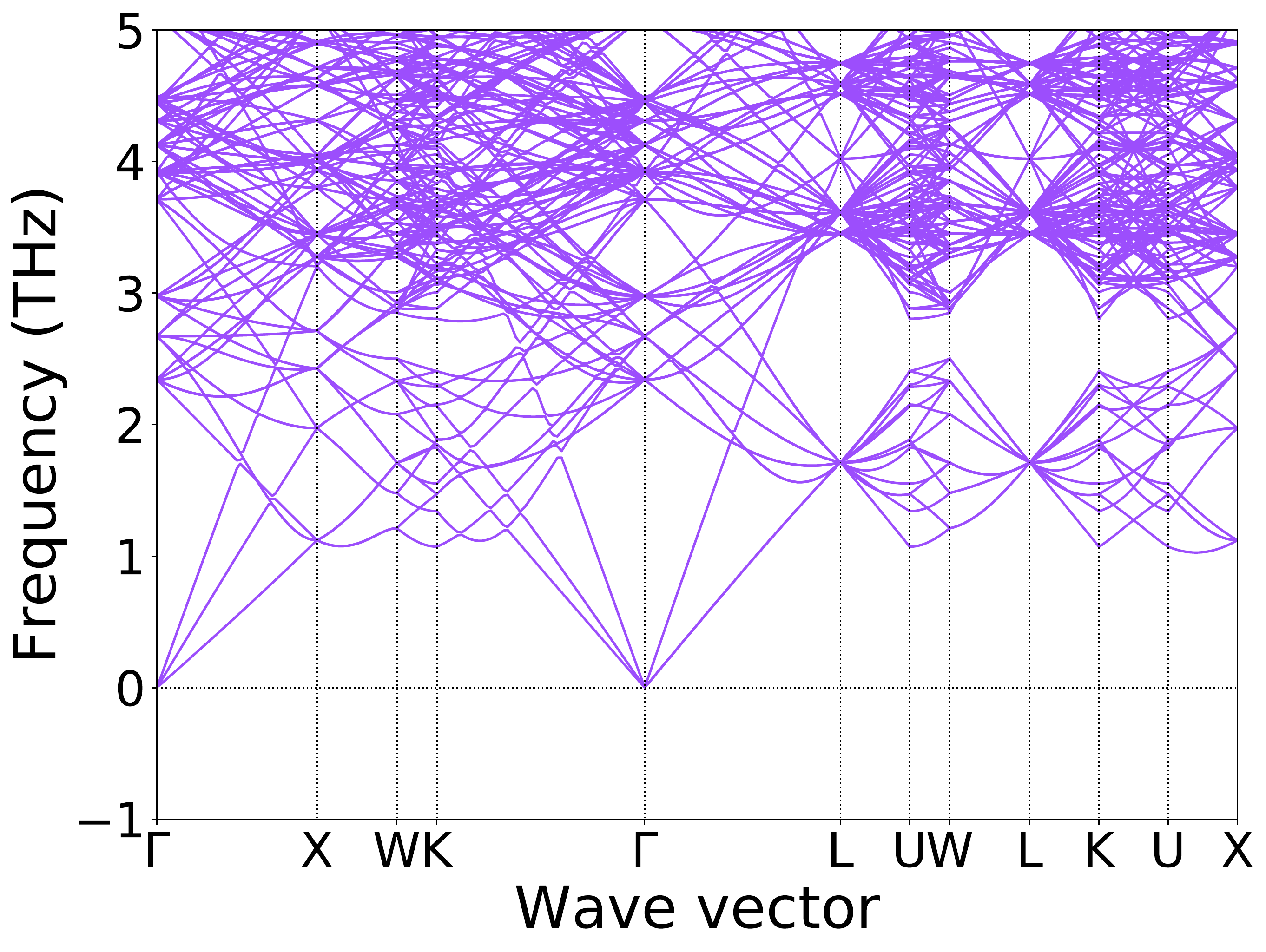}}
    \subfigure[MTP]{\label{fig:mtp_Cu_phonon}\includegraphics[width=0.37\textwidth]{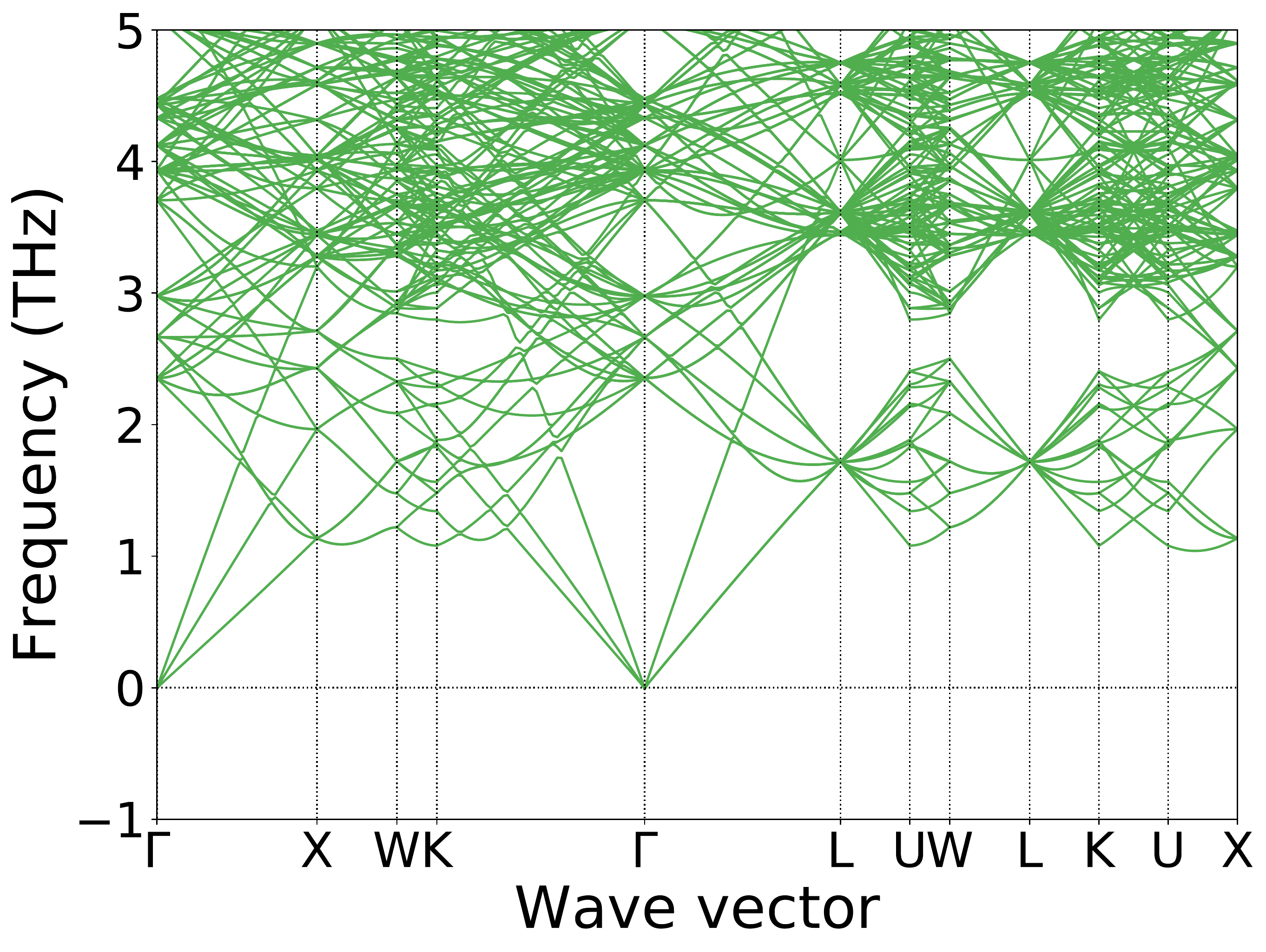}}\hspace{1cm}
    \subfigure[NNP]{\label{fig:nnp_Cu_phonon}\includegraphics[width=0.37\textwidth]{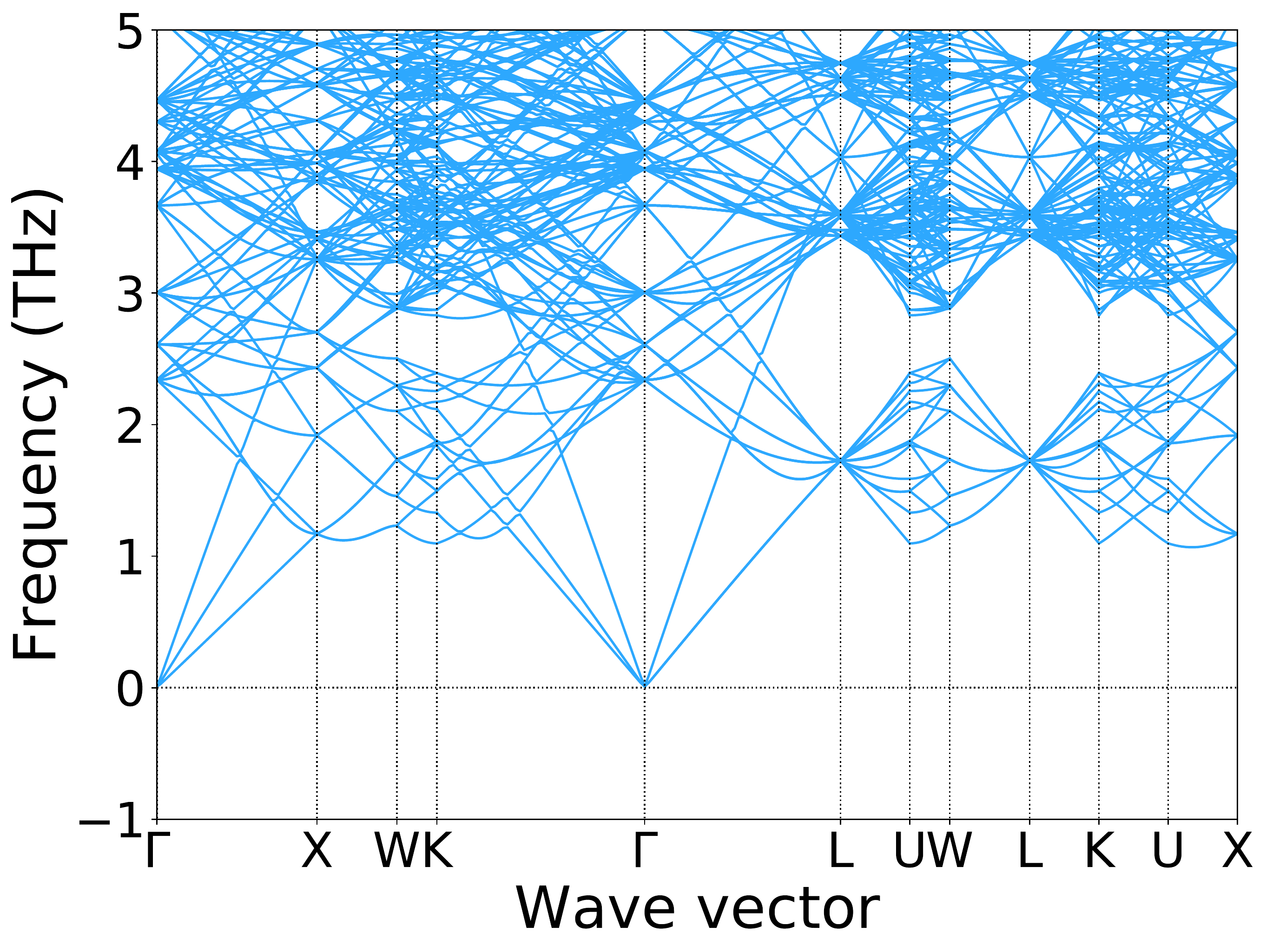}}
    \subfigure[SNAP]{\label{fig:snap_Cu_phonon}\includegraphics[width=0.37\textwidth]{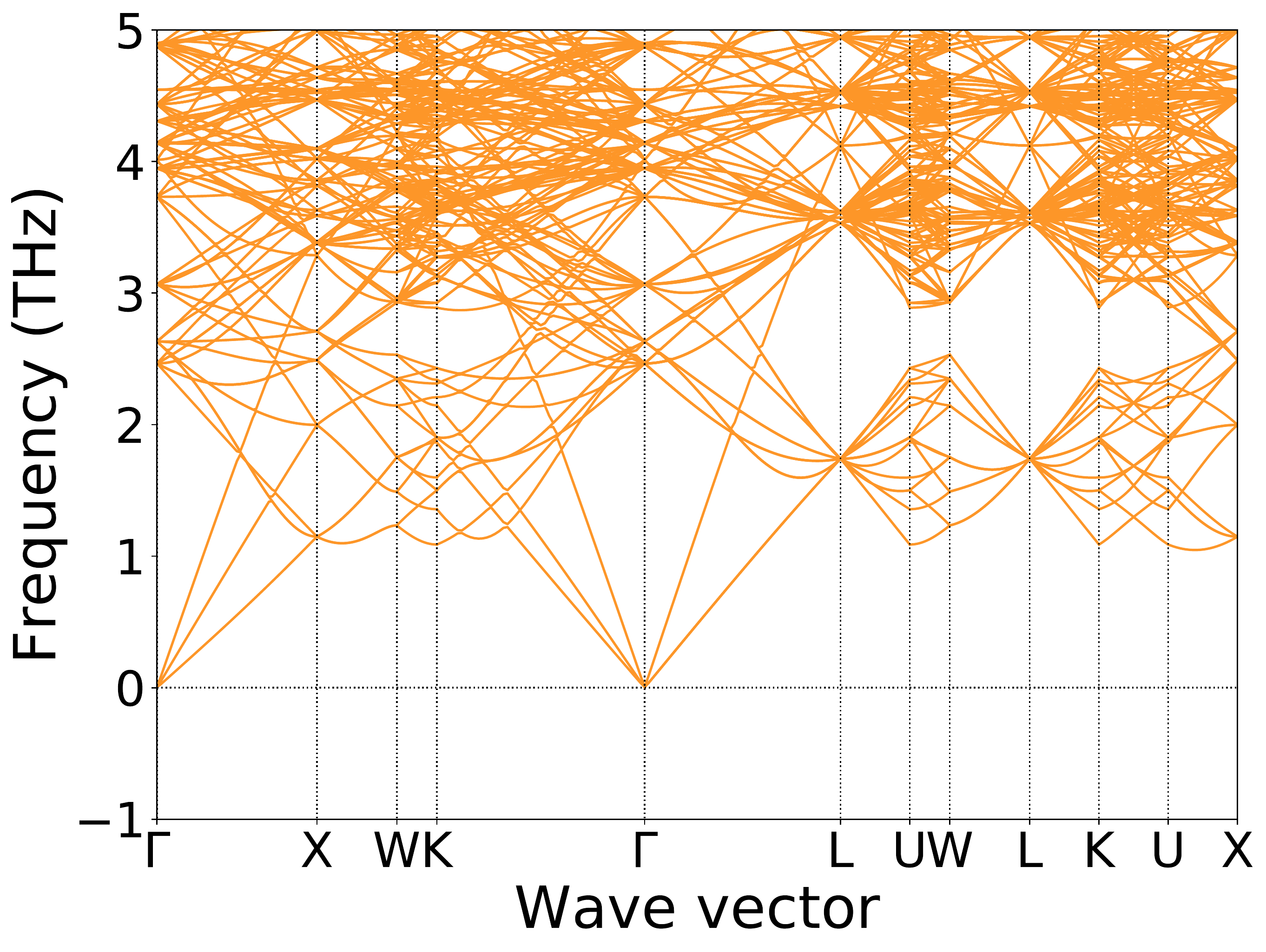}}\hspace{1cm}
    \subfigure[qSNAP]{\label{fig:q_snap_Cu_phonon}\includegraphics[width=0.37\textwidth]{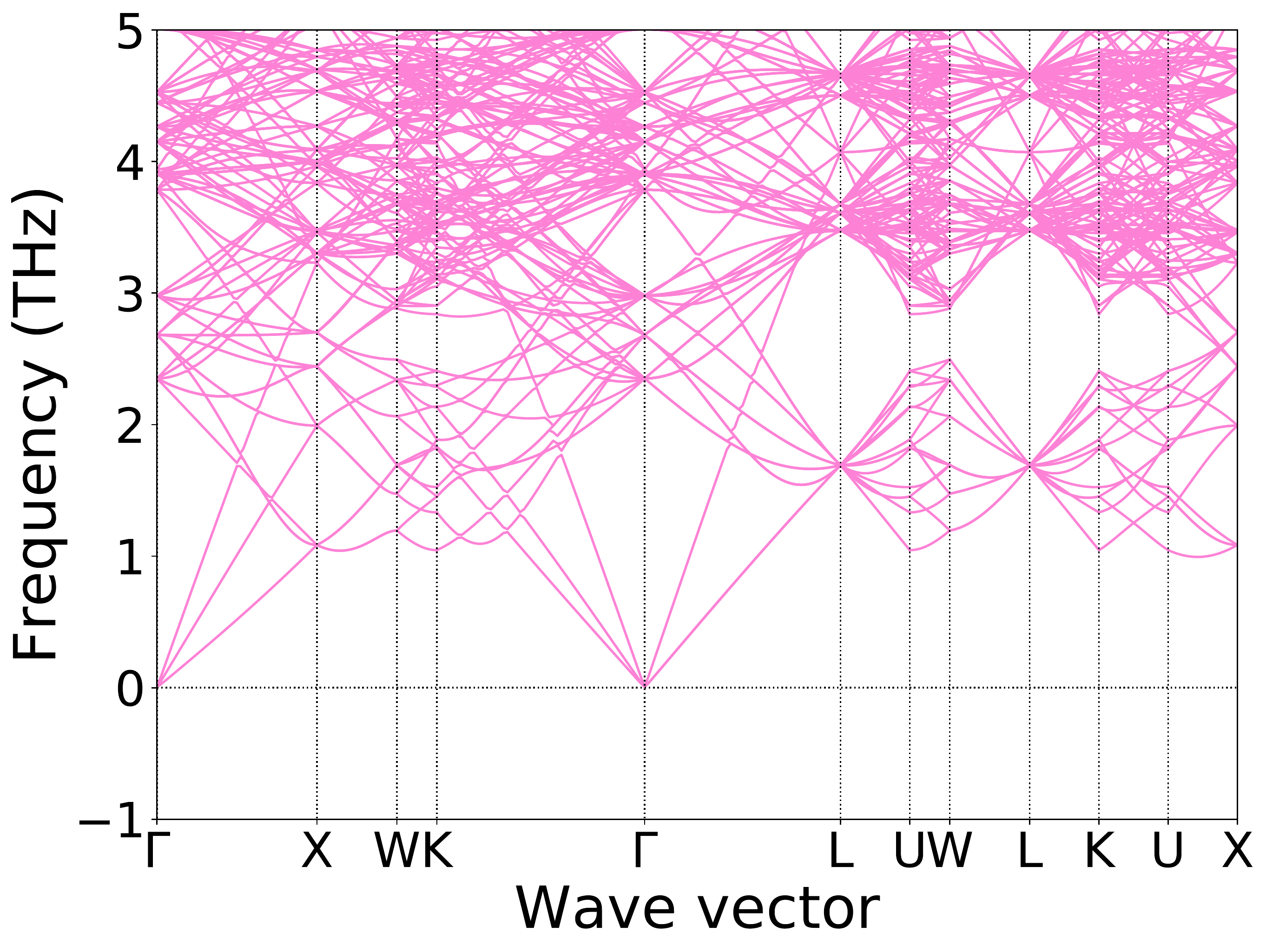}}
    \caption{Phonon dispersion curves for 108-atom fcc Cu supercell via (a) DFT (b) GAP (c) MTP (d) NNP (e) SNAP (f) qSNAP.}
    \label{fig:Cu_phonon}
\end{figure}

\clearpage

\begin{figure}
    \centering
    \subfigure[DFT]{\label{fig:dft_Si_phonon}\includegraphics[width=0.37\textwidth]{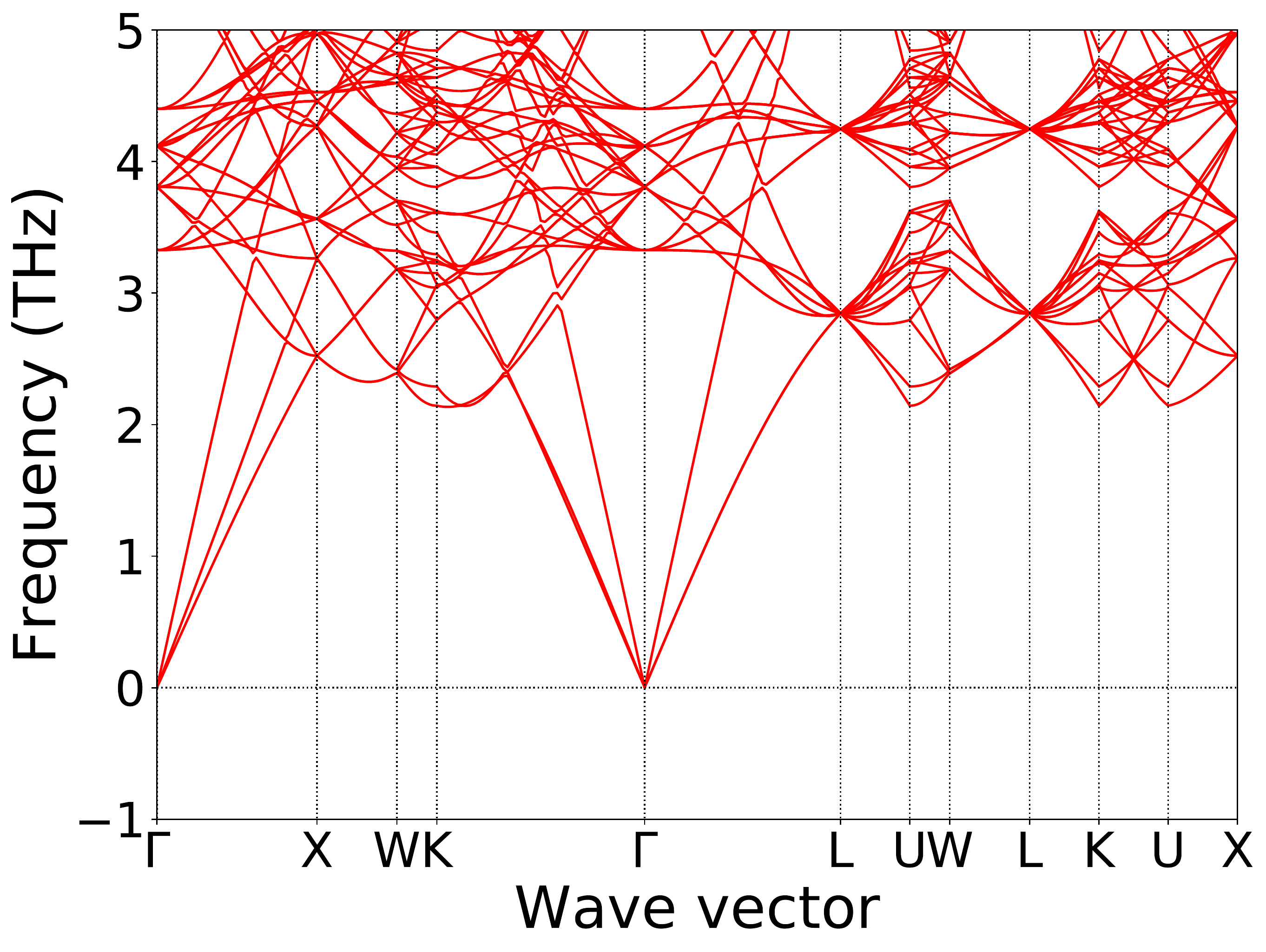}}\hspace{1cm}
    \subfigure[GAP]{\label{fig:gap_Si_phonon}\includegraphics[width=0.37\textwidth]{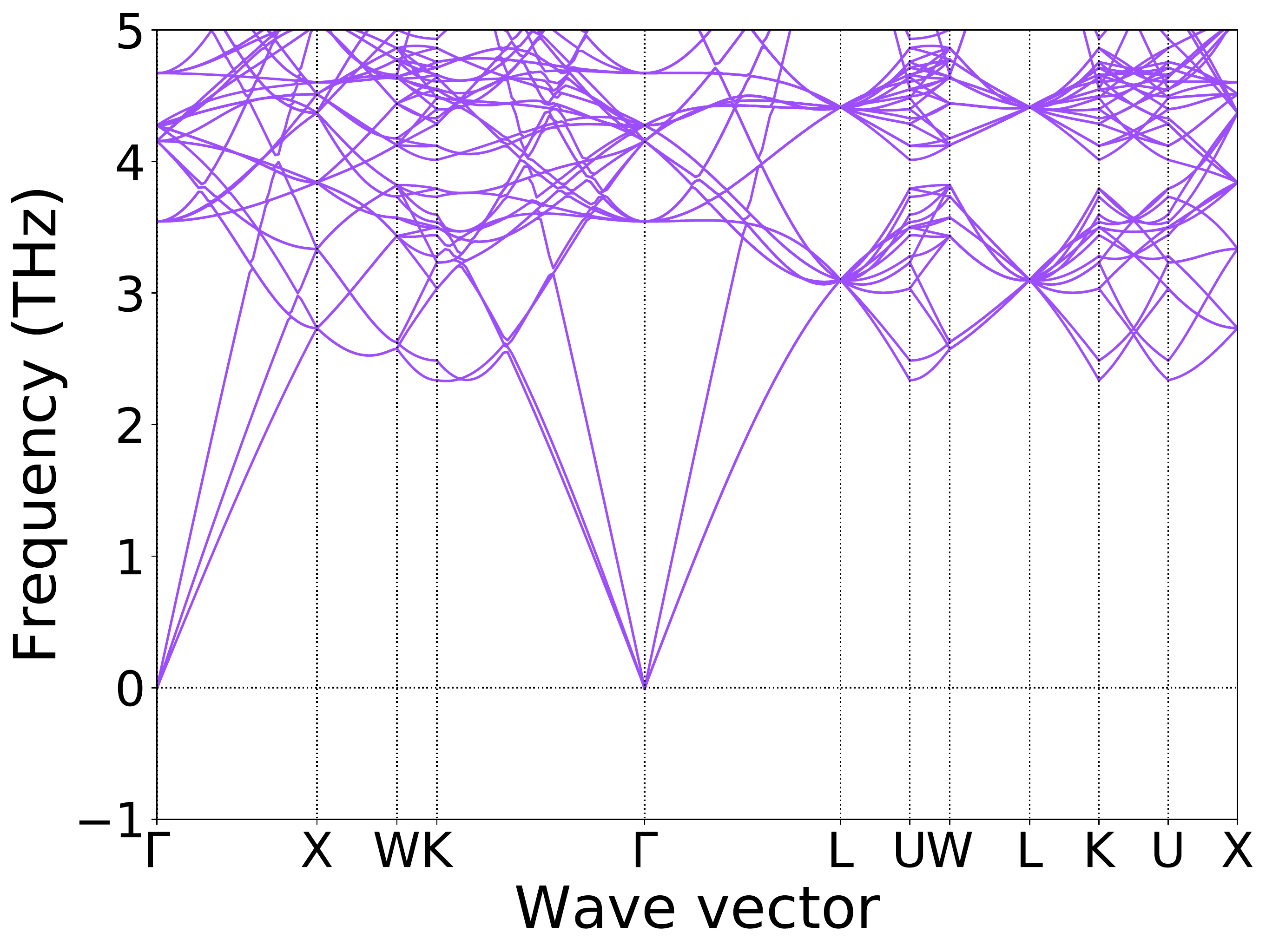}}
    \subfigure[MTP]{\label{fig:mtp_Si_phonon}\includegraphics[width=0.37\textwidth]{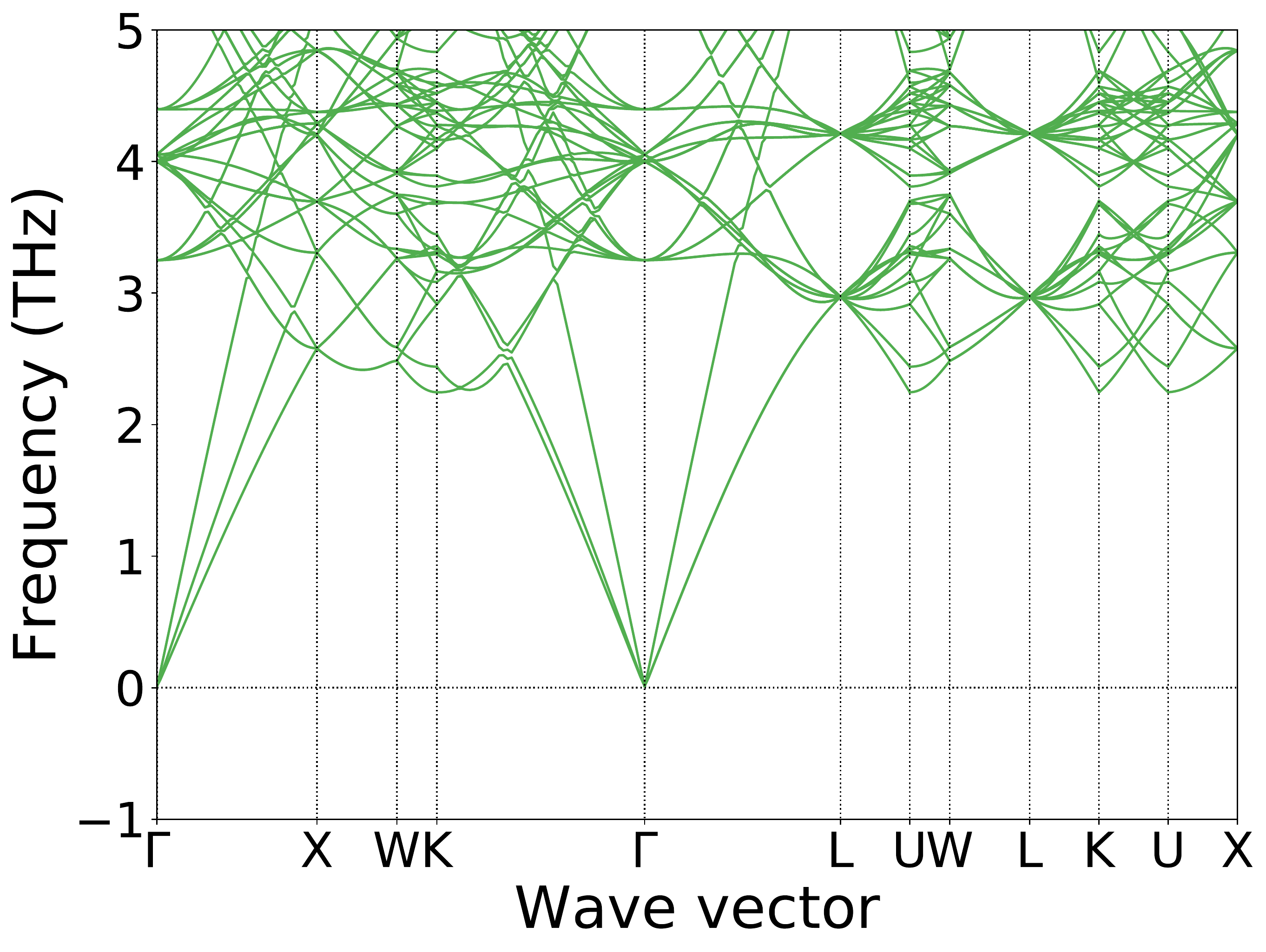}}\hspace{1cm}
    \subfigure[NNP]{\label{fig:nnp_Si_phonon}\includegraphics[width=0.37\textwidth]{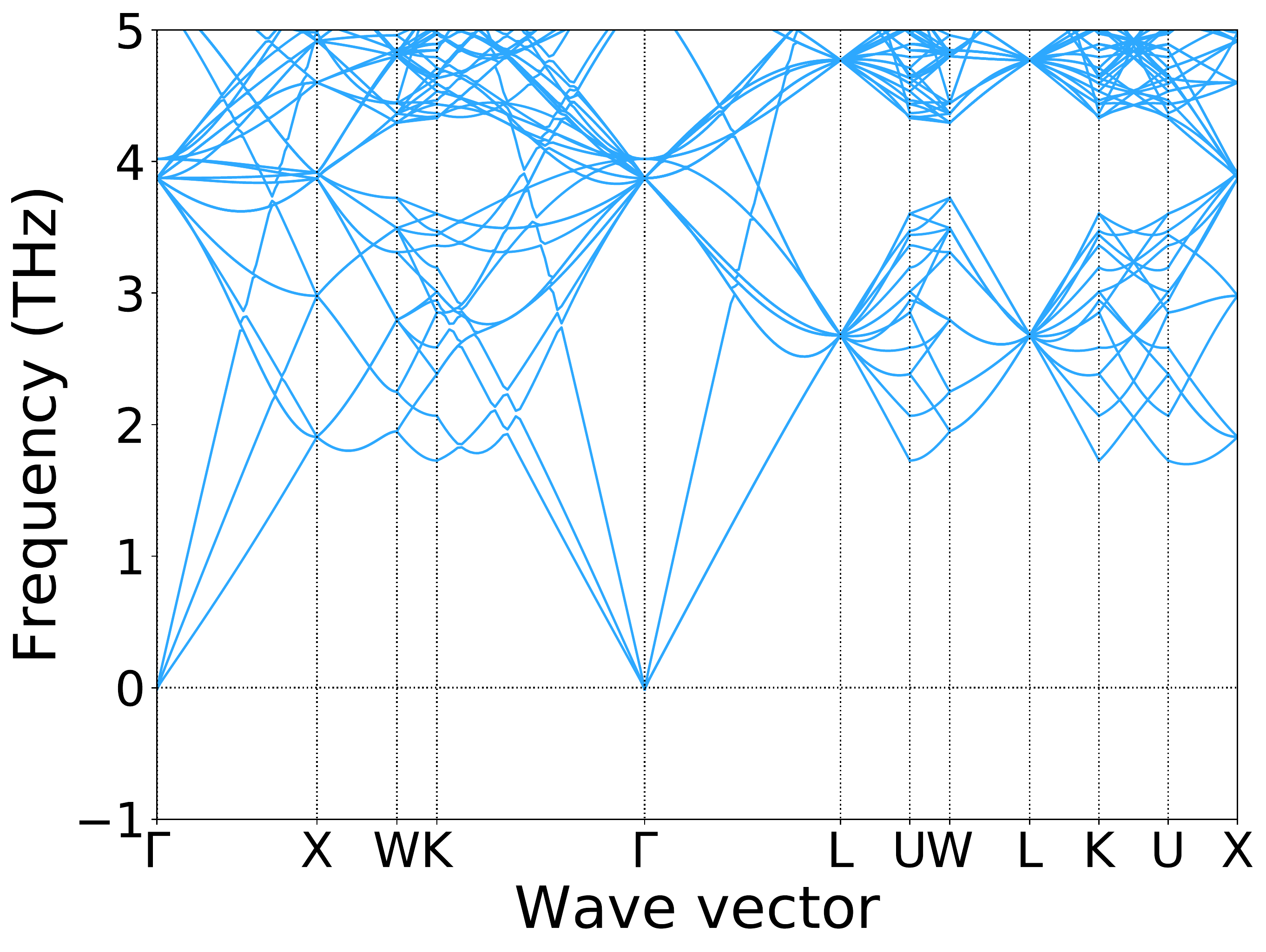}}
    \subfigure[SNAP]{\label{fig:snap_Si_phonon}\includegraphics[width=0.37\textwidth]{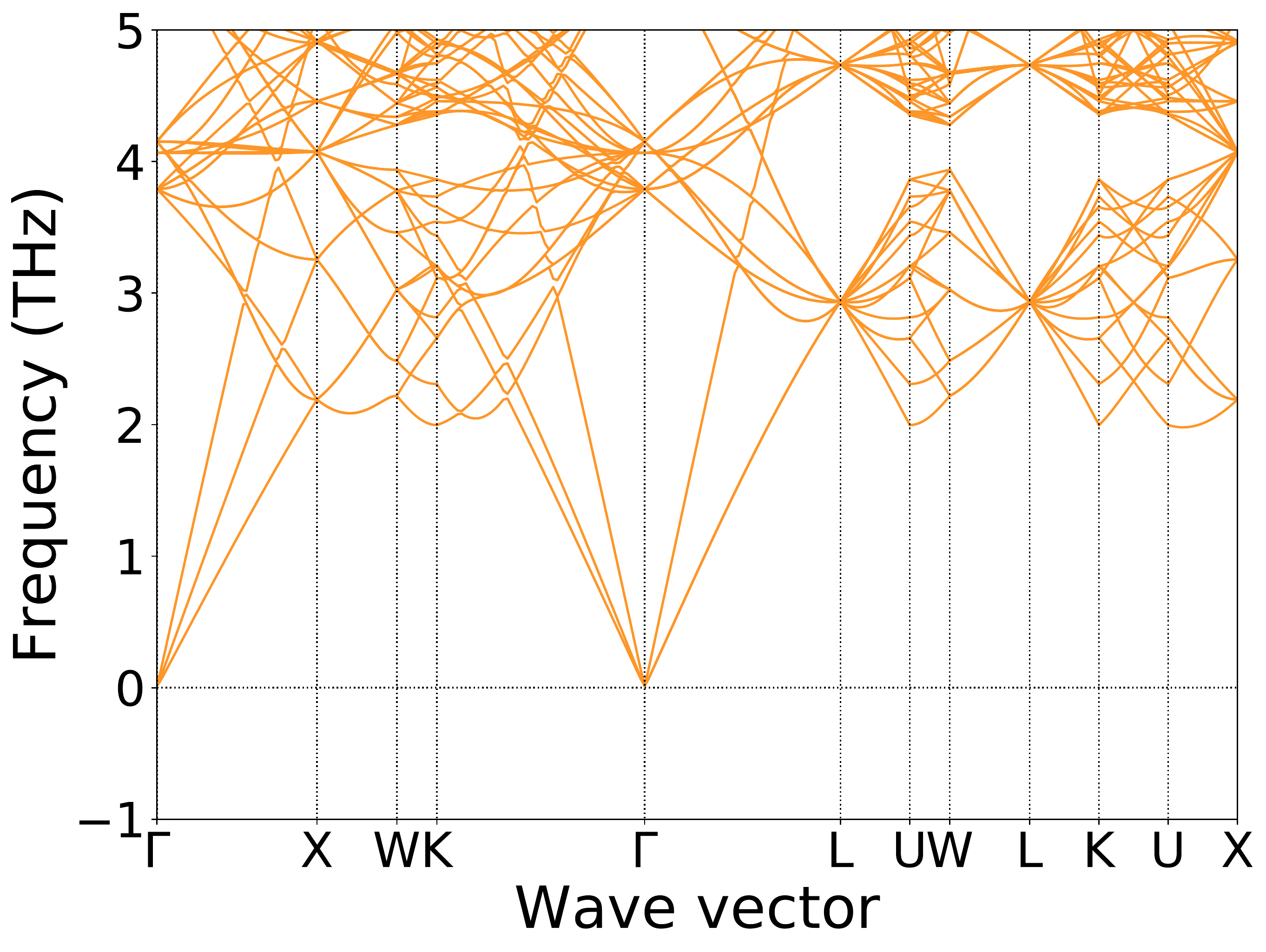}}\hspace{1cm}
    \subfigure[qSNAP]{\label{fig:q_snap_Si_phonon}\includegraphics[width=0.37\textwidth]{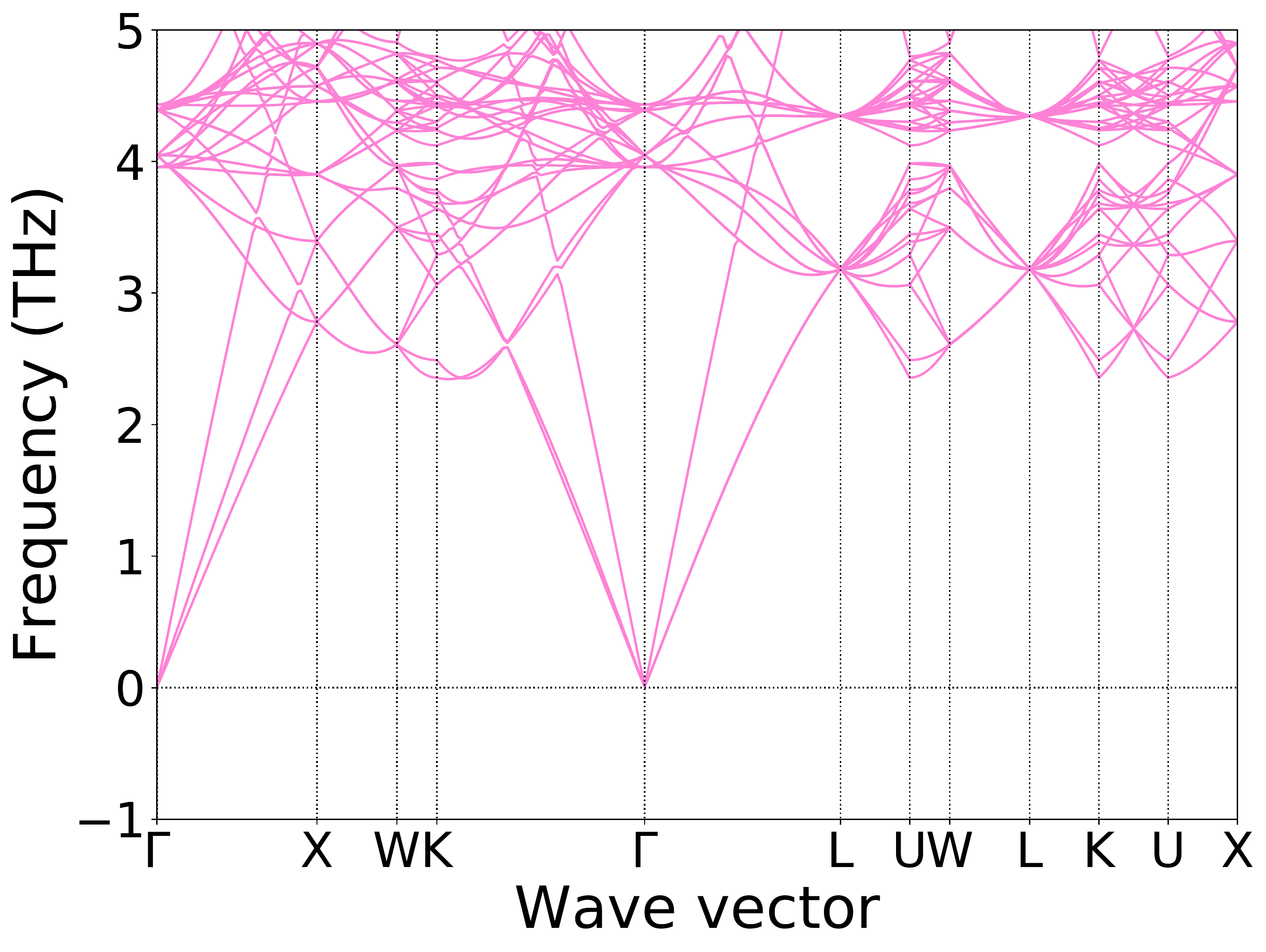}}
    \caption{Phonon dispersion curves for 64-atom diamond Si supercell via (a) DFT (b) GAP (c) MTP (d) NNP (e) SNAP (f) qSNAP.}
    \label{fig:Si_phonon}
\end{figure}

\clearpage

\begin{figure}
    \centering
    \subfigure[DFT]{\label{fig:dft_Ge_phonon}\includegraphics[width=0.37\textwidth]{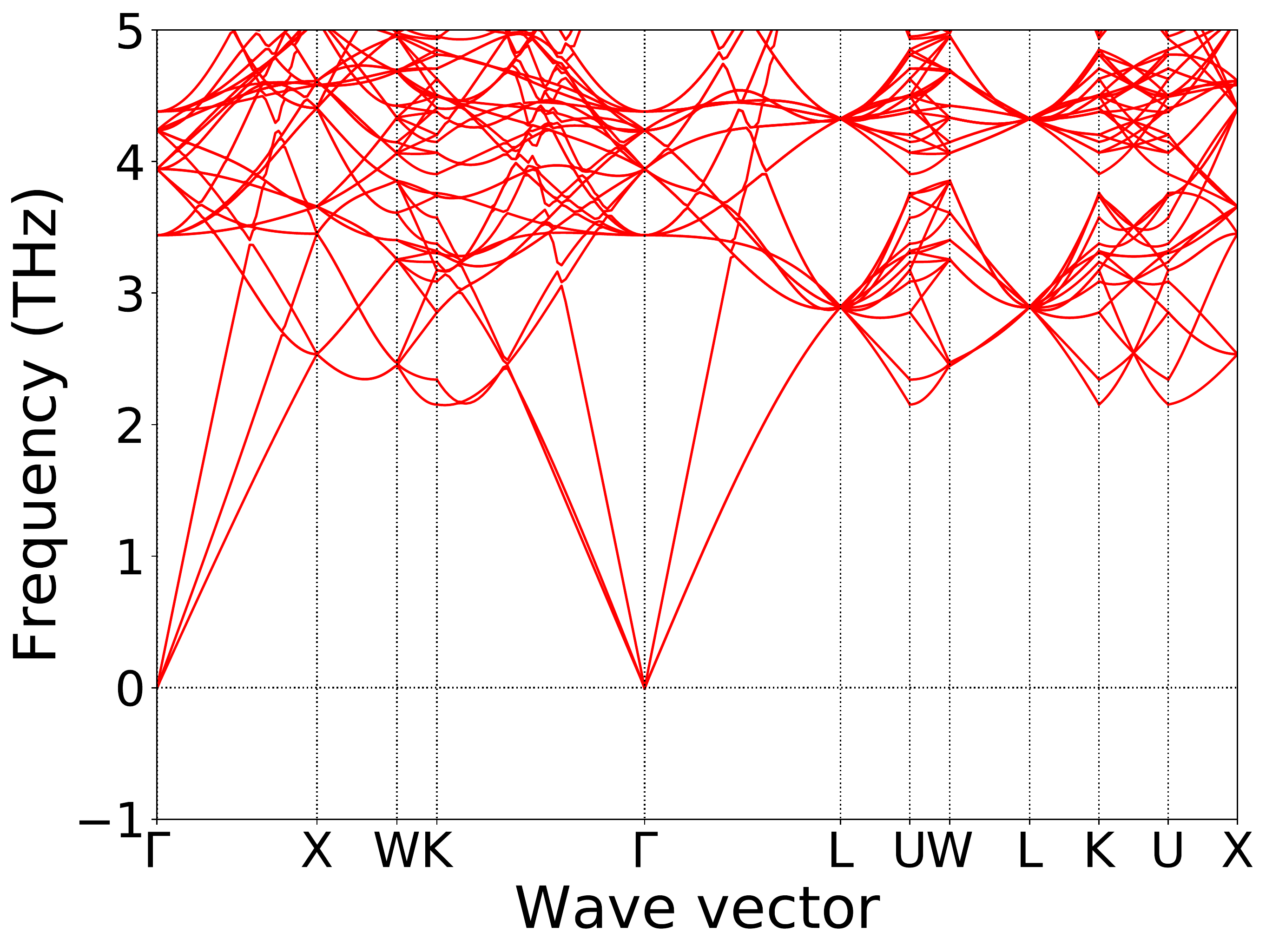}}\hspace{1cm}
    \subfigure[GAP]{\label{fig:gap_Ge_phonon}\includegraphics[width=0.37\textwidth]{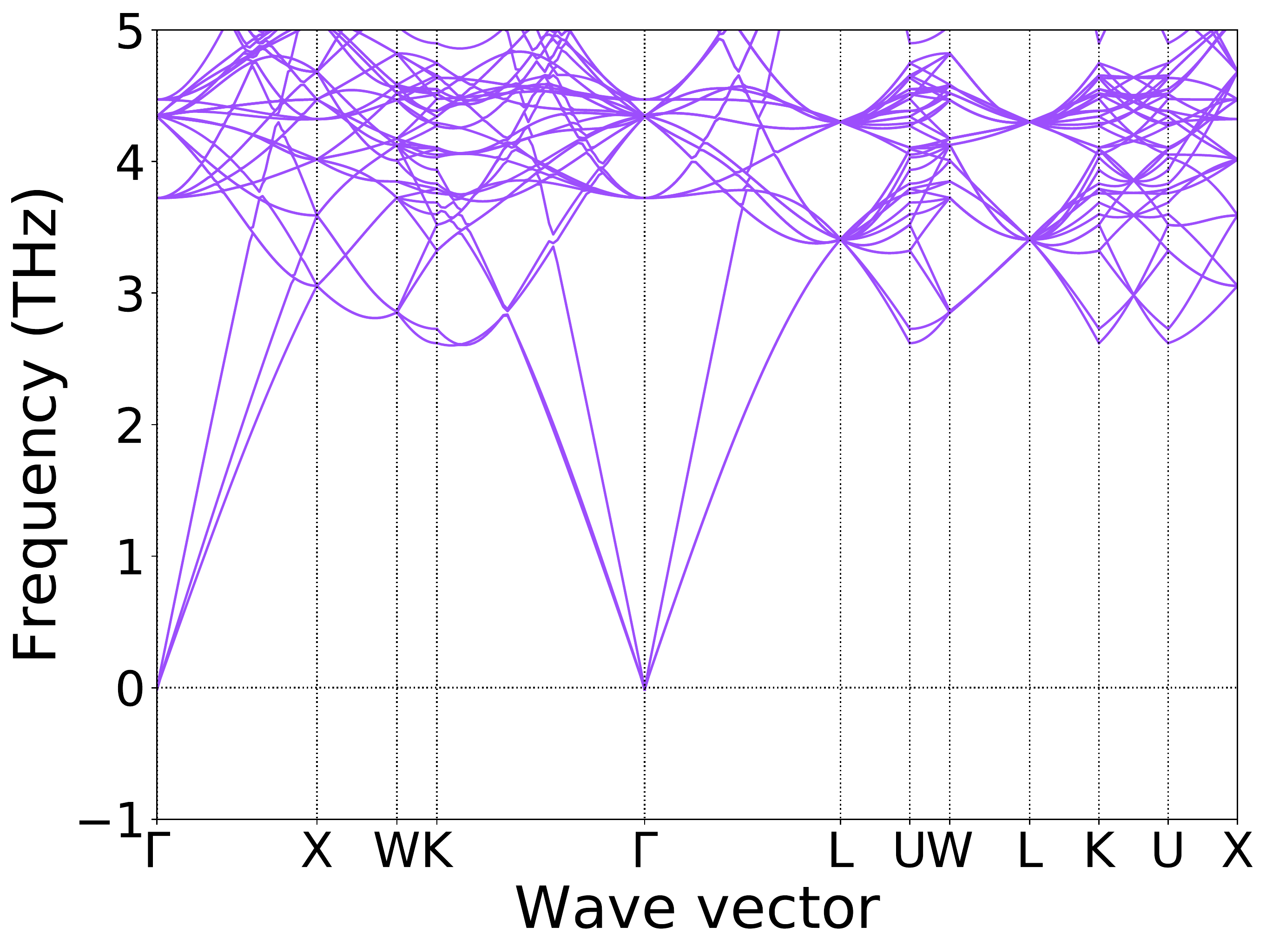}}
    \subfigure[MTP]{\label{fig:mtp_Ge_phonon}\includegraphics[width=0.37\textwidth]{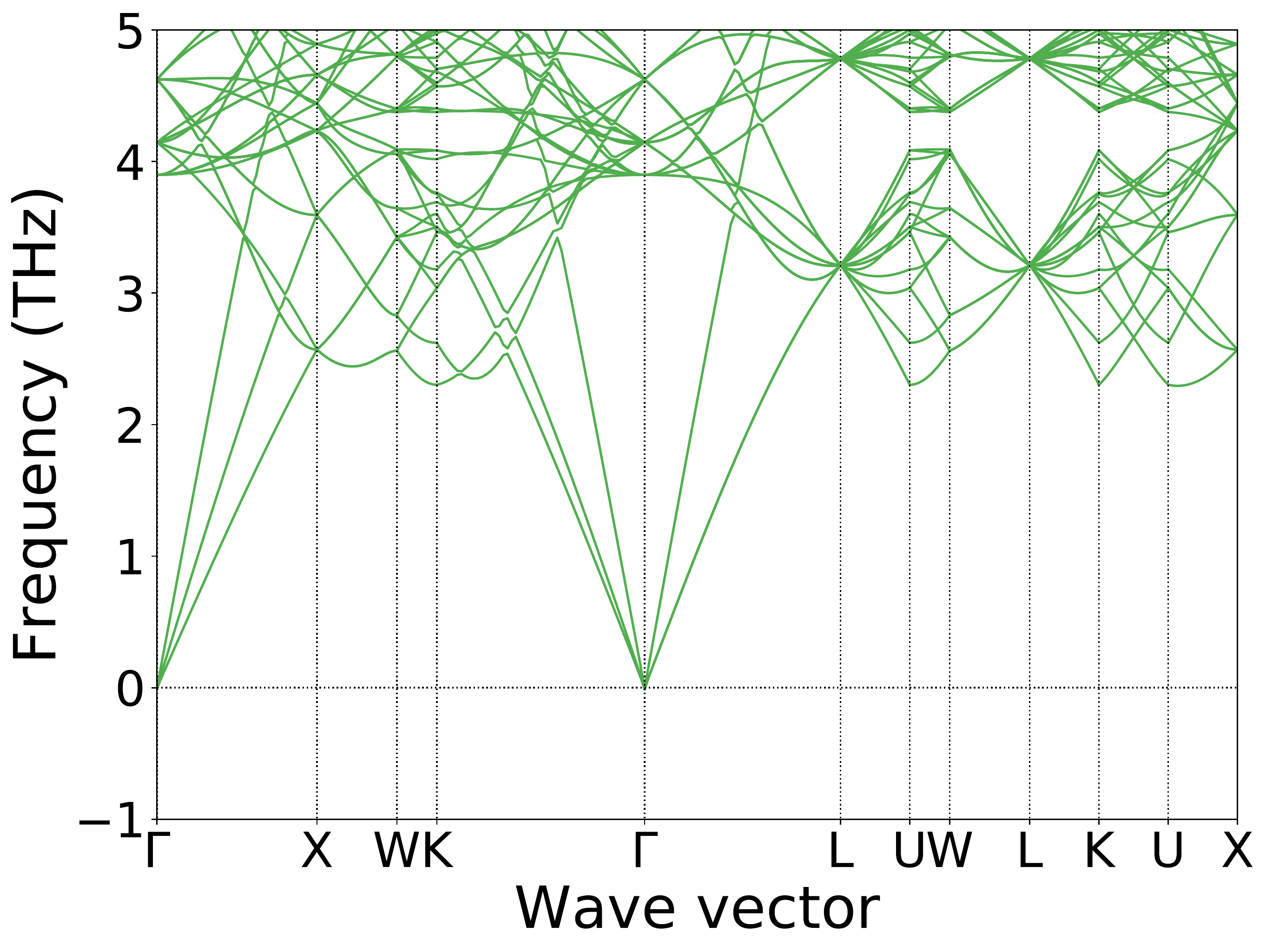}}\hspace{1cm}
    \subfigure[NNP]{\label{fig:nnp_Ge_phonon}\includegraphics[width=0.37\textwidth]{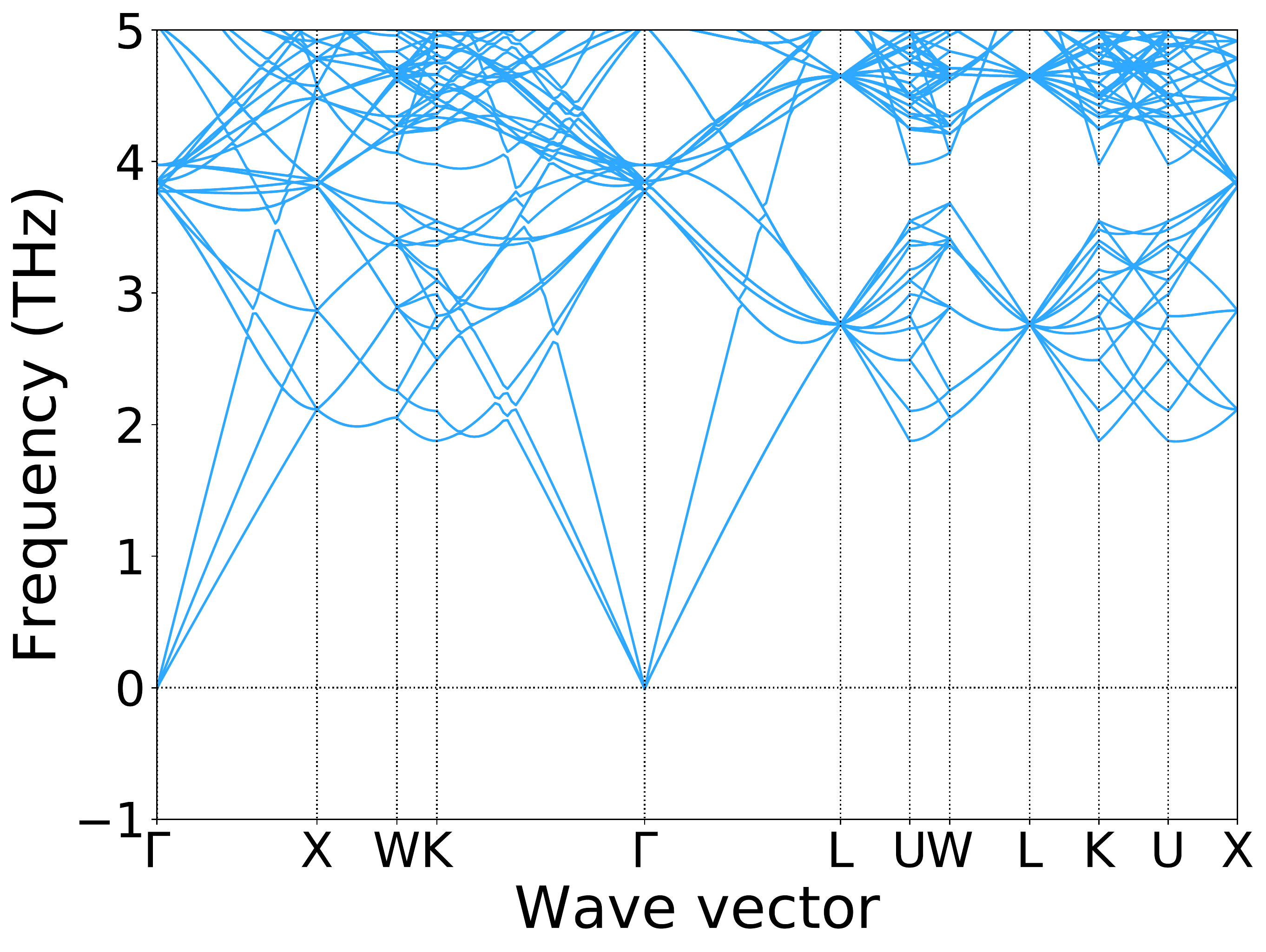}}
    \subfigure[SNAP]{\label{fig:snap_Ge_phonon}\includegraphics[width=0.37\textwidth]{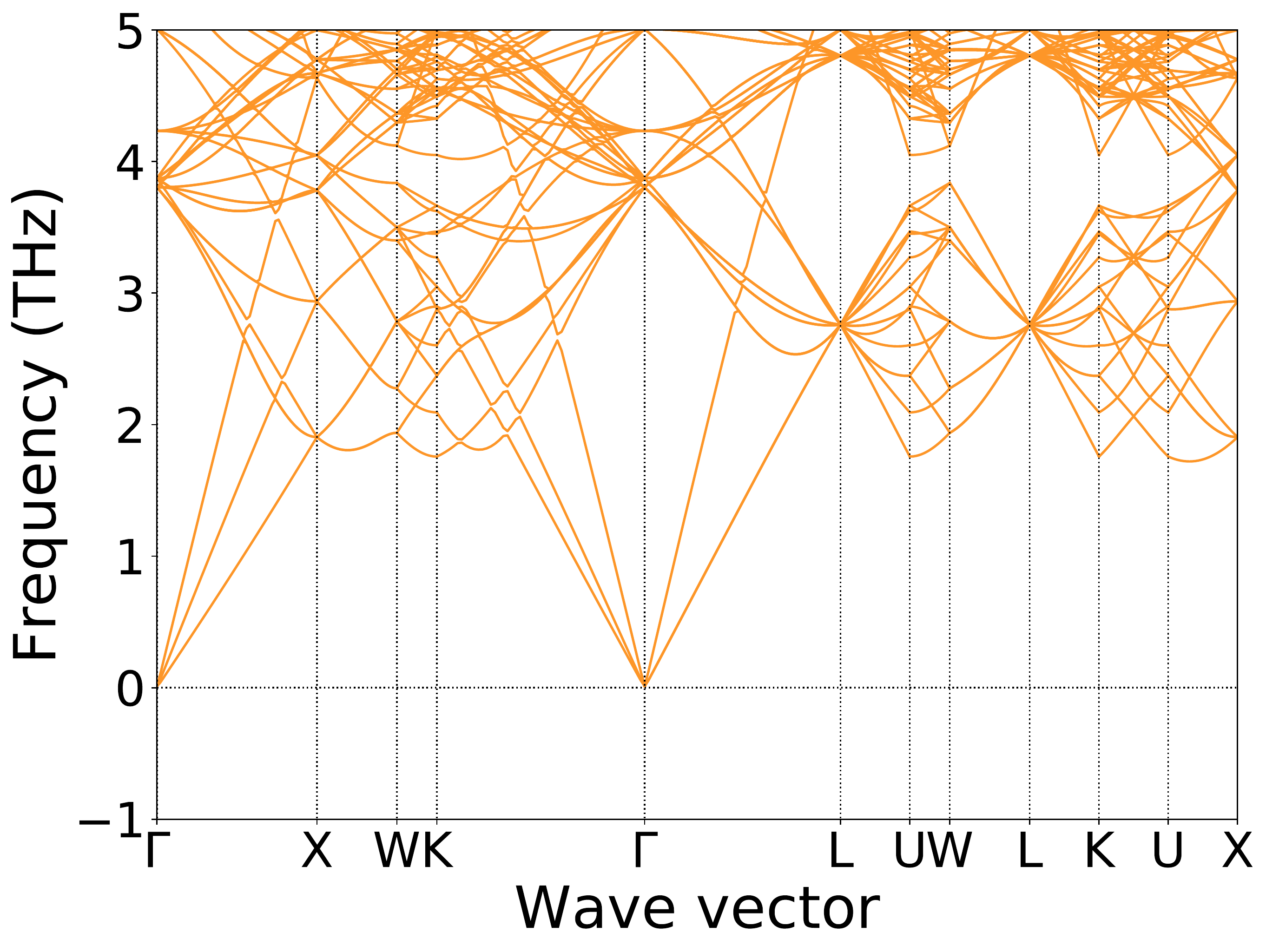}}\hspace{1cm}
    \subfigure[qSNAP]{\label{fig:q_snap_Ge_phonon}\includegraphics[width=0.37\textwidth]{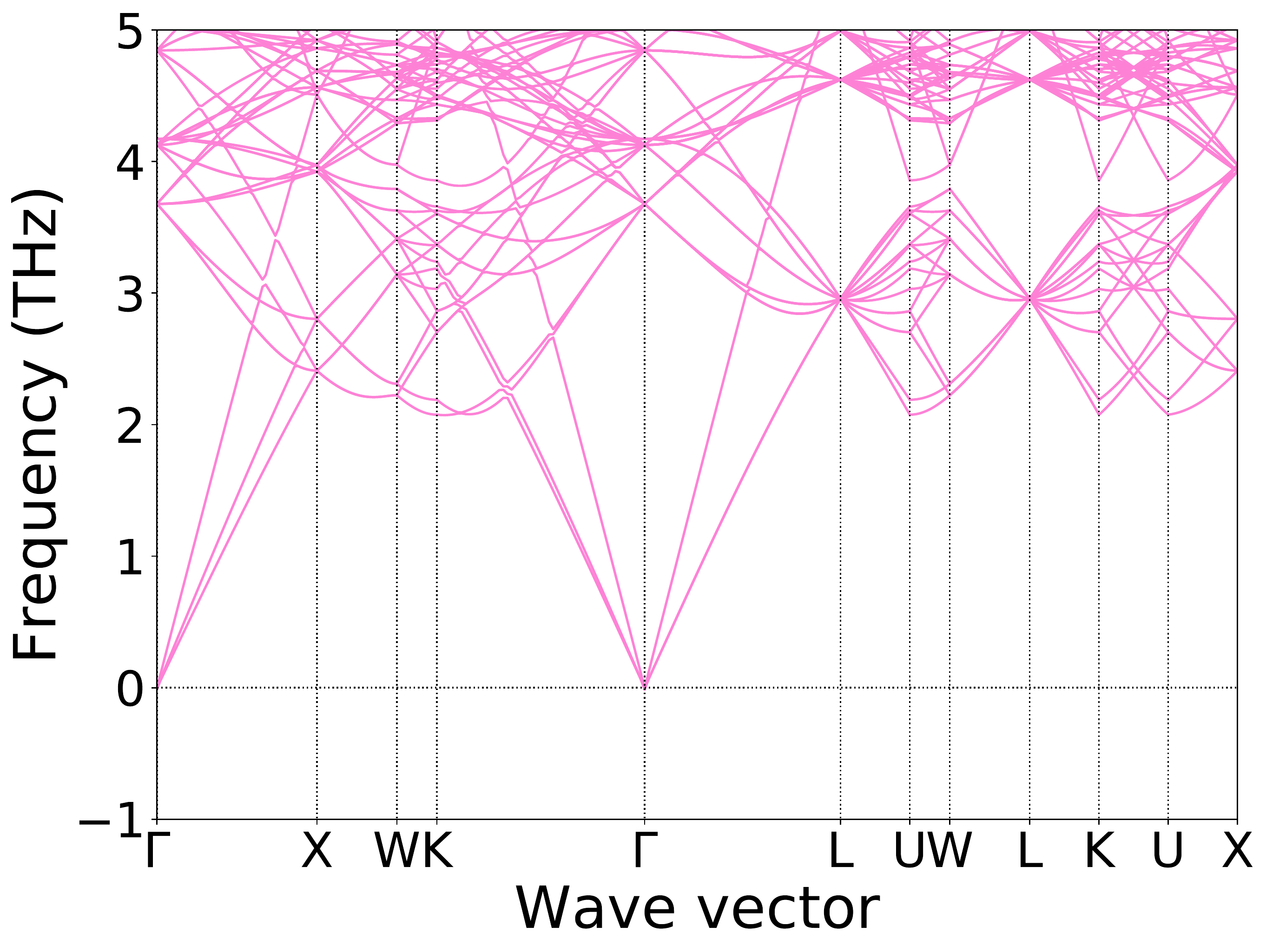}}
    \caption{Phonon dispersion curves for 64-atom diamond Ge supercell via (a) DFT (b) GAP (c) MTP (d) NNP (e) SNAP (f) qSNAP.}
    \label{fig:Ge_phonon}
\end{figure}

\clearpage

\section{Trade-off between prediction accuracy and computational cost}

All timings were performed using LAMMPS calculations on a single CPU core of Intel i7-6850k 3.6 GHz.

\begin{figure}[htp]
    \centering
    \includegraphics[width=0.65\textwidth]{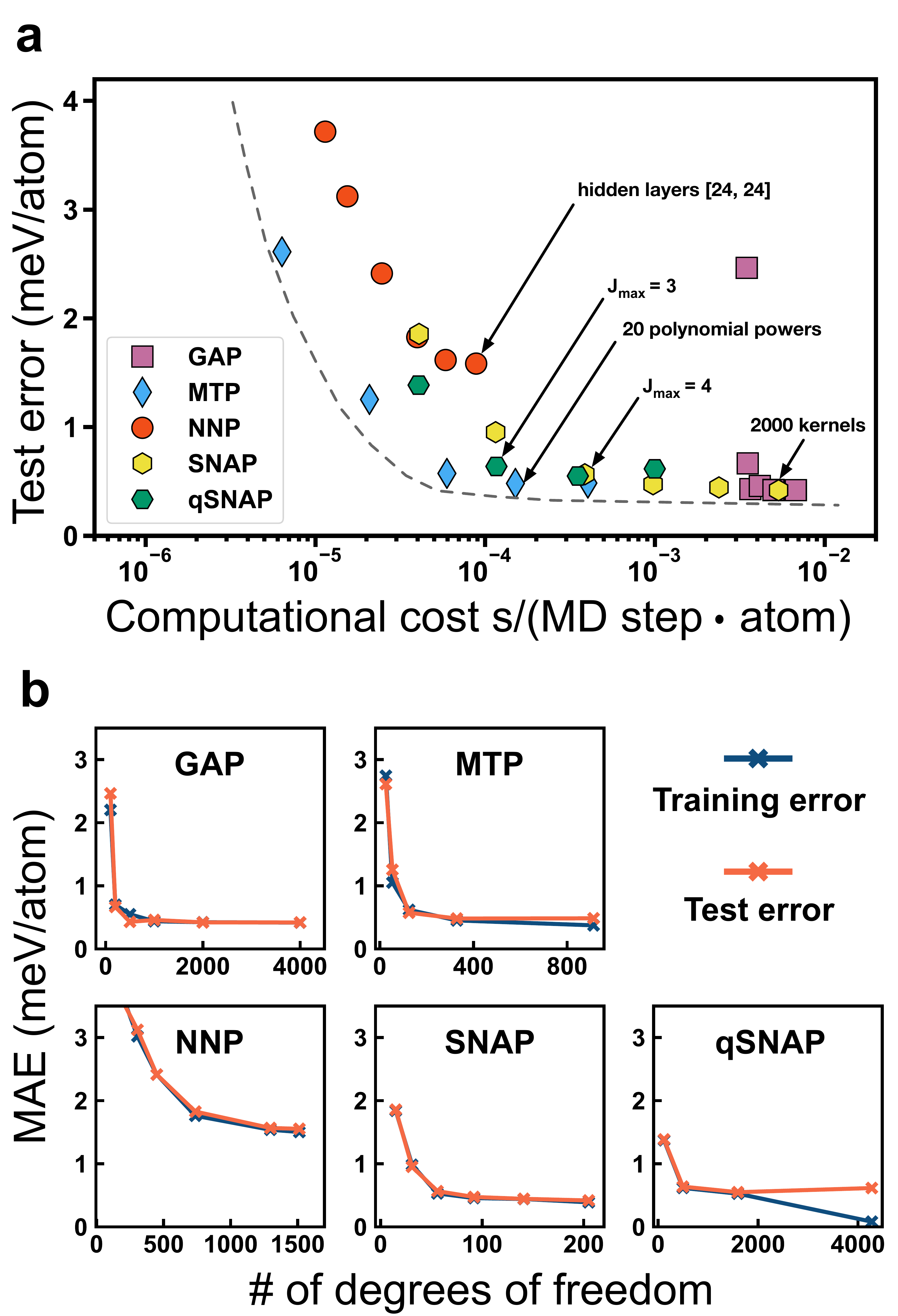}
    \caption{\textbf{a} Test error versus computational cost for fcc Ni system. An Pareto frontier represents boundary enclosing points of different ML-IAPs. \textbf{b} Comparisons between training error and test error versus the number of degrees of freedom for each ML-IAP.}
    \label{fig:Ni_time_benchmark}
\end{figure}

\pagebreak

\begin{figure}[htp]
    \centering
    \includegraphics[width=0.65\textwidth]{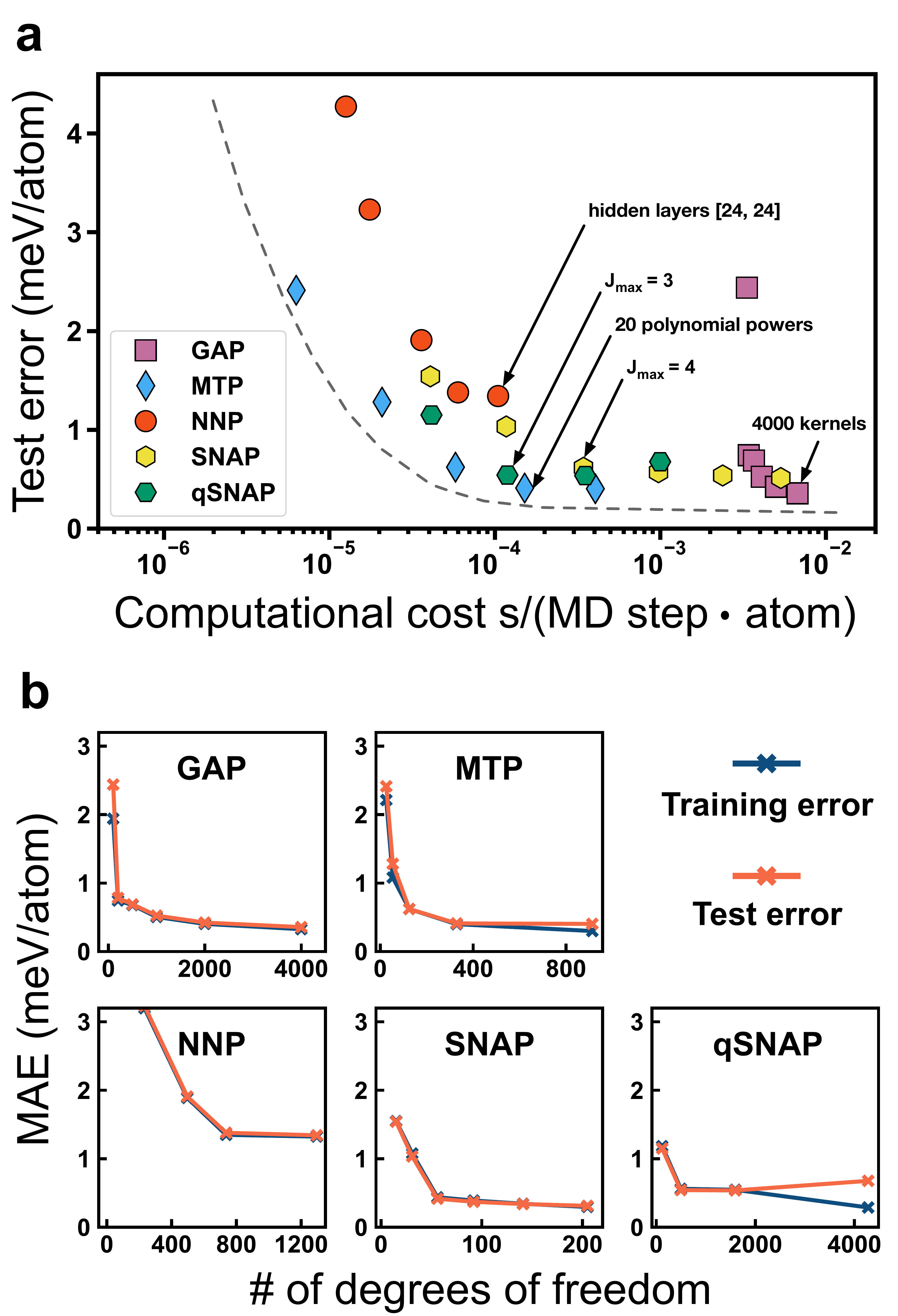}
    \caption{\textbf{a} Test error versus computational cost for fcc Cu system. An Pareto frontier represents boundary enclosing points of different ML-IAPs. \textbf{b} Comparisons between training error and test error versus the number of degrees of freedom for each ML-IAP.}
    \label{fig:Cu_time_benchmark}
\end{figure}

\pagebreak

\begin{figure}[htp]
    \centering
    \includegraphics[width=0.65\textwidth]{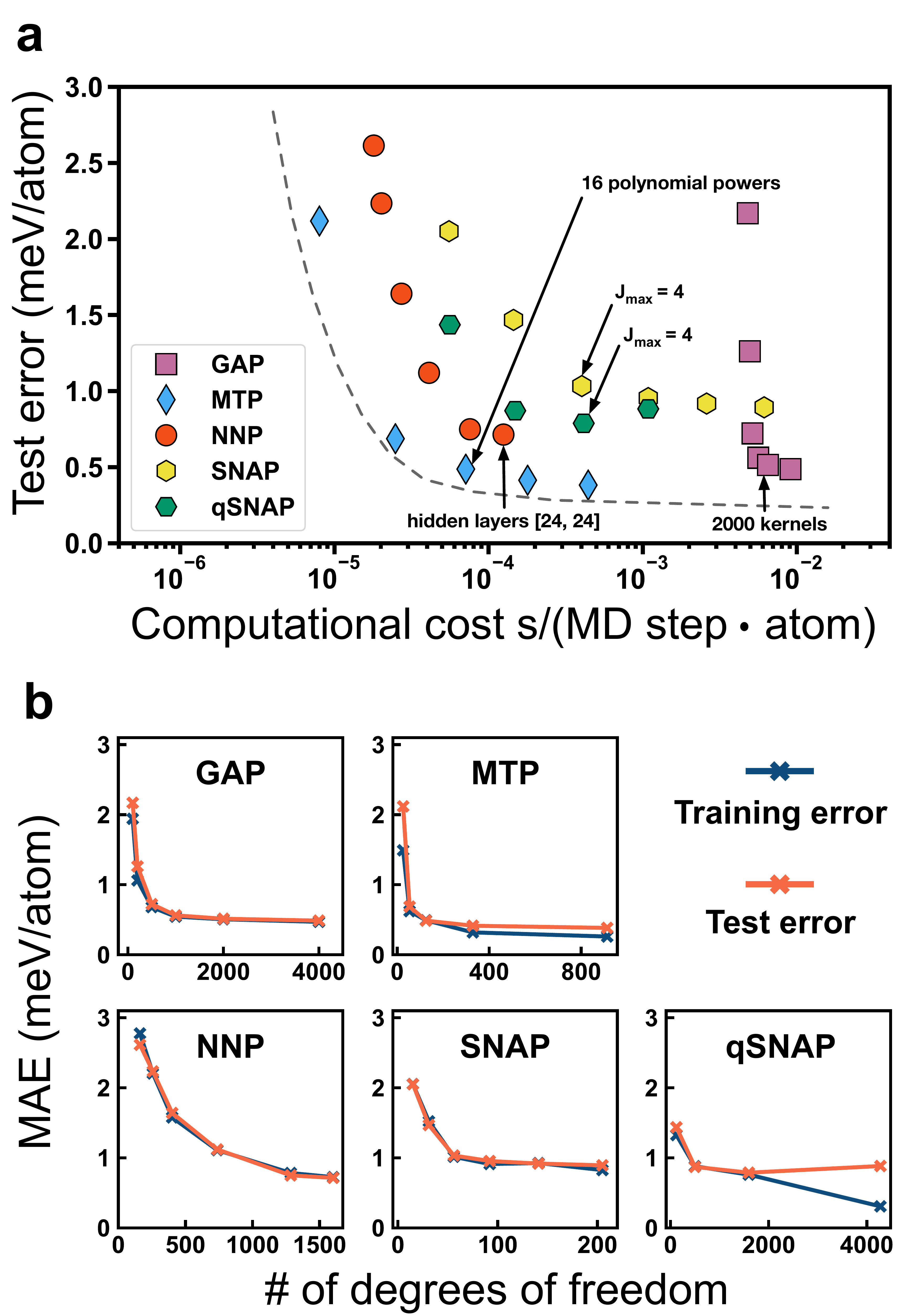}
    \caption{\textbf{a} Test error versus computational cost for bcc Li system. An Pareto frontier represents boundary enclosing points of different ML-IAPs. \textbf{b} Comparisons between training error and test error versus the number of degrees of freedom for each ML-IAP.}
    \label{fig:Li_time_benchmark}
\end{figure}

\pagebreak

\begin{figure}[htp]
    \centering
    \centering
    \includegraphics[width=0.65\textwidth]{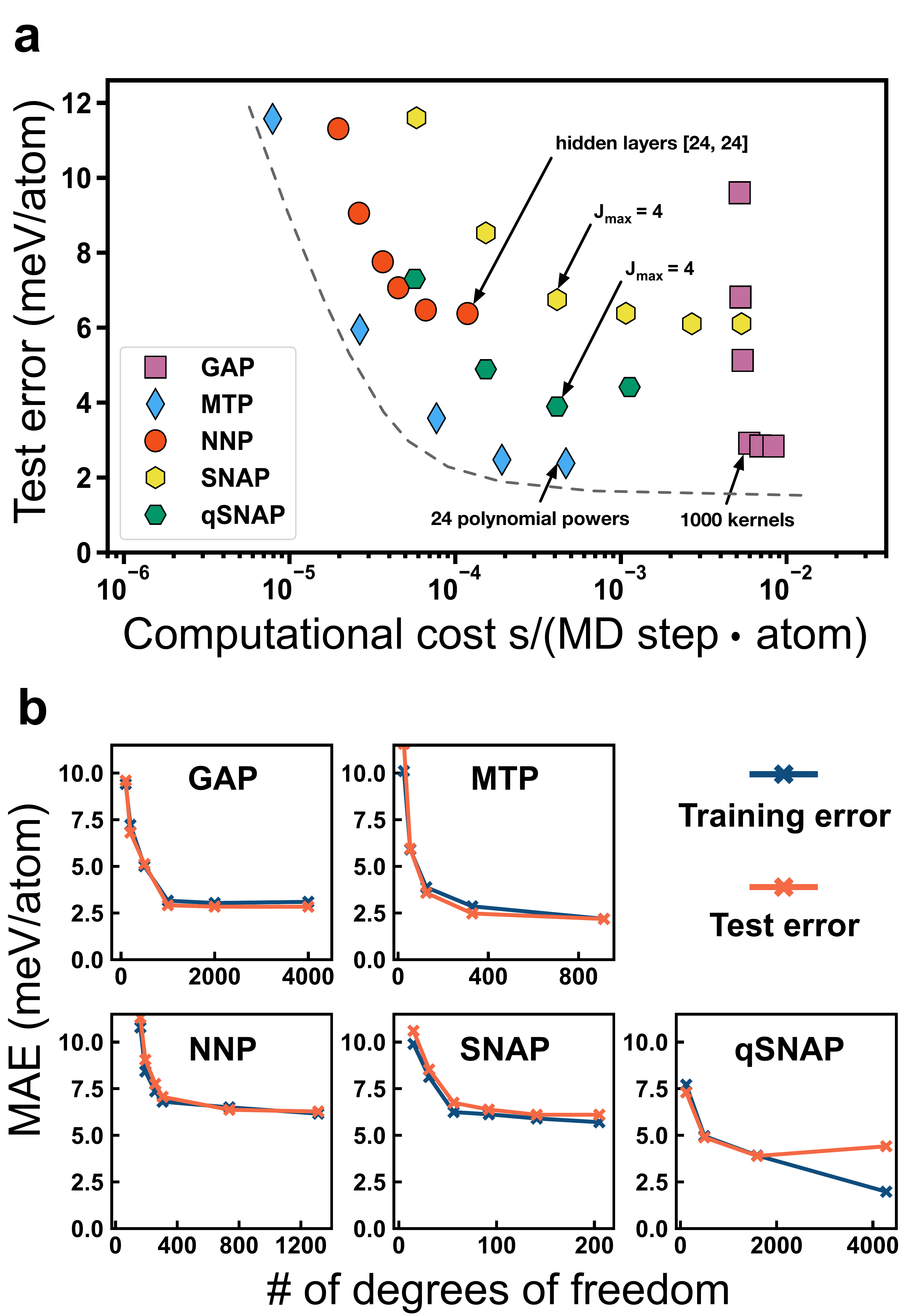}
    \caption{\textbf{a} Test error versus computational cost for diamond Si system. An Pareto frontier represents boundary enclosing points of different ML-IAPs. \textbf{b} Comparisons between training error and test error versus the number of degrees of freedom for each ML-IAP.}
    \label{fig:Si_time_benchmark}
\end{figure}

\pagebreak

\begin{figure}[htp]
    \centering
    \includegraphics[width=0.65\textwidth]{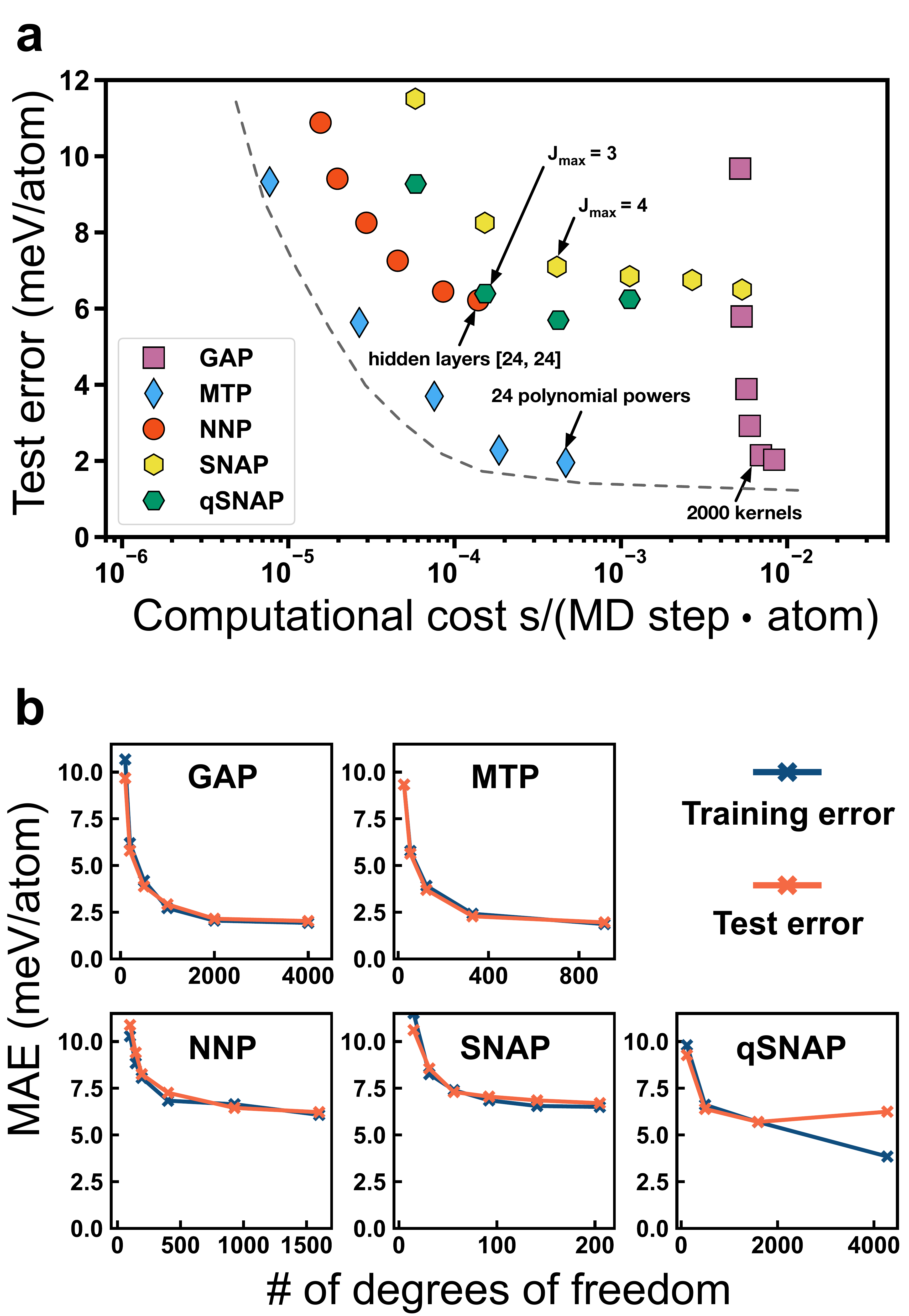}
    \caption{\textbf{a} Test error versus computational cost for diamond Ge system. An Pareto frontier represents boundary enclosing points of different ML-IAPs. \textbf{b} Comparisons between training error and test error versus the number of degrees of freedom for each ML-IAP.}
    \label{fig:Ge_time_benchmark}
\end{figure}

\clearpage

\section{Linear scaling relationship of ML-IAPs}

To investigate the scaling effect on the computational cost of different ML-IAPs, we perform npt MD simulations at a constant temperature of 300 K on structures with different size including $6\times6\times6$, $10\times10\times10$, $14\times14\times14$, and $18\times18\times18$ supercells, corresponding to 432, 2000, 5488, and 11664 atoms of bulk Mo, respectively. The computational costs are presented in Table \ref{table:scaling_effect}. All ML-IAPs show greatly linear scaling relationship with respect to the number of atoms.

\begin{center}
\begin{table}[h]
\begin{tabular}{  M{3.5cm}  M{1.5cm}  M{1.5cm}  M{1.5cm}  M{1.5cm} }
\hline
\hline
\noalign{\smallskip}
No. of atoms    &   432 &  2000  &   5488  &   11664  \\
\noalign{\smallskip}
\hline
\noalign{\smallskip}
GAP    &   1.455 &   6.711 &   18.519    &  38.461 \\
\hline
\noalign{\smallskip}
MTP    &   0.069 &   0.308 &   0.843    &  1.799 \\
\hline
\noalign{\smallskip}
NNP    &   0.052 &   0.236 &   0.674    & 1.439 \\
\hline
\noalign{\smallskip}
SNAP    &   0.063 &   0.294 &   0.809    &  1.701 \\
\hline
\noalign{\smallskip}
qSNAP    &   0.063 &   0.299 &   0.805    &  1.730 \\
\hline
\end{tabular}
\caption{The scaling effect on computational cost of different ML-IAPs. The unit of computational cost is seconds per MD step. Timings were performed by LAMMPS calculations on a single CPU core of Intel i7-6850k 3.6 GHz.}
\label{table:scaling_effect}
\end{table}
\end{center}

\pagebreak

\bibliography{new_collection}